\documentclass[%
 reprint,
 superscriptaddress,
nofootinbib,
 amsmath,amssymb,
 aps,
longbibliography,
prx,
]{revtex4-1}

\DeclareMathOperator{\Tr}{Tr}
\usepackage{graphicx}
\usepackage{subfigure}
\usepackage{multirow}
\renewcommand{\thesubfigure}{\alph{subfigure}}

\makeatletter
\renewcommand{\@thesubfigure}{(\thesubfigure)\space}

\makeatother
\usepackage{dcolumn}
\usepackage{bm}
\usepackage[colorlinks, linkcolor=blue, citecolor=blue, urlcolor=blue]{hyperref}

\usepackage{slashed}

\begin{document}


\title{Entanglement Entropy of $(2+1)$D Quantum Critical Points with Quenched Disorder: \\ Dimensional Reduction Approach}

\author{Qicheng Tang}
\email{tangqicheng@westlake.edu.cn}
\affiliation{%
	Department of Physics, Zhejiang University, Hangzhou 310058, China
}%
\affiliation{School of Science, Westlake University, Hangzhou 310024, China }

\author{W. Zhu}
\email{zhuwei@westlake.edu.cn}

\affiliation{School of Science, Westlake University, Hangzhou 310024, China }

\date{\today}


\begin{abstract}
A formidable perspective in understanding quantum criticality of a given many-body system is through its entanglement contents. Until now, most progress are only limited to the disorder-free case. 
Here, we develop an efficient scheme to compute the entanglement entropy of $(2+1)$-dimensional quantum critical points with randomness, from a conceptually novel angle where the quenched disorder can be considered as dimensionally reducible interactions. 
As a concrete example, we reveal novel entanglement signatures of  $(2+1)$-dimensional Dirac fermion exposed to a random magnetic field, which hosts a class of emergent disordered quantum critical points. 
We demonstrate that the entanglement entropy satisfies the area-law scaling, and observe a modification of the area-law coefficient that points to the emergent disordered quantum criticality. 
Moreover, we also obtain the sub-leading correction to the entanglement entropy due to a finite correlation length. 
This sub-leading correction is found to be a universal function of the correlation length and disorder strength. 
We discuss its connection to the renormalization group flows of underlying theories. 
%
%
\end{abstract}

\maketitle


\section{Introduction}

Entanglement expresses non-local connotations inherent to quantum mechanics, which has prompted remarkable insights into various fields of modern physics, 
bridging microscopic laws in quantum matters~\cite{Amico2008, Eisert2010} and macroscopic structure of space-time~\cite{Bombelli1986_area, Srednicki1993_area, Callan1994GeometricEntropy, KABAT1995, RTformula2006,  swingle2012holography, takayanagi2017holographic, Tatsuma2018}. 
Compared to the traditional methods, the study of many-body wave function from the perspective of quantum entanglement can unveil novel properties in a large variety of collective quantum phenomena, ranging from the presence of topological order~\cite{kitaev2006TopoEE, levin2006TopoEE,LiHaldane2008_ES, Qi2012ES} to the onset of quantum criticality~\cite{HOLZHEY1994443, vidal2003entanglement, Calabrese_2004,  fradkin2006EE2p1Critical, william2015cornerEE}. 
Indeed, the overwhelming majority of works done so far are in support of entanglement-based analysis as a profitable tool to diagnose strong correlations for both in and out-of equilibrium systems~\cite{Calabrese_2005, Calabrese_2009_review, laflorencie2016quantum, XiaoGang2017RevModPhys, pollmann2018_OTOC_rand, altman2019Review, NatRevPhys2019dynamics}. 

A simple way to analyze the entanglement structure  is to separate a target system into  subsystem $A$ and its complement $\overline A$, then a measure of the entanglement between $A$ and $\overline A$ is given by the von Neumann entropy associated with reduced density matrix $\rho_A$: $S=-\Tr [\rho_A \ln \rho_A]$, which is also referred as the \textit{entanglement entropy} (EE). 
Intriguingly, the EE is typically not an extensive quantity for many-body ground states, instead it usually satisfies an \textit{area-law}~\cite{Bombelli1986_area, Srednicki1993_area}. 
That is, the EE is proportional to the area of surface separating two subsystems, in sharp contrast with the thermal entropy that should obey the volume-law. 
The emergent area-law EE partially reflects a decay of correlation associated with quantum many-body states~\cite{nahum2020_majorana_defect, tang2021nonunitary}.
Especially, in $(1+1)$ dimension, the area-law is a character of massive theories with exponentially decaying correlations~\cite{hastings2007EE, cirac2008AreaLaw, brandao2013AreaLaw, cho2018AreaLaw}, and the logarithmic correction on the EE is expected for critical systems~\cite{HOLZHEY1994443, vidal2003entanglement, Calabrese_2004}. 
In higher dimensions, the area-law is believed to be generally hold in quantum field theories (QFTs), as a consequence of the locality of physical interactions~\cite{CasiniHuerta_2009, Eisert2010, Tatsuma2018}. 
Such strong restriction of the entanglement alludes a deep connection with black hole physics~\cite{RTformula2006, raamsdonk2010_spacetime_entanglement, faulkner2014_holographic_CFT, swingle2014universality}, and also offers crucial implications on the numerical computations on lattice models~\cite{white1992dmrg, verstraete2004PEPS, vidal2008EfficientSimulate, Schollwock2011_age_mps, laflorencie2016quantum}, thus it is of vital importance.

\begin{table*}
	\caption{\label{tab:method_EE}
		A summary of existing methods on calculating EE.}
	\begin{tabular}{|cc|c|c|}
		\hline
		\multicolumn{2}{|c|}{\multirow{2}{*}{\textbf{Method}}} & {\multirow{2}{*}{\textbf{Advantages}}} & {\multirow{2}{*}{\textbf{Limitations}}} \\ 
		\multicolumn{2}{|c|}{} & &  \\ \hline
		\multicolumn{1}{|c|}{\multirow{3}{*}{\begin{tabular}[c]{@{}c@{}}\\ \\ Real-Time\\ Approach\end{tabular}}} & \begin{tabular}[c]{@{}c@{}} \\[-1.5ex] Numerical \\ determination of $\rho_A$ \end{tabular} & \begin{tabular}[c]{@{}c@{}} \\[-1.5ex] Applicable to any theory with \\ discretization on a lattice \end{tabular}  & \begin{tabular}[c]{@{}c@{}} \\[-1.5ex] Exponential growth of computational \\ complexity in non-integrable systems\end{tabular} \\ \cline{2-4} 
		\multicolumn{1}{|c|}{} & \begin{tabular}[c]{@{}c@{}} \\[-1.5ex] Correlation matrix\\ technique~\cite{chung2001density, Peschel_2003calculation, Peschel_2009reduced}\end{tabular} & \begin{tabular}[c]{@{}c@{}} \\[-1.5ex] Polynomial computational \\complexity in system size\end{tabular} & Restricted to the Gaussian states \\ \cline{2-4} 
		\multicolumn{1}{|c|}{} & Resolvent technique~\cite{Casini_2009_resolvent_1p1dirac} & \begin{tabular}[c]{@{}c@{}} \\[-1.5ex] Applicable to analytical \\ solution with multi-regions\end{tabular} & \begin{tabular}[c]{@{}c@{}} \\[-1.5ex] Restricted to $(1+1)$D free massless \\ fermions and chiral bosons   \end{tabular}   \\ \hline
		\multicolumn{1}{|c|}{\multirow{-2}{*}{\begin{tabular}[c]{@{}c@{}} \\ \\[-1.5ex] \ Euclidean-Time \ \\ Approach\\ (Replica Trick) \\ \cite{Callan1994GeometricEntropy, Calabrese_2004, CasiniHuerta_2009} \end{tabular}}} & \begin{tabular}[c]{@{}c@{}} \\[-1.5ex] Heat-kernel technique \\ \cite{KABAT1995, casini2010_EE_sphere, Solodukhin2011, Myers2013_MassiveEE} \end{tabular} & \begin{tabular}[c]{@{}c@{}} \\[-1.5ex] Applicable to analytical solution \\   \end{tabular} & 
		\begin{tabular}[c]{@{}c@{}} \\[-1.5ex] Restricted to the quadratic order in \\ quantum fluctuations   \end{tabular}  \\ \cline{2-4} 
		\multicolumn{1}{|c|}{} & \begin{tabular}[c]{@{}c@{}} \\[-1.5ex] Green's function \\ technique~\cite{Calabrese_2004, CasiniHuerta_2009} \end{tabular} & \begin{tabular}[c]{@{}c@{}} \\[-1.5ex] \  Applicable to analytical solution; \ \\ \\[-2ex]  
			Applicable to higher dimensions \end{tabular} & 
		\begin{tabular}[c]{@{}c@{}} \\[-1.5ex] Capability and feasibility in interacting \\ theories are yet to be explored  \end{tabular}  \\ \hline
		\multicolumn{2}{|c|}{CFT Approach~\cite{HOLZHEY1994443, Calabrese_2004, cardy2008_form_factor, casini2010_EE_sphere, casini2011towards}} & \begin{tabular}[c]{@{}c@{}} \\[-1.5ex] \ Applicable to universal prediction \ \\ of EE in (1+1)D critical systems\end{tabular} & \begin{tabular}[c]{@{}c@{}} \\[-1.5ex] Hard to be extended into massive \\ theories and higher dimensions\end{tabular} \\ \hline
		\multicolumn{2}{|c|}{Holographic Approach~\cite{RTformula2006,RT2006Aspects,takayanagi2017holographic}} & \begin{tabular}[c]{@{}c@{}} \\[-1.5ex] Reduced to a geometric problem \end{tabular} & \begin{tabular}[c]{@{}c@{}} \\[-1.5ex] Limited by poor knowledge on the \\ gravitational dual of given QFTs \end{tabular} \\ \hline
		\multicolumn{2}{|c|}{ \begin{tabular}[c]{@{}c@{}} \\[-1.5ex] Extensive mutual information model~\cite{Casini_2009_EMI, pablo2021_EMI} \\ (a quasiparticle picture of entangled pairs)~\cite{nahum2020_majorana_defect, tang2021nonunitary} \end{tabular} } & \begin{tabular}[c]{@{}c@{}} \\[-1.5ex] Reduced to a geometric problem \\ (much simpler than holography) \end{tabular} & \begin{tabular}[c]{@{}c@{}} \\[-1.5ex] Does not correspond to an actual \\ CFT beyond $(1+1)$ dimensions \end{tabular} \\ \hline
		\multicolumn{1}{|c|}{\multirow{2}{*}{\begin{tabular}[c]{@{}c@{}} \\ Dimensional\\ Reduction\end{tabular}}} & \begin{tabular}[c]{@{}c@{}} \\[-1.5ex] \ Summation of $(1+1)$D EE \ \\ with (effective) mass~\cite{Ryu2006EEandBerryPhase}\end{tabular} & \begin{tabular}[c]{@{}c@{}}\\[-1.5ex] Quick evaluation of EE in \\  free theories \end{tabular} & 
		\begin{tabular}[c]{@{}c@{}}\\[-1.5ex] Assuming EE to be extensive  \end{tabular} \\ \cline{2-4} 
		\multicolumn{1}{|c|}{} & \begin{tabular}[c]{@{}c@{}} \\[-1.5ex]  Summation of $(1+1)$D \\  entropic-c function~\cite{CasiniHuerta2005MassiveScalar}\end{tabular} & \begin{tabular}[c]{@{}c@{}} \\[-1.5ex] Quick evaluation of EE in \\  free theories \end{tabular} & \begin{tabular}[c]{@{}c@{}} \\[-1.5ex]  
			Assuming EE to be extensive \end{tabular} \\ \hline
	\end{tabular}
\end{table*}

Moreover, in addition to the area-law contribution, the EE may host a sub-leading correction that encodes universal constraints of underlying theories free of ultraviolet cutoffs. 
It gives a unique measure of the effective degrees of freedom of the theory, which should monotonically decrease along the renormalization group (RG) flows. 
This motivates an idea to inspect irreversible renormalization group (RG) flows in general dimensions from the viewpoint of quantum entanglement~\cite{CASINI2004_1p1_EE_RG, Myers2010_holographic_c_theorem, Myers2011_holographic_c_theorem, Klebanov2012_3dEE, Casini2012_RG_EE_2p1, liu2013RefinementEE, Fei2015_generalized_F_theorem}. 
To be specific, this is related to a proposal of the irreversibility theorem under RG transformations in general dimensions, dubbed by the \textit{$F$-theorem}~\cite{Myers2010_holographic_c_theorem, Jafferis2011_F_theorem, Myers2011_holographic_c_theorem}.  
In this regard, the exact form of the EE, including the area-law and sub-leading terms, is quite informative for understanding the quantum criticality of underlying theories.

Nevertheless, to rigidly compute the EE of QFTs is challenging.
The existing methods have various restrictions (please see Tab.~\ref{tab:method_EE} and Sec.~\ref{sec:review} for a summary), most of them are only limited to space-time $(1+1)$-dimension~\cite{HOLZHEY1994443, Calabrese_2009_review, CasiniHuerta_2009}. 
For higher-dimensional theories, despite of the significant progress on studying clean systems~\cite{Calabrese_2004, Metlitski2009_ON, casini2010_EE_sphere, casini2011towards,  Hertzberg2011_FreeScalar, klebanov2012,  Hertzberg2012_InteractingScalar, Myers2013_MassiveEE, Whitsitt2017_LargeN_WF, Hung2017_ON, Hung2017_Instanton, chen2020EE_interacting_harmonic, iso2021_EE_ZM, iso2021_EE_composite}, the entanglement properties of quantum critical points with the quenched disorder are yet mostly unexplored~\cite{Rafael2004, lin2007EE_InfRand, rong2008_rand_ising}. 
To date, it remains elusive if or not the $(2+1)$D disordered quantum critical points~\cite{Harris_1974, Pruisken1983Delocalization, Daniel1992randomTFI, wenger1994_disorder_Dwave, Ludwig1994_IQH_transition, goldman2017QED3, sachdev2017QED3, yerzhakov2018disordered, goldman2020interplay} share the same entanglement characteristics as the clean ones, or to what extent randomness affects the entanglement scaling law. 
These questions are important, since the disorder inevitably exists in realistic physical systems and possibly changes the critical scaling exponents~\cite{narovlansky2018_disorder_RG}.
However, the randomness and imperfection generally lower global symmetries, therefore many well-established tools such as the celebrated conformal field theory (CFT) and/or heat-kernel techniques cannot be applied straightforwardly. 
To understand the entanglement in disordered quantum critical points, it is highly desired to develop an innovative approach that works efficiently in the space-time dimension higher than $(1+1)$-dimension.

In this paper, to fill this blank, 
we explore a \textit{dimensional reduction} approach to analytically compute the EE for $(2+1)$D theories, without resorting to global conformal symmetry. 
This scheme allows an explicit evaluation of the EE for theories exposed to static potentials. 
As a concrete example of the disordered quantum critical point, we investigate the case of a $(2+1)$D Dirac field exposed to a random static magnetic field, which includes a non-trivial critical line as varying the randomness strength~\cite{Ludwig1994_IQH_transition, Wen1996conformal, Wen1996Multifractality, Mirlin2001Multifractality, Ryu2009Multifractal}. 
In particular, we analytically derive an area-law scaling of the EE, which signals the critical behavior of the ground state. 
This analytical solution is in line with the numerical simulation on the corresponding lattice model. 
Last but not least, we demonstrate that  by considering a finite correlation length away from the criticality,  there is a universal sub-leading correction to the EE. 
Its connection with the $F$-theorem is discussed. 
%
In short, our work not only offers a tool for faithfully calculating the EE of general quantum theories, but also provides for the first time a systematic investigation of the entanglement properties of  $(2+1)$D  disordered quantum critical points.

This paper is structured as follows. Sec.~\ref{sec:review} summarizes the existing methods of calculating EE in QFTs. 
We then discuss the general strategy of the dimensional reduction scheme in Sec.~\ref{sec:strategy}, and show how our idea is developed. 
As a benchmark, we apply our method to $(2+1)$D free scalar field in Sec.~\ref{sec:free_scalar} and free Dirac field in Sec.~\ref{sec:free_dirac}, which faithfully recover the previously known area-law behavior of the EE. 
The calculation is further extended into $(2+1)$D Dirac fermion exposed to a random magnetic field in Sec.~\ref{sec:gauge_dirac}, with an introduction to the background of investigating this model presented in Sec.~\ref{sec:gauge_dirac_intro}. 
The derivation of an analytical solution of the EE in this disordered theory is addressed in Sec.~\ref{sec:gauge_dirac_calc}, which is validated by the corresponding lattice simulation in Sec.~\ref{sec:gauge_dirac_numeric}. 
We then discuss a quasiparticle picture to understand the observed area-law in the point of view of correlations in Sec.~\ref{sec:quasiparticle}. 
At last, by connecting with the irreversibility of RG flows, we point out the physical meaning of computing the universal sub-leading term of EE in Sec.~\ref{sec:EE_RG}. 
%
%
These results are concluded in Sec.~\ref{sec:conclud}, with outlooks for some open questions. 
Appendices contain technical details about the current calculation and known results of the investigated model, with a short discussion on the effect of many-body interactions.

\section{Technical Overview}

In this work, we focus on the EE of a pure ground state. 
It is expected that the EE of a $(d+1)$D QFT ($d > 1$) satisfies an area-law scaling~\cite{Calabrese_2004, CasiniHuerta_2009}
\begin{equation}\label{eq:area_law}
S \sim c_{\rm cut-off} \mathcal{A}/\epsilon^{d-1} + \gamma_d, 
\end{equation}
where $\mathcal{A}$ is the area of the codimension-one entangling surface, and $\epsilon$ is a microscopic cut-off. 
Here, the leading term of EE depends on the UV cut-off, and its coefficient $c_{\rm cut-off}$ is sensitive to the choice of regularization scheme.  
It reflects the intrinsic nature of the system only when it becomes a function of the coupling constants. 
The second sub-leading $\gamma$ term is expected to provide universal information of underlying theories. 
In particular, when perturbing away from a quantum critical point by a finite correlation length $\xi$, one expect the $\gamma$ term behaves as
\begin{equation}\label{eq:finite_mass_term}
\gamma_d 
\sim r_d \mathcal{A} / \xi^{d-1},
\end{equation} 
where we consider the theory lives in even spatial dimension $d$, and a smooth boundary is assumed for the entanglement cut.  
The coefficient $r_d$ is expected to be finite, and might provide useful information for characterizing the universality~\cite{Metlitski2009_ON, Casini2012_RG_EE_2p1, Myers2013_MassiveEE, Tatsuma2018}.

As mentioned in the introductory part, it is generally hard to determinate the form of Eq.~\eqref{eq:area_law} for a general theory with quenched disorder by using the currently existing methods. 
This motivates us to develop a novel scheme for calculating it. 
Here we present a brief review of existing methods on calculating EE in QFTs (see Table.~\ref{tab:method_EE}). 
Based on this, we will show how the previous investigations inspire us to propose an exact dimensional reduction method.
The connection to and distinction from the existing studies will be also addressed in detail.

\subsection{Existing methods of calculating EE}\label{sec:review}

\subsubsection{Real-time approach}

By definition, the calculation of EE requires the spectrum information of the reduced density matrix $\rho_A$. The most straightforward way is to diagonalize it directly in Minkowski spacetime, which is so-called \textit{real-time approach}. 
In principle, numerical methods (e.g. exact diagonalization technique) can determine the spectrum of $\rho_A$ for any discretized system (lattice model). 
However, due to the exponentially growing Hilbert space, the computationally accessible size (typically about $10-20$ qubits) is extremely small comparing with the realistic systems. 

For free theories, the full information of their ground state is encoded in two-point correlators $C_{ij} = \langle \Psi | c_i^\dagger c_j | \Psi \rangle$. 
This fact leads to the implementation of \textit{correlation matrix} method~\cite{chung2001density, Peschel_2003calculation, Peschel_2009reduced} for calculating EE
\begin{equation}\label{eq:correlation_matrix}
S = - \Tr \left[ C_A \ln C_A + (1 - C_A) \ln (1 - C_A) \right] ,
\end{equation}
where $C_A$ is the correlation matrix for subsystem $A$. It only requires diagonalization of a $N \times N$ matrix and $N$ is number of lattice sites. This method has been widely used in numerical simulations. 

Notably, for certain cases, the correlation matrix method can give an analytical solution of EE and entire spectrum of the reduced density matrix~\cite{Casini_2009_resolvent_1p1dirac}.
By taking Eq.~\eqref{eq:correlation_matrix} as an integral operator with kernel $C_A$ inside certain intervals, the EE can be written in terms of a contour integral of its resolvent. 
This technique is valuable to determine multi-interval EE of $(1+1)$D free massless fermions and chiral bosons, however it is restricted to these cases due to the mathematical difficulty on calculating the exact resolvent. 

\subsubsection{Euclidean approach: Replica trick} 

Direct calculation of the EE in Minkowski spacetime is mainly limited to finite-size numerical simulation for discrete lattices instead of continuous spacetime. This leads to the difficulty on determining the scaling behavior of EE. By contrast,
the Euclidean approach via replica trick~\cite{Callan1994GeometricEntropy, Calabrese_2004, CasiniHuerta_2009}, is powerful for solving EE analytically. 
The replica trick is introduced to avoid the difficulty of taking logarithm to the reduced density operator $\rho_A$. With introducing a replica index of $n$, the EE can be rewritten as 
\begin{equation}\label{eq:EE_replica}
S = \left. - \frac{\partial}{\partial n} 
\ln \Tr \left( \rho_A^n \right) \right|_{n \to 1} .
\end{equation}
The physical meaning of the index $n$ is to make $n$ decoupled identical copies of the theory. Analytic continuation of $n$ is then assumed before taking the replica limit $n \to 1$. 

Since we are interested in the case of ground state, the trace of $\rho_A^n$ has a natural Euclidean path integral representation~\cite{Calabrese_2004, Calabrese_2009_review, Tatsuma2018}
\begin{equation}
\Tr \left( \rho_A^n \right) = \frac{Z^{(n)}}{\left[ Z^{(1)} \right]^n} , 
\end{equation} 
where $Z^{(n)}$ represents the partition function defined on the $n$-fold replica spacetime manifold with the entanglement cut along $A$. 
The calculation of EE is then reduced to the problem of solving the partition function $Z^{(n)}$ on a certain $n$-fold non-smooth manifold as
\begin{equation}\label{eq:replica_EE}
S = - \left. \frac{\partial}{\partial n} 
\left[ \ln Z^{(n)} - n \ln Z^{(1)} \right] \right|_{n\to1} .
\end{equation} 
Geometrically, the manifold is equivalent to an Euclidean spacetime with conical singularities at coincident points that is described by the metric~\cite{KABAT1995, Calabrese_2004, Tatsuma2018} 
\begin{equation}\label{eq:metric_conical}
ds^2 = d\rho^2 + n^2 \rho^2 d\theta^2 + \sum_{i=3}^{d_{\mathcal{M}}} dx_i^2 ,
\end{equation}
where $n$ is the replica index that characterizes this metric, $d_\mathcal{M}$ is the spacetime dimension, and the $(x_1,x_2)$ plane is written in terms of the polar coordinates $(\rho, \theta)$.
In this paper, we focus on the case of an infinite cone, i.e. $\rho \in [0, \infty)$, $\theta \in [0, 2\pi)$ and $\{ x_i \}$ in the whole space. 
For solving the functional integral and differential equations, this metric can be described by changing boundary condition from the ordinary period of $\theta \sim \theta + 2\pi$ to $\theta \sim \theta + 2\pi n$.  

For free theories, the partition function is one-loop divergent, so that the \textit{heat-kernel} technique is quite standard for calculating it~\cite{VASSILEVICH2003HeatKernel}. 
Several models, including free scalar, Dirac and Maxwell fields with/without curvature coupling, were investigated in previous works~\cite{Callan1994GeometricEntropy, KABAT1995, Nesterov2010_modified_HeatKernel, Solodukhin2011, Hertzberg2011_FreeScalar, Myers2013_MassiveEE}. 
However, the heat-kernel technique meets difficulty when dealing with generic interacting theories on a manifold with conical singularities, since it captures only the quadratic order of quantum fluctuations (effective action at one-loop level) and there is no closed analytical expression for higher-order heat-kernel coefficients on replica manifold with conical singularities.

Another possible way to estimate the partition function $Z^{(n)}$ is through the Green's function $G^{(n)}$ on replica manifold~\cite{Calabrese_2004, CasiniHuerta_2009, Metlitski2009_ON, Hung2017_ON}. They are related by taking a derivative with respect to the mass\footnote{This is only for free scalar field, and for free Dirac field this relation becomes $\partial_m \ln Z^{(n)} = - \Tr G^{(n)}$.} 
\begin{equation}\label{eq:free_ZG}
\frac{\partial}{\partial m^2} \ln Z^{(n)} = - \frac{1}{2} \Tr G^{(n)} 
\end{equation}
Here the concept of the Green's function is not limited to 
its original meaning in solving differential equations, 
but is extended to the two-point correlation function of QFTs. 
This is important for calculating EE in the theories with no direct field-equation representation, e.g. disordered systems.    
However, unlike the universal expansion procedure in heat-kernel technique, there is no general way for calculating replicated Green's function in higher-dimensional interacting theories. 
Fortunately, the calculation of Green's function of QFTs in curved spacetime with conical singularities has been attracted considerable attentions in various contexts, such as scattering of electromagnetic waves~\cite{Dowker_1977, dowker1978, Linet1994_scalar, Linet1995_spinor} and orbifold conformal field theory ~\cite{dowker1992, CHANG1993, fursaev1994, Cognola1994, functional_determinant}. 
These studies provide valuable knowledge for calculating EE in QFTs.

\subsubsection{Conformal field theory approach}

For critical systems described by CFT, 
there are some universal behaviors of the EE that are analytically accessible. 
In $(1+1)$ dimension, CFT techniques (combine with the replica trick) have received great achievement of calculating the EE in critical systems, demonstrating a logarithmic divergent EE with a prefactor of central charge $c$ that characterizes universality of the quantum criticality~\cite{HOLZHEY1994443, Calabrese_2004, cardy2008_form_factor}
\begin{equation}\label{eq:2dCFTEE}
S_{2{\rm D \ CFT}} = \frac{c}{3} \ln \frac{l}{\epsilon} + c' ,
\end{equation} 
where $l$ is the size of a single-interval subsystem in an infinite chain, $\epsilon$ is a UV cut-off of lattice constant, and $c'$ is a non-universal finite term. 

For higher dimensions, the conformal symmetry is generally not so strong as $2$D to fully determine the scaling behavior of the EE. 
For spherical entangling surface in $\mathbb{R}^{1,d}$ flat Minkowski spacetime, the problem of the EE of a CFT can be conformally mapped to the solution of thermal entropy in a $\mathbb{R} \times \mathbb{H}^{d}$ hyperbolic space~\cite{bisognano1975, bisognano1976, hislop1982_modular_scalar_CFT, casini2010_EE_sphere, casini2011towards}, where an infrared (IR) cutoff leads to the area-law EE of the quantum fields in $\mathbb{R}^{1,d}$ at UV. 
However, this approach cannot be extended to generic geometries, where the local form of modular Hamiltonian is unknown.

\subsubsection{Holographic approach}

The difficulty of calculating the EE in higher dimensional QFTs motivates a holographic interpretation of the EE based on the conjecture of AdS/CFT correspondence, which bridges the $(d+2)$D AdS space and a $(d+1)$D CFT~\cite{Maldacena1999AdSCFT}. 
It was proposed that the calculation of EE can be reduced to the problem of finding extreme surface inside the AdS space, for which a Bekenstein-Hawking-like formula (the RT formula) naturally gives an area-law~\cite{RTformula2006, RT2006Aspects}. 
Nevertheless, the RT formula is still far away from the answer to entanglement in QFTs. 
The use of RT formula requires the dictionary between field theories and its gravitational dual, however, only few cases are known. 
Meanwhile, in the AdS calculation, although solving the extreme surface is a classical task, in most cases we can only perform a numerical estimation on it. 
More importantly, there is no rigorous proof of the holographic principle, and the sufficient condition for the establishment of RT formula remains an open question.

\subsubsection{Quasi-particle picture and extensive mutual information model}

The area-law EE can be understood within a quasiparticle picture, which assumes that the entanglement is made of the correlations between entangled quasiparticles in the system~\cite{Calabrese_2005, nahum2020_majorana_defect, tang2021nonunitary}. 
This assumption reduces the calculation of EE to a simple geometric problem of summing up the distribution of these quasiparticle pairs. 
In parallel to the aforementioned quasiparticle picture that comes from a dynamical diffusion-annihilation process of free fermions, the observation of extensive mutual information in the ground state of $(1+1)$D massless Dirac fermions~\cite{CasiniHuerta2005MassiveScalar} motivates investigations on an ``extensive mutual information'' model~\cite{Casini_2009_EMI}, which has been used for understanding the entanglement structure with various applications~\cite{casini2015_EMI, bueno2019_EMI, bueno2021_EMI}.  
Recently, it is proven that the extensive mutual information model does not correspond to an actual CFT beyond $(1+1)$ dimensions, so that fail to be an exact solution of EE in higher dimensions~\cite{pablo2021_EMI}. 
However, it does capture the leading scaling behavior of entanglement and provide significant understanding in fermionic scale-invariant systems.

\subsection{General strategy of dimensional reduction}\label{sec:strategy}

The general idea of dimensional reduction is to use low-dimensional results (which is known) to calculate higher-dimensional results (which is hard to know). 
For non-interacting cases, one can consider that the higher-dimensional theories are constructed by infinite many $(1+1)$D modes with an effective mass that is associated with its momentum.
This fact motivates a direct reduction of higher-dimensional entropy to a sum of $(1+1)$D entropic $c$-function (defined as $c(L) = L \frac{dS(L)}{L}$ for the subsystem with spatial size $L$)~\cite{CasiniHuerta2005MassiveScalar, Ryu2006EEandBerryPhase, CasiniHuerta_2009, swingle2012_fermi_liquid_EE, Murciano_2020}. 
These calculations are quite simple and provide an intuitive picture on entanglement structure of (2+1)D many-body states. 
However, this procedure has two drawbacks. 
First, this calculation of the EE ~\cite{CasiniHuerta2005MassiveScalar, Ryu2006EEandBerryPhase} requires the additivity of the entropic function, which is mathematically less evident. 
Second, this method is hard to be extended into generic models for exact results, so the dimensional reduction scheme in previous works only has phenomenological meaning. 
Therefore, seeking for other possible (exact) dimensional reduction approach on a firm ground is highly desired.

In order to solve the aforementioned problems, let us consider one question first: Which physical quantity (that we are familiar with) is capable of giving the scaling of EE?  
Apparently, the most suitable one for QFTs is the Green's function on replica manifold $G^{(n)}$, 
which has great advantages on computation with the help of tools from conventional perturbation theory such as diagram technique and renormalization group analysis~\cite{Metlitski2009_ON}.  
%
In this work, we explore an efficient framework of dimensional reduction to obtain $G^{(n)}$ for calculating EE.

Here, we explain how this works for the simplest case of constructing $(d+1)$D Green's function of free scalar field in usual flat Minkowski spacetime from its $(d+0)$D reduction. 
Start from its action
\begin{equation}\label{eq:action_scalar}
	\begin{aligned}
		I^{[d+1]} & = \int dt \int d^d x \left( 
		-\frac{1}{2} \partial^\mu \phi \partial_\mu \phi - \frac{m^2}{2} \phi^2 
		\right) \\
		& = \int dt \int d^d x \int \frac{d\omega}{2\pi} e^{-i \omega t} 
		\left[ \mathcal{L}^d_0 + \mathcal{L}^d_{\rm int}(\omega) \right]
	\end{aligned}
\end{equation}
where $\mathcal{L}^d_0 = -\frac{1}{2} 
\partial^i \phi(\omega) \partial_i \phi(\omega) - \frac{m^2}{2} \phi^2(\omega)$ 
is the free Lagrangian density
and $\mathcal{L}^d_{\rm int}(\omega) = \frac{\omega^2}{2} \phi^2(\omega)$ 
is the interacting term in $(d+0)$D in a quadratic form, 
with the index $\mu$ runs over the spacetime dimensions and $i$ only for spatial. 
This action has the exact solution of the Green's function as 
$G^{\mathtt{[d+1]}}_0(k, \omega) = \left(-\omega^2+k^2+m^2\right)^{-1}$, 
and can be represented as a sum of all tree-level diagrams with respect to $\omega$
\begin{equation}\label{eq:dimred_scalar}
	G^{\mathtt{[d+1]}}_0(k,\omega) = g^{[d+0]}_0(k) \sum_{l=0}^\infty \left[ \omega^2 g^{\mathtt{[d+0]}}_0(k) \right]^l . 
\end{equation}
Here the $(d+0)$D Green's function $g^{\mathtt{[d+0]}}_0(k,0)  $ is regarded as the ``free'' solution of $\mathcal{L}^d_0 $, and Eq.~\eqref{eq:dimred_scalar} actually defines an alternative approach of dimensional reduction, with the quadratic construction as an inherent regulator. 
From here on, we use the upper index in $[...]$ to represent the spacetime dimension, while that in $(...)$ to denote the replica index. For simplicity, $g^{[d+0]}\equiv g$ stands for the Green's function in $(d+0)$ dimension, and symbol $G^{[d+1]}\equiv G$ is used for the full Green's function in $(d+1)$ dimension.

The advance of the above dimensional reduction of Green's function can also deal with possible interactions, at least in the perturbative region. 
In particular, for the case of adding a static potential without dynamics, i.e. the interaction with an field that does not depend on time, the extension can be made by considering the effect of interactions as quantum corrections to the $(d+0)$D Green's function $g^{\mathtt{[d+0]}}(k)$ as 
\begin{equation}\label{eq:dimred_int}
	G^{\mathtt{[d+1]}}(k,\omega) = g^{\mathtt{[d+0]}}(k) 
	\sum_{l=0}^\infty \left[ \omega^2 g^{\mathtt{[d+0]}}(k) \right]^l .
\end{equation}
Here we note that this formula is just a formal representation, and we are not limited to this concrete construction. 
A direct application is the system with quenched disorders, of which the random effect can be absorbed into the lower-dimensional Green's function as a dimensional reducible interaction. 
%
More importantly, the above discussions are not restricted to the flat spacetime, but can be directly extended to the case of curved spacetime with certain singularities. 
In other words, it works for the Green's function on replica manifold $G^{(n)}$ that is able to give the EE. 
Therefore, our task of calculating higher-dimensional EE turns into the calculation of lower-dimensional interacting Green's functions on the replica spacetime manifold. 
This is of course a difficult problem, however, we will show that the calculation at replica limit $n\to1$ is very much similar to the usual perturbation theory in flat spacetime, but with a term of the conical singularity.

At the end of this section, let us emphasize the motivation to apply the dimensional reduction scheme, instead of calculating perturbative expansions directly in high dimensions~\cite{Hertzberg2012_InteractingScalar}. 
Generally speaking, one important consequence of interactions is the breakdown of  Gaussianality of the many-body ground state, which usually requires a higher-order (multi-loop) calculation to capture the non-Gaussian features. 
Unfortunately, a proper renormalization scheme for these higher-order corrections in $(2+1)$D is less known in the context of calculating the EE, since the fields are living on the replica manifold with conical singularities instead of the usual flat spacetime~\cite{chen2020EE_interacting_harmonic, iso2021_EE_ZM, iso2021_EE_composite}. 
Moreover, it is worth noting that there are celebrated approaches to access non-perturbative properties of the ground state in $(1+1)$D, such as the $(1+1)$D CFT. The further construction of an approximate (effective) theory based on these non-perturbative results has shown a powerful perspective in understanding low-energy collective excitations in condensed matter~\cite{shankar1994RG}. 
The key of our dimensional reduction method is to access the entanglement structure  via  a similar manner. 
Later we will see that this construction does reproduce the explicit form of $(2+1)$D results through conventional field theory techniques.

\section{ $(2+1)$D free scalar field}\label{sec:free_scalar}

We start the calculation of EE in $(2+1)$D QFT by the free scalar field using the above dimensional reduction method, as a benchmark. 
This process is instructive and provides insights on the further calculation for $(2+1)$D Dirac field.  

\subsection{Direct solution}

Let us start from a brief review of the area-law EE of a free scalar field living on a ``waveguide'' geometry $\mathbb{R}^2 \times \mathbb{I}$, where the wavefunction propagates as a plane wave on the finite interval $\mathbb{I}$~\cite{Calabrese_2004, Hertzberg2011_FreeScalar, Myers2013_MassiveEE}. 
The non-interacting Green's function of scalar fields $G^{(n)}$ satisfies the Helmholtz equation that is defined on the corresponding $3$D replica spacetime manifold $\mathcal{M}^{(n)}$
\begin{equation}\label{eq:Helmholtz}
	(\nabla^2 - m^2) G^{(n)}(\mathbf{r, r'}) 
	= - \delta^{[3]} (\mathbf{r, r'}) ,
\end{equation}
where $\mathbf{r,r'}$ are $3$D vectors and $\delta^{[3]}$ is the Dirac-delta function in $3$D. 
In our case, the replica manifold follows the waveguide construction $\mathcal{M}^{(n)} = \mathcal{C}^2 \times \mathbb{I}$ as the product of a $2$D cone $\mathcal{C}^2$ and the interval $\mathbb{I}$, with the metric in Eq.~\eqref{eq:metric_conical}. 
The solution of $G^{(n)}$ in cylindrical coordinates $\mathbf{r} = (\rho, \theta, r_\perp)$ is then given by 
\begin{equation}
	\begin{aligned}
		& G^{(n)}(\mathbf{r,r'}) = 
		\int \frac{d k_\perp}{2\pi} e^{i k_\perp (r_\perp-r_\perp')} 
		\frac{1}{2\pi n}  
		\sum_{q=0}^{\infty} d_q \\
		& \qquad \cos \left[ \frac{q}{n} (\theta-\theta') \right] 
		\int_0^\infty \frac{J_{q/n}(\lambda \rho) J_{q/n}(\lambda \rho')}
		{\lambda^2 + m^2 + k_\perp^2} \lambda d\lambda
	\end{aligned}
\end{equation}
where $k_\perp$ is the momentum of the translation-invariant $r_\perp$-direction that perpendicular to the plane of polar coordinates $\mathbf{r}_\parallel = (\rho, \theta)$, $q$ is the angular momentum in the $(\rho,\theta)$ plane that takes integer values, $d_0 = 1$, $d_{q>0} = 2$, and $J_q(\lambda \rho)$ is the Bessel function of first kind at $q$-th order with the eigenvalue $\lambda$ in the radial equation. 

Taking trace of $G^{(n)}$ requires the information at coincident points $(\rho, \theta) \to (\rho', \theta')$, where the Green's function is generally UV divergent. 
In our case, the divergence comes from the sum over angular momentum $q$, and can be regularized in the calculation of the normalized partition function $Z^{(n)}/\left[ Z^{(1)} \right]^n$. 
Mathematically, this is achieved by using the Euler-Maclaurin formula that translates the summation to an improper integral with remaining terms (see Appendix~\ref{app:EMformula}). It gives 
\begin{equation}\label{eq:scalar_Gn_3D}
	G^{(n)} - G^{(1)}
	= \int \frac{dk_\perp}{2\pi} 
	\frac{1-n^2}{12\pi n^2} 
	\left[ K_0(\sqrt{k_\perp^2+m^2}\rho) \right]^2 , 
\end{equation}
where $K_0(x)$ is the zero-th order modified Bessel function of the second kind, and all the higher-order remaining terms in Euler-Maclaurin expansion vanish at the coincident points. 
Then it leads to 
\begin{equation}\label{eq:direct_scalar_trG}
	\begin{aligned}
		\frac{\partial}{\partial m^2} \ln \frac{Z^{(n)}}{\left[ Z^{(1)} \right]^n} 
		& = - \frac{1}{2} \Tr^{(n)} \left[ G^{(n)} - G^{(1)} \right] \\
		& = - \int dr_\perp \int \frac{dk_\perp}{2\pi} 
		\frac{1-n^2}{24n(k_\perp^2+m^2)} ,
	\end{aligned}
\end{equation}
where $\Tr^{(n)}$ represents that the integral over full spacetime is taking on the $n$-fold manifold.
The integral $\int dr_\perp = \mathcal{A}$ gives the area of the entangling surface in the finite interval $\mathbb{I}$. 
Here the integral over $k$ will lead to logarithmic divergence that requires a cut-off of $\epsilon^{-1}$ ($\epsilon \ll 1$ plays the role of lattice constant). 
These give the regularized area-law EE 
\begin{equation}\label{eq:freescalar_EE}
	S = -\frac{\mathcal{A}}{12} \int_{-\infty}^\infty \frac{dk_\perp}{2\pi} 
	\ln \frac{k_\perp^2 + m^2}{k_\perp^2 + \epsilon^{-2}} 
	= \frac{\mathcal{A}}{12} \left( \epsilon^{-1} - m \right) .
\end{equation}
For the massless case $m=0$, we simply have $ S = \frac{\mathcal{A}}{12 \epsilon} $ as the leading UV-divergent area-law scaling.

\subsection{Dimensional reduction calculation}

\begin{figure}\centering
	\includegraphics[width=\columnwidth]{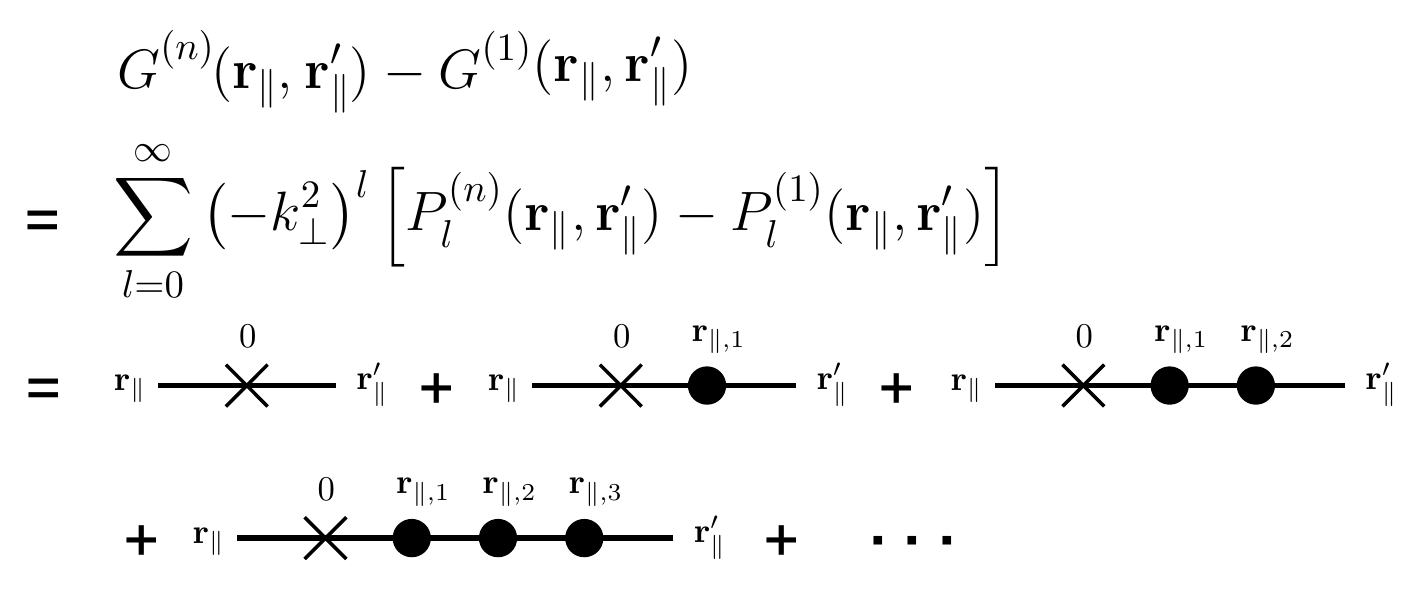}
	\caption{
		\label{fig:diagram}
		The diagram representation of the replicated Green's function of $(2+1)$D free scalar field via the dimensional reduction method, with ignoring higher-order  terms of $\mathcal{O}((1-n)^2)$. 
		Here the lines are the usual flat Green's function in its real-space representation, the dot with label $\mathbf{r}_{\perp, l}$ represents a vertex of $-\omega^2 \int^{(1)} d^2 \mathbf{r}_{\perp, l}$, and the x at original point denotes a factor of $2\pi \frac{1-n^2}{6n^2}$. 
	}
\end{figure}

In this section, we show that the above result of EE can be reproduced through the dimensional reduction method. 
As we have introduced in Sec.~\ref{sec:strategy}, for dimensional reduction we need to calculate the products of the Green's function on $2$D replica manifold. In real-space representation, they are 
\begin{equation}
	\begin{aligned}
		P^{(n)}_l ( \mathbf{r}_{\parallel}, \mathbf{r}'_{\parallel} ) 
		& = \int^{(n)} d^2 \mathbf{r}_{\parallel, 1} 
		\cdots \int^{(n)} d^2 \mathbf{r}_{\parallel, l} \\
		& \qquad g^{(n)}( \mathbf{r}_{\parallel}, \mathbf{r}_{\parallel, 1} ) 
		\cdots g^{(n)}( \mathbf{r}_{\parallel, l}, \mathbf{r}'_{\parallel} ) ,
	\end{aligned}
\end{equation}
which gives the $l$-order perturbation $ (-k_\perp^2)^l P^{(n)}_l $. 
Here $\int^{(n)}$ represents that the integral is performed on the $n$-fold manifold. 
The exact calculation of these products are hard, since for $g^{(n)}$ at general two points we do not have a simple relation as Eq.~\eqref{eq:scalar_Gn_3D}. 
However, if one consider the approximate expansion near the coincident points that contributes to the entanglement entropy~\cite{Hertzberg2012_InteractingScalar}, the calculation can be simplified again by using the Euler-Maclaurin formula, which gives 
\begin{equation}\label{eq:Euler-Maclaurin}
	g^{(n)} (\mathbf{r}_\parallel, \mathbf{r}'_\parallel)
	\sim 
	g^{(1)} (\mathbf{r}_\parallel, \mathbf{r}'_\parallel) 
	+ \frac{1-n^2}{12\pi n^2} K_0(m\rho) K_0(m\rho')  . 
\end{equation}
This relation reduces the product of $g^{(n)}$ into
\begin{equation}\label{eq:scalar_G_product}
	\begin{aligned}
		& \quad P^{(n)}_{l} 
		( \mathbf{r}_{\parallel}, \mathbf{r}'_{\parallel} ) \\
		& = (l+1) 
		\int^{(1)} d^2 \mathbf{r}_{\parallel, 1} 
		\cdots \int^{(1)} d^2 \mathbf{r}_{\parallel, l} 
		g^{(n)}( \mathbf{r}_{\parallel}, \mathbf{r}_{\parallel, 1} ) \\
		& \qquad g^{(1)}( \mathbf{r}_{\parallel, 1}, \mathbf{r}_{\parallel, 2} ) 
		\cdots g^{(1)}( \mathbf{r}_{\parallel, l}, \mathbf{r}'_{\parallel} ) 
		+ \mathcal{O}((1-n)^2),
	\end{aligned}
\end{equation}
where the factor of $(l+1)$ is the symmetry factor,  
and the higher-order terms of $\mathcal{O}((1-n)^2)$ vanish in the EE as taking the derivative and the replica limit ${n \to 1}$. 
The Eq.~\eqref{eq:scalar_G_product} actually defines an expansion of the products $P^{(n)}_l$ around $(1-n)$, which leads to the simplification of calculating $P^{(n)}_l$ via conventional diagram techniques as we will demonstrate below. 
Here $g^{(n)}$ should be written in terms of $g^{(1)}$ and the remaining term, since we have changed the integral measure onto a single copy instead of the entire $n$-fold manifold. 
It gives 
\begin{equation}\label{eq:complex_Gn_G1_scalar}
	\begin{aligned}
		& \quad P^{(n)}_l 
		( \mathbf{r}_{\parallel}, \mathbf{r}'_{\parallel} ) 
		- P^{(1)}_l 
		( \mathbf{r}_{\parallel}, \mathbf{r}'_{\parallel} ) \\
		& = (l+1) \frac{1-n^2}{12\pi n^2} K_0(m\rho) 
		\int^{(1)} d^2 \mathbf{r}_{\parallel, 1} 
		\cdots 
		\int^{(1)} d^2 \mathbf{r}_{\parallel, l}  \\
		& \qquad \qquad \quad K_0(m\rho_1) 
		g^{(1)}( \mathbf{r}_{\parallel, 1}, \mathbf{r}_{\parallel, 2} ) 
		\cdots g^{(1)}( \mathbf{r}_{\parallel, l}, \mathbf{r}'_{\parallel} ) .
	\end{aligned}
\end{equation}
It is important to notice that elements in the above expansion contains only the usual flat Green's function of free scalar field in $2$D spacetime, 
\begin{equation}
	g^{(1)}(\mathbf{r}_\parallel, \mathbf{r}_\parallel') 
	= \int \frac{d^2 \mathbf{k}_\parallel}{(2\pi)^2}
	\frac{e^{i \mathbf{k}_\parallel (\mathbf{r}_\parallel - \mathbf{r}_\parallel')}}
	{{k}_\parallel^2 + m^2} 
	= \frac{1}{2\pi} K_0(m |\mathbf{r}_\parallel - \mathbf{r}_\parallel'| ) .
\end{equation}
This means that we are dealing with nothing unusual but the ordinary diagrams with a non-trivial additional vertex that comes from the conical singularity, see Fig.~\ref{fig:diagram}. 
Now the calculation of EE is fully reduced to conventional perturbation theory that we are familiar with,  
Eq.~\eqref{eq:complex_Gn_G1_scalar}
is then simplified as 
\begin{equation}\label{eq:momentum_Gn_G1_scalar}
	\begin{aligned}
		& \quad P^{(n)}_l 
		( \mathbf{r}_{\parallel}, \mathbf{r}'_{\parallel} ) 
		- P^{(1)}_l 
		( \mathbf{r}_{\parallel}, \mathbf{r}'_{\parallel} ) \\
		& = \frac{l+1}{\Gamma(l+1)} 
		\frac{1-n^2}{12\pi n^2} K_0(m\rho) 
		\left( \frac{\rho'}{2m} \right)^l  K_l(m\rho') .
	\end{aligned}
\end{equation}
In practice, we find that it is more convenient to trace over the reduced two-dimensions before taking the summation over the perturbation levels $l$, which gives
\begin{equation}
	\left( -k_\perp^2 \right)^l  
	\Tr^{(n)}_{2D} \left[ P^{(n)}_l - P^{(1)}_l \right] \\
	=  \frac{1-n^2}{12n m^2} 
	\left( - \frac{k_\perp^2}{m^2} \right)^l .
\end{equation}
Its sum over $l$ is just a geometric sequence, and gives to the higher-dimensional Green's function as shown diagrammatically in Fig.~\ref{fig:diagram}. This leads to
\begin{equation}
	\begin{aligned}
		\Tr^{(n)}_{3D} \left[ G^{(n)} - G^{(1)} \right]  
		= \frac{1-n^2}{12n} \int d r_\perp 
		\int \frac{dk_\perp}{2\pi} 
		\frac{1}{k_\perp^2 + m^2} ,
	\end{aligned}
\end{equation}
which is identical to the calculation in Eq.~\eqref{eq:direct_scalar_trG}, 
and  leads the same result of EE in Eq.~\eqref{eq:freescalar_EE}. 
Here we have tested that exchanging the order of  $\Tr^{(n)}_{2D}$ and sum over $l$ does not influence on the result (see details in Appendix~\ref{app:change_order}).

Here, we stress that the key step in the above calculation is an integral of the replicated Green's function $G^{(n)}$ over the cone $\mathcal{C}^2$, as shown in Eq.~\eqref{eq:complex_Gn_G1_scalar} and~\eqref{eq:momentum_Gn_G1_scalar}. It is usually UV divergent and requires introducing a microscopic cutoff for accessing its finite contribution.
Fortunately, to distinguish these singularities on the replica n-fold manifold is quite straightforward in dimensional reduction scheme, with the aid of experience in $(2+0)$D~\cite{Calabrese_2004}. In a word, this example shows the proposed dimensional reduction scheme correctly captures the singularity contributed to the EE.


\section{ $(2+1)$D free Dirac field}\label{sec:free_dirac}

In previous section we have shown that our proposed dimensional reduction method faithfully recovers the area-law EE for free scalar field. 
Now we will present an exact derivation of the area-law EE in $(2+1)$D free Dirac field in a similar manner. 

The action of free Dirac field in $2$D Euclidean space is
\begin{equation}
	I^{[2]}_D = \int d^2x \overline{\Psi} (\gamma^\mu \partial_\mu + m) \Psi ,
\end{equation}
and the corresponding spinor Green's function satisfies 
\begin{equation}
	(\gamma^\mu \partial_\mu + m) g^{(n)}_D 
	(\mathbf{r}_\parallel, \mathbf{r}_\parallel') 
	= - \delta^{[2]} (\mathbf{r}_\parallel, \mathbf{r}_\parallel') ,
\end{equation}
where the index of spacetime dimensions $\mu = 1,2$ with $\gamma^1 = \sigma_1$ and $\gamma^2 = \sigma_2$. 
The solution on polar coordinates $\mathbf{r}_\parallel = (\rho, \theta)$ is 
\begin{equation}\label{eq:Green_Dirac}
	\begin{aligned}
		& \quad 
		g^{(n)}_D (\mathbf{r}_\parallel, \mathbf{r}_\parallel') 
		= \frac{1}{4\pi n} \sum_{q=-\infty}^{\infty} 
		e^{i\frac{q}{n}(\theta-\theta')} \int_0^\infty 
		\frac{\lambda d\lambda}{\lambda^2+m^2} \\
		&    \left(\begin{matrix}
			m J_{\frac{q}{n}}(\lambda \rho) J_{\frac{q}{n}}(\lambda \rho')  
			& i \lambda e^{-i\theta'} 
			J_{\frac{q}{n}}(\lambda \rho) J_{\frac{q}{n}+1}(\lambda \rho')  \\
			i \lambda e^{i\theta} 
			J_{\frac{q}{n}+1}(\lambda \rho) J_{\frac{q}{n}}(\lambda \rho') 
			& m e^{i(\theta-\theta')} 
			J_{\frac{q}{n}+1}(\lambda \rho) J_{\frac{q}{n}+1}(\lambda \rho')  
		\end{matrix}\right) .
	\end{aligned}
\end{equation}
As taking $n=1$, it reduces to the usual spinor Green's function with the difference on a global factor of $\frac{1}{2}$ that comes from the choice of normalizing the entire spinor (see details in Appendix~\ref{app:1p1_free_dirac}). 
This ensures that the EE of each spinor component of the free fermion is the half of the scalar case in $2$D.

For constructing the $3$D spinor Green's function, we need to introduce an additional Dirac-$\gamma$ matrix in higher dimension $\gamma^0 = \sigma_3$.
Similar to the case of scalar field, the $3$D spinor Green's function is represented as
\begin{equation}
	G^{(n)}_D ( \mathbf{r}_{\parallel}, \mathbf{r}'_{\parallel} ; k_\perp )
	= \sum_{l=0}^{\infty} (i k_\perp)^l 
	P^{(n)}_{D,l} ( \mathbf{r}_{\parallel}, \mathbf{r}'_{\parallel} ) 
\end{equation}
with the $l$-product of $2$D functions $g^{(n)}_D$ 
\begin{equation}
	\begin{aligned}
		P^{(n)}_{D,l} ( \mathbf{r}_{\parallel}, \mathbf{r}'_{\parallel} ) 
		& = \int^{(n)} d^2 \mathbf{r}_{\parallel, 1} 
		\cdots \int^{(n)} d^2 \mathbf{r}_{\parallel, l} 
		g_D^{(n)}( \mathbf{r}_{\parallel}, \mathbf{r}_{\parallel, 1} ) \\
		& \quad 
		\left[ \gamma^0 g_D^{(n)}( \mathbf{r}_{\parallel, 1}, \mathbf{r}_{\parallel, 2} ) \right]
		\cdots \left[ \gamma^0 
		g_D^{(n)}( \mathbf{r}_{\parallel, l}, \mathbf{r}'_{\parallel} ) 
		\right] . 
	\end{aligned}
\end{equation}
Analog to the free scalar field, here we would like to transform the real-space Green's function into momentum representation. 
The off-diagonal components in replicated spinor Green's function $G_D^{(n)}$ is generally hard to be dealt with, due to the non-trivial spin structure. 
However, it is important to notice that the double product of the spinor function is diagonal and identical to the scalar case~\cite{Myers2013_MassiveEE}. 
Meanwhile, the odd-order terms all vanish in the later trace of the higher dimension (the integral over $k_\perp$), since they are odd functions of $k_\perp$. 
These facts lead to the simplification of 
\begin{equation}
	G^{(n)}_{D} - G^{(1)}_{D} 
	= \sum_{l=0}^\infty (-k_\perp^2)^l
	\left[ P^{(n)}_{D,2l} - P^{(1)}_{D,2l} \right] 
\end{equation}
with
\begin{equation}
	\begin{aligned}
		& \quad  P^{(n)}_{D,2l}
		( \mathbf{r}_{\parallel}, \mathbf{r}'_{\parallel} )
		- P^{(1)}_{D,2l}
		( \mathbf{r}_{\parallel}, \mathbf{r}'_{\parallel} ) \\
		& = (l+1) \frac{1-n^2}{6n^2} 
		\frac{m}{2\pi} K_0(m\rho) 
		\int \frac{d^2\mathbf{k}_\parallel}{2\pi} 
		\frac{e^{i \mathbf{k}_\parallel \mathbf{r}_\parallel'} \ \textbf{Id}} 
		{({k}_\parallel^2 + m^2)^{l+1}} ,
	\end{aligned}
\end{equation}
where $\textbf{Id}$ is a two-by-two identity matrix, and the higher-order terms of $\mathcal{O}((1-n)^2)$ are ignored. 
Analog to the free scalar field, it leads to the trace on $3$D replica manifold 
\begin{equation}
	\begin{aligned}
		\Tr^{(n)}_{3D} \left[ G^{(n)}_D - G^{(1)}_D \right] 
		= \frac{1-n^2}{12n}  
		\int d r_\perp    \int \frac{dk_\perp}{2\pi}
		\frac{2m}{k_\perp^2 + m^2} 
	\end{aligned}
\end{equation}
and the corresponding normalized partition function 
\begin{equation}\label{eq:PartFunc_Dirac}
	\begin{aligned}
		& \quad 
		\ln \frac{Z^{(n)}}{\left[ Z^{(1)} \right]^n} 
		= -\int dm \Tr^{(n)}_{3D} \left[ G^{(n)}_D - G^{(1)}_D \right] \\
		& = -\frac{1-n^2}{12n} \mathcal{A} 
		\int \frac{dk_\perp}{2\pi}  
		\int dm^2 \frac{1}{k_\perp^2 + m^2} .
	\end{aligned} 
\end{equation}
Finally, we have the EE in $(2+1)$D free Dirac field 
\begin{equation}\label{eq:EE_dirac_0}
	S = \frac{1}{6} \mathcal{A} \left( \epsilon^{-1} - m \right) .
\end{equation}
Comparing with the free scalar case, we obtain
\begin{equation}
	r_{\rm dirac} = 2 r_{\rm scalar} ,
\end{equation}
where $r$ is the coefficient of the mass scaling in Eq.~\eqref{eq:finite_mass_term}. 
Here we see, through the dimensional reduction calculation, the EE of $(2+1)$D free Dirac field is observed to exhibit an area-law behavior, consistent with the previous results of calculating the entropic $c$-function~\cite{CasiniHuerta2005MassiveScalar, RT2006Aspects, CASINI2007_2p1scalar, CASINI2009_2p1dirac} and the heat-kernel on replica manifold~\cite{Myers2013_MassiveEE}.

At last, we would like to comment on the difficulty of performing a direct calculation of solving the eigenvalue problem on replica manifold. 
Opposite to the scalar case, the spinor wavefunction on replicated waveguide geometry $\mathcal{C}^{2} \times \mathbb{I}$ cannot be separated into the product of two individual eigenfunctions on $\mathcal{C}^2$ and $\mathbb{I}$. 
However, analog to the previous investigation on the heat kernel~\cite{Myers2013_MassiveEE}, we find that the dimensional reduction of the spinor Green's function does not require the separation of eigenfunctions (see Appendix~\ref{app:1p1_free_dirac}).

\section{ $(2+1)$D Dirac fermions exposed to a random magnetic field}\label{sec:gauge_dirac}

After recovering the known results of $(2+1)$D free scalar and Dirac fields as a benchmark, in this section, we move onto the case of $(2+1)$D Dirac fermions exposed to a random magnetic field (static gauge field)~\cite{Ludwig1994_IQH_transition, Wen1996conformal, Wen1996Multifractality, Mirlin2001Multifractality, Ryu2009Multifractal, Ye1999}:
\begin{equation}\label{eq:model_gauge}
\mathcal{L} = \overline{\Psi} \gamma^\mu 
(\partial_\mu + i \sqrt{g_A} A_\mu) \Psi 
+ \overline{\Psi} (i \omega \gamma^0) \Psi ,
\end{equation} 
where $A_\mu$ describes the random gauge field (vector potential).  
For simplicity, here $A_\mu$ is chosen to be Gaussian-distributed 
\begin{equation}
\mathcal{P}(A_\mu) \propto 
e^{-\frac{1}{2} \int d^2\mathbf{r}_\parallel 
	A_\mu^2(\mathbf{r}_\parallel)} ,
\end{equation}
with vanishing gauge flux on average. 
By absorbing the coupling constant $\sqrt{g_A}$ into the gauge field $A_\nu$, it is clear that $g_A$ plays the role of the variance of the disorders.

We would like to highlight that this example is quite meaningful. 
First, in the presence of randomness, one cannot exactly solve the eigenvalue problem of Dirac spinor due to the lack of a straightforward field-equation description. 
One may consider an average of the random field, however, this will lead to a certain type of effective interaction of the Dirac field $g_A \left(\overline{\Psi} \gamma^\mu \Psi\right)^2$~\cite{fradkin1986disorder, wenger1994_disorder_Dwave} that is hard to be dealt with by many established tools such as heat-kernel technique. 
Second, instead of a model with explicit interactions, the EE of this model can be numerically calculated up to $\sim 10^4$ lattice sizes (see Sec.~\ref{sec:gauge_dirac_numeric}), which provides an unbiased way to validate our analytical results. As a comparison, for a model with explicit interactions, the numerical calculation of the EE may suffer from strong finite-size effect.
Third, it is conjectured that the random magnetic field leads to a multifractal critical ground state~\cite{Ludwig1994_IQH_transition, Wen1996conformal, Wen1996Multifractality, Fradkin1997multifractal}, based on the traditional numerical/theoretical methods. We anticipate to uncover this criticality from its internal entanglement structure.
In a word, this is a good example to demonstrate the power of our dimensional reduction method.

\subsection{Preliminary results}\label{sec:gauge_dirac_intro}

The study on the localization-delocalization transition induced by disorder is a central subject in condensed matter physics~\cite{anderson1958local, wegner1979disorder, abrahams1979localization, Kramer1993Localization, mirlin2008review_AndersonTransition}. 
It is well known that localization property depends on the dimensionality and underlying symmetry~\cite{wegner1979disorder, abrahams1979localization, kramer1981localization, Pruisken1981Anderson}. 
In history, $(2+1)$D Dirac fermion exposed to a random magnetic field or transverse gauge-field randomness received much attention, which is expected to describe the universality class of the metal-insulator transition in the integer
quantum Hall effect~\cite{Pruisken1983Delocalization, Ludwig1994_IQH_transition, Wen1996conformal, Mirlin2001Multifractality, Ryu2009Multifractal}, the quantum fluctuations in quantum spin
liquids~\cite{savary2016_spin_liquid}, and disordered graphene~\cite{guinea2010graphene}.  
Interestingly, it has been proposed that this problem has an exactly solvable zero-energy wavefunction with multifractal critical scaling behaviors~\cite{Wen1996conformal, Wen1996Multifractality, Fradkin1997multifractal}, which could be immune to randomness and thus escape from localization.

Especially, when $\omega = 0$ in Eq.~\eqref{eq:model_gauge} the random gauge field preserves the chiral symmetry, so that the zero-energy wavefunction of this model remains critical under the perturbation. 
It can be exactly solved within a non-unitary CFT, 
and the multifractal scaling exponents of zero-energy state is determined to be $\Delta = 1 - \frac{g_A}{2\pi}$ as a consequence of negative dimensional operators~\cite{Wen1996conformal}. 
The exponent is continuously tuneable as changing the randomness strength $g_A$, and it becomes negative at $g_c = 2\pi$, indicating a spontaneous symmetry breaking. 

For solving the zero-mode, it is beneficial to apply the Hodge decomposition to the $2$D gauge field 
\begin{equation}
A_\mu = \epsilon_{\mu \nu} \partial_\nu \Phi_1(x) + \partial_\mu \Phi_2(x)
\end{equation}
and introducing the axial gauge transformation 
\begin{equation}
\overline{\Psi} = \overline{\Psi}' e^{\gamma^5 \sqrt{g_A} \Phi_1 + i \sqrt{g_A} \Phi_2} 
, 
\Psi = e^{\gamma^5 \sqrt{g_A} \Phi_1 - i \sqrt{g_A} \Phi_2} \Psi' .
\end{equation}
The original Lagrangian density becomes 
\begin{equation}\label{eq:Lagrangian_axial}
\mathcal{L} = \overline{\Psi}' (\gamma^\mu \partial_\mu + M) \Psi' 
+ i\omega \overline{\Psi}' e^{2 \gamma^5 \sqrt{g_A} \Phi_1} \gamma^0 \Psi' .
\end{equation}
Here we impose a ``mass'' term $M \overline{\Psi}' \Psi' $ into the theory, which measures the gap between ground state and the first excited state in the chiral representation. 
Rather than dealing with the real mass of the original Dirac field, this treatment does not break the chiral symmetry of the fixed points. 
This leads to a simple calculation of the partition function and a reasonable estimation on the scaling behavior with respect to the finite correlation length, which is important for further analysis on the RG flows (see Sec.~\ref{sec:EE_RG}). 
The first term is just a free theory of the axial spinor filed $\{ \Psi', \overline{\Psi}' \}$, and the second term can be calculated perturbatively. 
Since there is no dynamical term of the gauge field, the components after Hodge decomposition can be treated as real scalars. 
Our choice of the Gaussian-distributed probability $\mathcal{P}(A_\mu)$ leads to the equivalence with a massless free scalar theory for both of $\Phi_1$ and $\Phi_2$. 
This ensures the exact representation of the axial transformation and leads to non-perturbative solution of the zero-mode.

\subsection{Explicit derivation of the EE for Dirac fermions exposed to a random magnetic field}\label{sec:gauge_dirac_calc}

Here we calculate the EE in this model by using the dimensional reduction method. 
To achieve this, we need to solve the replicated Green's function $g^{(n)}_{D,{\rm gauge}}(\mathbf{r}_\parallel, \mathbf{r}_\parallel')$ for the Lagrangian in Eq.~\eqref{eq:Lagrangian_axial}. 
The situation is more complicated than the previous free cases, since now we have to deal with an additional random field. 
Fortunately, the non-perturbative solution of zero-mode in $(2+0)$-dimension provides a suitable starting point to apply our dimensional reduction scheme.

To be specific, we construct the approximated excited states
from the exact zero-mode by using the perturbation theory. 
The replicated Green's function of massive free Dirac theory has been shown in Eq.~\eqref{eq:Green_Dirac}, we now consider the effect of random magnetic field as the correction to internal lines in the construction of higher-dimensional theory, which appears in the form of an additional vertex correlator of the longitudinal axial field $\Phi_1$. 
It gives the following perturbation expansion of the replicated Green's function for Dirac field with random static gauge:
\begin{equation}
G^{(n)}_{D,\text{gauge}}(\mathbf{r}_\parallel, \mathbf{r}_\parallel', \omega) 
= \sum_{l=0}^\infty  (-\omega^2)^l
P^{(n)}_{D,2l}
(\mathbf{r}_\parallel, \mathbf{r}_\parallel') , 
\end{equation}
with the $2l$-th order product of $(2+0)$D Green's function
\begin{equation}\label{eq:gauge_expansion_Gn}
\begin{aligned}
& \quad 
P^{(n)}_{D,2l}(\mathbf{r}_\parallel, \mathbf{r}_\parallel')  
=
\int d^2 \mathbf{r}_{\parallel,1} 
\cdots 
\int d^2 \mathbf{r}_{\parallel,2l} 
g^{(n)}_{D}(\mathbf{r}_\parallel, \mathbf{r}_{\parallel,1}) \\ 
& \qquad 
\gamma^0 
g^{(n)}_{D}(\mathbf{r}_{\parallel,1}, \mathbf{r}_{\parallel,2}) 
\mathcal{V}_{g_A}^{(n)}
\gamma^0 
g^{(n)}_{D}(\mathbf{r}_{\parallel,2}, \mathbf{r}_{\parallel,3}) 
\cdots \\
& \qquad 
\gamma^0 
g^{(n)}_{D}(\mathbf{r}_{\parallel,2l-1}, \mathbf{r}_{\parallel,2l}) 
\mathcal{V}_{g_A}^{(n)}
\gamma^0 
g^{(n)}_{D}(\mathbf{r}_{\parallel,2l}, \mathbf{r}_{\parallel}') , 
\end{aligned}
\end{equation}
where $g_{D}^{(n)}$ is the $(2+0)$D replicated Green's function of Dirac field, and
$\mathcal{V}_{g_A}^{(n)} 
= \left\langle 
e^{\sqrt{g_A} \Phi_1(\mathbf{r}_{\parallel,1})} 
e^{-\sqrt{g_A} \Phi_1(\mathbf{r}_{\parallel,2})}
\right\rangle_{\mathcal{R}^{(n)}} $
is the vertex correlator of scalar $\Phi_1$ on $2$D replica manifold that appears as the consequence of disorder averaging. 
Here all odd-order terms are ruled out due to the vanishing vertex correlator. 
To further evaluate the full replicated Green's function, one can simplify the calculation by noticing that only conical singularity of order $(1-n)$ contributes to the EE. 
Meanwhile, due to its non-trivial vanishing at the coincident points, the vertex function can be approximated as $\mathcal{V}_{g_A}^{(n)} \sim \mathcal{V}_{g_A}^{(1)} = |\mathbf{r}_{\parallel, 1} - \mathbf{r}_{\parallel, 2}|^{\frac{g_A}{2\pi}}$ without counting its conical contribution.  
These leads to 
\begin{equation}
\begin{aligned}
& \quad 
P^{(n)}_{D,2l}(\mathbf{r}_\parallel, \mathbf{r}_\parallel') 
- P^{(1)}_{D,2l}(\mathbf{r}_\parallel, \mathbf{r}_\parallel')  \sim (l+1) 
\frac{1-n^2}{6n^2} 
\frac{M}{2\pi} \\ 
& \qquad 
K_0(M\rho)  
\int \frac{d^2 \mathbf{k}}{2\pi}  
e^{i \mathbf{k} \mathbf{r}_\parallel'} 
\frac{ \left[ C(g_A) \right]^{2l} }
{\left( k^2 + M^2 \right)^{l(1+\frac{g_A}{2\pi})+1}} ,
\end{aligned}
\end{equation}
where $C(g_A) = 2^{\frac{g_A}{2\pi}}
\frac{\Gamma(1+\frac{g_A}{4\pi})}{\Gamma(1-\frac{g_A}{4\pi})}$, 
and we have ignored the terms of $\mathcal{O}((1-n)^2)$ that vanish in the EE as taking the replica limit $n \to 1$ of Eq.~\eqref{eq:replica_EE}.

With the above expansion in hand, a resummation of perturbative order $l$ reproduces an approximated $(2+1)$D replicated Green's function $G_{D,\text{gauge}}^{(n)}$, and consequently the partition function $Z^{(n)}$. 
By skipping the sophisticated algebra (see Appendix~\ref{app:gauge_dirac}), the leading term of EE for Dirac fermions exposed to a random static gauge field (magnetic field) is given by 
\begin{equation}\label{eq:EE_dirac_randomgauge}
\begin{aligned}
S_{\rm gauge} 
& = \frac{1}{6} ( 1 - \frac{g_A}{2\pi} \mu_{\rm gauge} ) \frac{\mathcal{A}}{\epsilon^{1+\frac{g_A}{2\pi}}} \\
& \approx \frac{1}{6} \frac{\mathcal{A}}{\epsilon} \left[ 1 - \frac{g_A}{2\pi} \mu_{\rm gauge} + \frac{g_A}{2\pi} \ln \epsilon^{-1}  \right] ,
\end{aligned}
\end{equation}
where $\mathcal{A}$ is the sub-system boundary length, $\mu_{\rm gauge} = \ln 2 - \gamma$ is a positive constant, and $\epsilon$ is an UV cutoff. 
%
Eq.~\eqref{eq:EE_dirac_randomgauge} is the main result of this work. 
It shows that the EE of $(2+1)$D Dirac fermions under random magnetic field remains the area-law scaling. 
The disorder effect only modifies the area-law coefficient. 
Here we only keep the lowest-order correction that is linear in $g_A$ in the current calculation. As we will show below,  by comparing with the numerical simulation the main feature of a random magnetic field is well captured in the current construction.

\subsection{Area-law scaling in the lattice realization}\label{sec:gauge_dirac_numeric}

\begin{figure}\centering
	\includegraphics[width=\columnwidth]{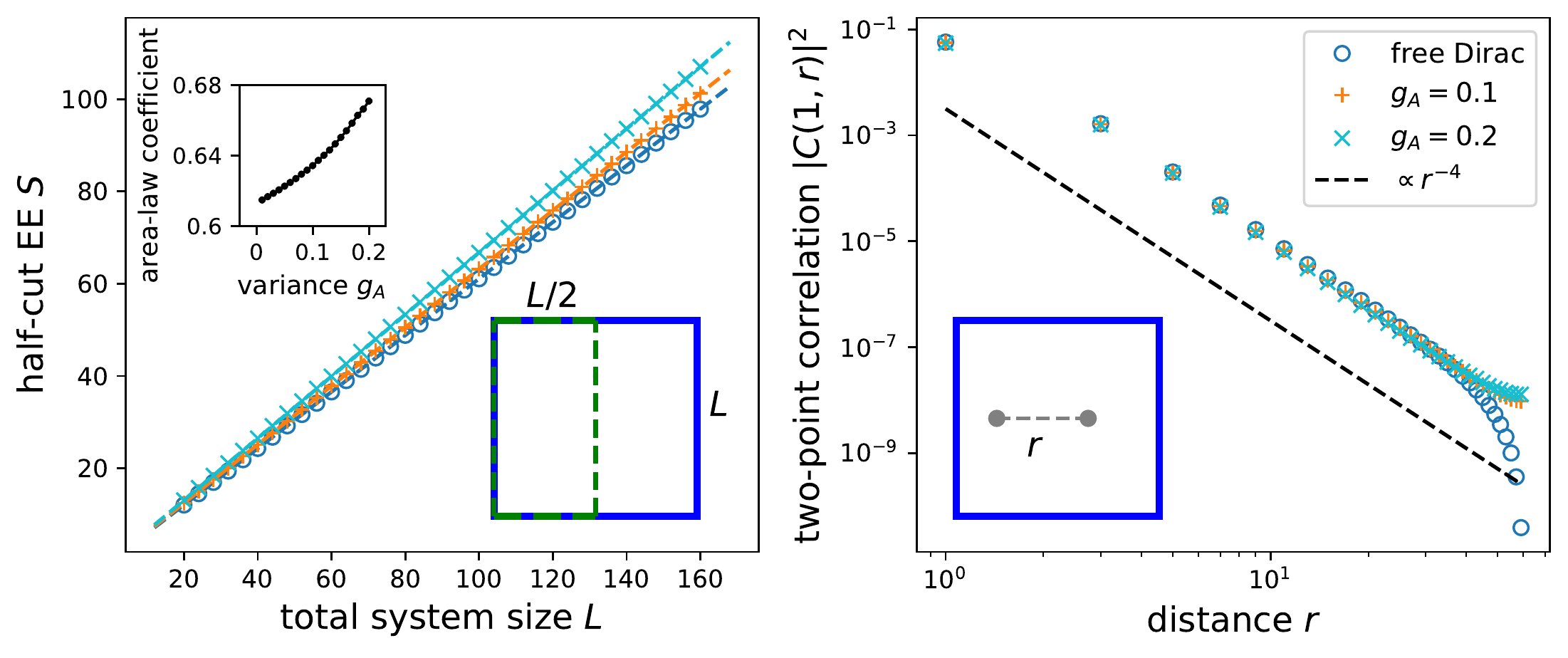}
	\caption{
		\label{fig:HalfCutEE_PiFlux_Square}
		Numerical simulation of the ground state properties of the random flux model on $L \times L$ square lattice. 
		(Left panel) The half-cut EE as a function of the total system size $L$ for randomness strength $g_A=0$ (blue circle), $0.1$ (orange plus) and $0.2$ (cyan crossed). (upper inset) The area-law coefficient as a function randomness strength $g_A$, obtained by a linear scaling of the half-cut EE for system size $L \in [20, 160]$. 
		(Right panel) Double-log plot of two-point correlation $|C(1,r)|^2$ as a function of the distance $r$, for various randomness strength $g_A$ with system size $L=160$. 
		The red dashed line is a power-law scaling of $ |C(1,r)|^2 \sim  r^{-4} $, which is the theoretical value for the free Dirac fermions. 
		Numerically, we have observed that the scaling behavior of $|C(1,r)|^2$ nearly unchanged in the region $g_A \in [0, 0.2]$. 
		For each value of $g_A$, the presented results are averaged over $200$ random realizations. 
	}
\end{figure}

To validate the EE dependence on the subsystem size, we perform a large-scale numerical simulation on the $\pi$-flux square lattice as a typical lattice realization of the Dirac fermion, with implementing the random magnetic field as~\cite{Ludwig1994_IQH_transition, mudry2003_random_hopping, ryu2001_random_flux}:  
\begin{equation}
H = - \sum_{\langle i,j \rangle} 
(-1)^{2 i_x + i_y} e^{i A_{ij}} 
c^\dagger_{i_x,i_y} c_{j_x,j_y} ,
\end{equation}
where $c^\dagger, c$ are the creation and annihilation operators of spinless fermions,  $\langle i,j \rangle$ represents the nearest-neighboring sites, and the random phase factor $A_{ij}$ is set to be Gaussian-distributed with the randomness strength (variance) $g_A$. 
At low-energy limit, this lattice Hamiltonian leads to the massless theory of Eq.~\eqref{eq:model_gauge}.

We numerically calculate the EE of the ground state by using the correlation matrix technique~\cite{chung2001density, Peschel_2003calculation, Peschel_2009reduced}. 
As shown in the left panel of Fig.~\ref{fig:HalfCutEE_PiFlux_Square}, the half-cut EE exhibits a linear growth with the boundary size $L$. 
We confirm that, for a moderate value of randomness strength (variance) $g_A \in[0, 0.5]$, the linear area-law scaling behavior is robust. 
Meanwhile, in the upper inset of Fig.~\ref{fig:HalfCutEE_PiFlux_Square}, we also investigate the dependence of the area-law coefficient on the randomness strength $g_A$. 
It slightly increases with $g_A$, which could be understood by analytical prediction Eq.~\eqref{eq:EE_dirac_randomgauge} if we assume $\epsilon$ as a small number. 
Moreover, numerically we have tested that choosing different kinds of randomness does not lead to qualitative change in these results. 
Therefore, we conclude that our dimensional reduction scheme fairly captures the main features of the EE for $(2+1)$D Dirac fermions exposed to a random magnetic field.

\subsection{Correlation, entanglement and criticality}\label{sec:quasiparticle}

We now turn to discuss the entanglement structure in the $(2+1)$D Dirac field under random magnetic field, and its relation to quantum correlation of the field operator.  
In the right panel of Fig.~\ref{fig:HalfCutEE_PiFlux_Square}, we show the squared two-point correlator after average, which exhibits a power-law scaling at long distance $|C(1,r)|^2 \sim r^{-k}$. Moreover, we numerically  find that the power-law correlation has little change when adjusting the strength $g_A$ of random magnetic field. This motivates us to think about some universal connections between the EE and intrinsic correlations. 

Here, we adopt a quasiparticle picture to describe the EE in scale-invariant fermionic systems~\cite{nahum2020_majorana_defect, tang2021nonunitary}, where the entanglement is considered to be produced by quasiparticle entangled-pairs in the system. 
The only control parameter in this picture is the distribution function of those pairs $P(r)$, which gives the EE 
\begin{equation}\label{eq:quasi_particle}
S_A \sim \int_A dV_A \int_{\overline{A}} dV_{\overline{A}} P(r_{A, \overline{A}}) , 
\end{equation}
where $A$ and $\overline{A}$ are complementary to the total system, and $r_{A, \overline{A}}$ is the distance between the (lattice) points in the two subsystems $A$ and $\overline{A}$. 
Although the current case is a ground state that different from the dynamical steady state with excitations, it is still naturally to understand $P(r_{A, \overline{A}})$ as the squared two-point correlation function of the fermion operator, which gives a power-law decay of $P(r_{A, \overline{A}}) \propto r_{A, \overline{A}}^{-k}$ for scale-invariant systems. 
An estimation of the integral in Eq.~\eqref{eq:quasi_particle} indicates that an area-law EE occurs when $k > 3$ for the spatial dimension $d=2$ (see details in Appendix~\ref{app:quasiparticle}). 
It turns out that the exponent of $k$ determines the scaling behavior of EE, 
so that a numerical estimation of the power-law scaling becomes much more meaningful than the ordinary detection of the scale invariance. 

For $(2+1)$D free Dirac field,  the asymptotic behavior of two-point correlator at the long-distance limit, as $|C(1,r)|^2 \propto r^{-4}$. 
In our finite-size numerics on a lattice model with size $160 \times 160$, we find a close power-law scaling of  $|C(1,r)|^2 \propto r^{-4.3}$. 
When varying the randomness strength (the variance of the random gauge field) $g_A$, the exponent is found to be almost unchanged. 
Plugging in these observations into the quasiparticle picture, it indicates a robust area-law scaling of the EE. 
This is exactly what we have observed in both the field theory calculation and numerical lattice simulation. 
In this context, the current model is one more example that can be understood in the quasiparticle picture phenomenologically. 
Moreover, we also would like to point out that the quasiparticle picture fails to capture the area-law coefficient, as discussed in the previous literature~\cite{pablo2021_EMI}.

\section{Entanglement and renormalization group}\label{sec:EE_RG}

 Besides the area-law scaling of the EE, our scheme is also capable of deriving the sub-leading term of the EE that is relevant with the dynamics of RG flow. 
 The RG flow serves as a coarse-graining of the microscopic degrees of freedom of a physical system, so that it is expected to be an irreversible process between the fixed points. 
 For $(1+1)$D QFTs, the irreversibility theorem of RG flows is known as the famous Zamolodchikov's $c$-theorem~\cite{Zamolodchikov1986Irreversibility}, which proves the existence of a $c$-function (the central charge of CFT that describes the fixed points) that monotonically decreases during RG flows. 
 Seeking for possible extensions of the $c$-theorem to generic dimensions is a long-standing challenge, especially for odd spacetime dimensions  without the concept of central charges~\cite{CARDY1988_4d_c_theorem, Myers2010_holographic_c_theorem, Jafferis2011_F_theorem, Myers2011_holographic_c_theorem, Klebanov2012_3dEE, Casini2012_RG_EE_2p1, Fei2015_generalized_F_theorem, Giombi_2016}. 
 
  \begin{figure}\centering
 	\includegraphics[width=0.66\columnwidth]{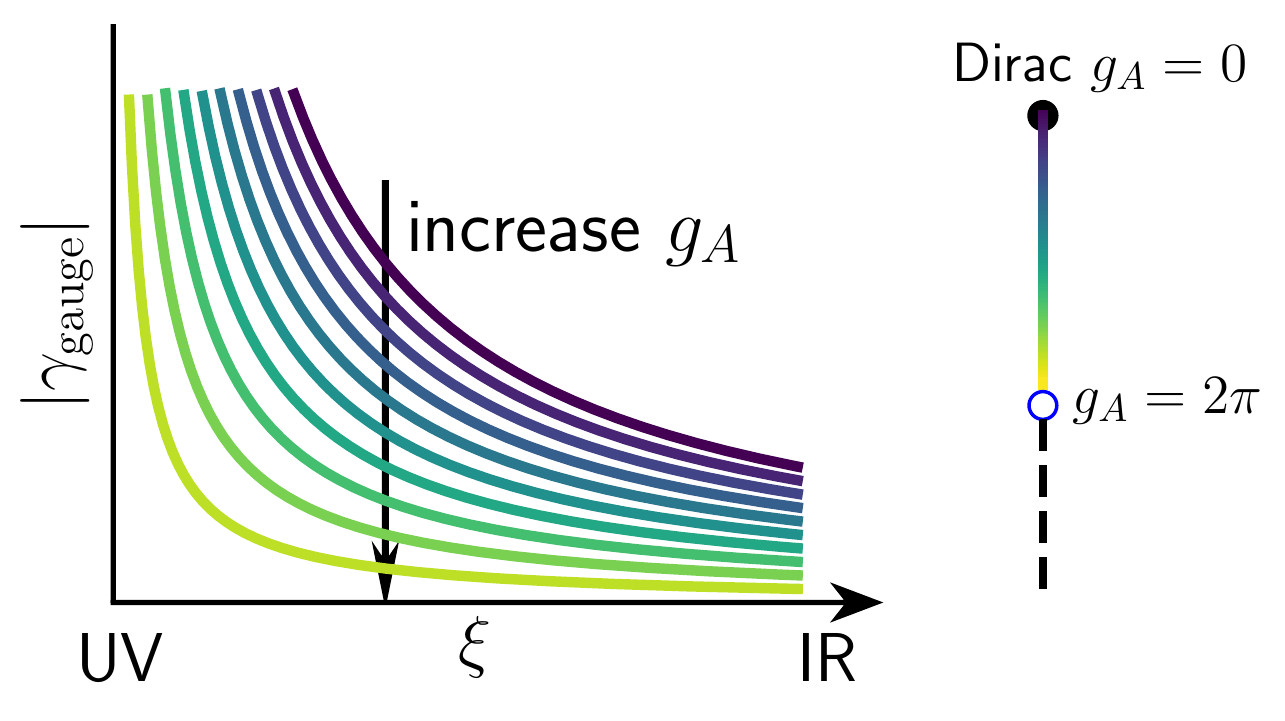}
 	\caption{
 		\label{fig:RGflow}
 		(Left) The sub-leading term $|\gamma_{\mathtt{gauge}}|$ of the EE under the RG flow, where $\xi$ is the correlation length.  
 		(Right) Schematic plot of the critical line including disordered fixed points within $0<g_A<2\pi$, and a spontaneous symmetry breaking occurs at $g_A = 2\pi$~\cite{Ludwig1994_IQH_transition}. 
 	}
 \end{figure}
 
 As mentioned in the introductory part, an important attempting is to understand the irreversibility of RG flows from EE. 
 In $(1+1)$D CFT, the EE is fully determined by the central charge, therefore it is natural to construct an \textit{entropic $c$-function} that points to the irreversibility of RG flows~\cite{CASINI2004_1p1_EE_RG}. 
 Furthermore, this idea is extended into higher dimensions, for which the universal finite term in EE (the $\gamma$ term in Eq.~\eqref{eq:area_law}) is expected to be an analog of the $c$-function~\cite{Myers2010_holographic_c_theorem, Myers2011_holographic_c_theorem, Casini2012_RG_EE_2p1, liu2013RefinementEE, Fei2015_generalized_F_theorem}. 
 Specifically, for $(2+1)$D QFTs, the $\gamma$ term is expected to be negative and satisfying the irreversibility relation~\cite{Casini2012_RG_EE_2p1, liu2013RefinementEE, Grover2011_3dEE_topo, Grover2014_3dEE}
 \begin{equation}\label{eq:F_therem}
 |\gamma_{\rm UV}| \geq |\gamma_{\rm IR}| , 
 \end{equation}
 which serves as a concrete construction of the \textit{$F$-function} that is expected to exist in the $F$-theorem.

In our derivation, this sub-leading term takes the form 
\begin{equation} \label{eq:subleading_EE}
\begin{aligned}
& \gamma_{\text{gauge}} 
\approx  r_{\text{gauge}} (g_A) \mathcal{A} M^{1+\frac{g_A}{2\pi}}
\sim r_{\text{gauge}} (g_A) \frac{\mathcal{A}}{\xi} , 
\end{aligned}
\end{equation}
where $r_{\text{gauge}} =  - \frac{1}{6} ( 1 - \frac{g_A}{2\pi} \mu_{\rm gauge} )$, 
and $\xi \sim M^{1+\frac{g_A}{2\pi}}$ is a finite correlation (see details in Appendix~\ref{app:dirac_randomgauge_full_form}). 
Compared with the leading contribution Eq.~\eqref{eq:EE_dirac_randomgauge}, the sub-leading term is independent of UV cut-off. 
Here, several remarks are given in order.
First, $\gamma_{\text{gauge}}$ is negative, as required by the $F$-theorem in $(2+1)$-dimension~\cite{Tatsuma2018}.
Second,  as one can see in Fig.~\ref{fig:RGflow}(left), the absolute value of $|\gamma_{\text{gauge}}|$ reduces monotonically by approaching the IR limit (increasing $\xi$), which again agrees with the $F$-theorem~\cite{Tatsuma2018}. 
Third, if we focus on the dependence of randomness strength $g_A$, we observe that $|\gamma_{\text{gauge}}|$ monotonically reduces as increasing $g_A$. 
By endowing the dynamical aspect to the EE along RG flows~\cite{Tatsuma2018}, 
i.e. the irreversibility relation in Eq.~\ref{eq:F_therem}, 
the present result of Eq.~\eqref{eq:subleading_EE} indicates a possible dynamic RG flow related to these disorder critical points~\cite{narovlansky2018_disorder_RG} in the perspective of quantum entanglement.

\section{CONCLUSIONS AND OUTLOOKS}\label{sec:conclud}

For a $(2+1)$D QFT with quenched disorders, analytical calculation of the EE is generally difficult. 
In the present work, we have developed a dimensional reduction approach to calculate the entanglement entropy (EE), which is able to deal with the presence of quenched disorders. 
In particular, we transform the $(2+1)$D replicated Green's function to infinite series of the $(2+0)$D (interacting) replicated Green's function, which can be calculated via conventional field theory techniques. 
The derivation can be greatly simplified in the replica limit, 
albeit difficult in evaluation of the quenched disorder on the $n$-fold replica manifold.

We first benchmark it on the free scalar field and Dirac fermion field.
As a non-trivial example, we consider Dirac fermions exposed to a random magnetic (gauge) field, where the traditional methods (in Tab.~\ref{tab:method_EE}) fail to give a straightforward derivation of the EE. 
Based on the proposed approach, we explicitly derive the area-law EE and observe an enhancement of quantum entanglement by the disorder. 
This nontrivial observation indicates the delocalization nature of applying a random static magnetic field to Dirac fermions, which is in contrast with the decaying quantum correlation in ordinary disordered systems. 
To our best knowledge, it has not been studied before.
We further utilize numerical simulation on the lattice model to validate our analytical solution. 
Additionally, we attempt to understand the emergent area-law EE from the microscopic details of quantum correlation, pointing to the critical scaling behavior of the ground state. 
Moreover, we give affirmative evidences that the sub-leading term of the EE monotonically reduces under the renormalization group flow. 
It provides the first piece of evidence to validate the $F$-theorem in $(2+1)$D disordered quantum critical point. 

Here we would like to stress that, the current dimensional reduction scheme is distinct from the existing literature. 
In existing works~\cite{CasiniHuerta2005MassiveScalar, Ryu2006EEandBerryPhase, CasiniHuerta_2009, swingle2012_fermi_liquid_EE, Murciano_2020}, the starting point relies on the known EE function in $(1+1)$D, summation of which gives the EE in $(2+1)$D. This process is conceptually intuitive, however, it is in against to the fact that the EE (of a many-body ground state) is not an extensive quantity, so that it is difficult to be extended into generic cases. 
In this work, to overcome this issue, we explore a distinguished path, based on constructing the $(2+1)$D Green's function using the dimensional reduction method. Compared to the aforementioned methods~\cite{Ryu2006EEandBerryPhase,CasiniHuerta2005MassiveScalar} (see Tab.~\ref{tab:method_EE}), this Green's function based scheme is quite feasible, without any prior knowledge about the EE. 
In addition, a series of works~\cite{Metlitski2009_ON, Whitsitt2017_LargeN_WF, Hung2017_ON}  consider the EE of CFT fixed points with dimension regularization scheme. 
The treatment of quantum corrections distinguishes our method from these calculations. 

Finally, we expect the current methodology advance will inspire fresh perspectives on the study of entanglement structure in $(2+1)$D critical systems. 
In particular, dimensional reduction allows to start with non-perturbative results in low-dimension, which leads to the great advantage in the study of  non-conformal field theories.  
Through an estimation of the scaling behavior of EE based on such a construction, one could provide valuable information for understanding low-energy collective behaviors of the system. 
Furthermore, there are some more extensions of the proposed dimensional reduction method. For example, a similar scheme could be extended to calculate the mutual information, which is another important entanglement measure that provides an upper bound for correlations in quantum theories.

\begin{acknowledgments}
	We thanks Yusen An, Xiao Chen, Shao-Kai Jian, Yin Tang and Zixia Wei for fruitful discussion. 
	Q.T. and W.Z. are supported by ``Pioneer" and ''Leading Goose" R\&D Program of Zhejiang (2022SDXHDX0005) and the foundation from Westlake University.
\end{acknowledgments}

\appendix
\begin{widetext}

\section{Regularization of the UV divergent replicated Green's function at coincident points}\label{app:EMformula}

In this appendix, we show how the Euler-Maclaurin formula gives a regularization of the UV divergent replicated Green's function at coincident points. 
Here the derivation follows the previous work~\cite{Calabrese_2004} by Calabrese and Cardy. 

The $2$D Green's function of the free scalar field on replica manifold is
\begin{equation}
\begin{aligned}
g^{(n)}(\rho, \theta; \rho', \theta') = 
\frac{1}{2\pi n}  
\sum_{q=0}^{\infty} d_q 
\cos \left[ \frac{q}{n} (\theta-\theta') \right] 
\int_0^\infty \frac{J_{q/n}(\lambda \rho) J_{q/n}(\lambda \rho')}
{\lambda^2 + m^2 + k_\perp^2} \lambda d\lambda ,
\end{aligned}
\end{equation}
where $d_0 = 2$ and $d_q = 1$ for $q \geq 1$. 
At coincident points, it becomes
\begin{equation}
g^{(n)}(\rho, \theta; \rho, \theta) 
= \frac{1}{2\pi n}  
\sum_{q=0}^{\infty} d_q 
\int_0^\infty \frac{J_{q/n}(\lambda \rho) J_{q/n}(\lambda \rho)}
{\lambda^2 + m^2 + k_\perp^2} \lambda d\lambda 
= \frac{1}{2\pi n} \sum_{q=0}^{\infty} d_q 
I_{q/n} (m\rho) K_{q/n} (m\rho) ,
\end{equation}
which shows a UV divergence due to summation over infinite many modes labeled by the angular momentum $q$. To regularize it we use the Euler-Maclaurin formula 
\begin{equation}
\int_0^\infty f(q) dq = h \left\{ \frac{f(0)}{2} + f(h) +  f(2h) + ...  \right\} + \sum_{k=1}^\infty
\frac{h^{2k} B_{2k}}{(2k)!} [- (\partial_q)^{2k-1}f(0)] ,
\end{equation}
where $B_k$ is Bernoulli number. We are interested in the case of $h = 1$ and $f(q) = I_{q/n}(m\rho) K_{q/n}(m\rho)$, which is divergent under an integral over $q$. 
For regularization, we insert a function $F(\frac{q}{n\Lambda})$ into $f(q)$, i.e. let $f(q)=I_{q/n}(m\rho) K_{q/n}(m\rho) F(\frac{q}{n\Lambda})$. The function $F(\frac{q}{n\Lambda})$ is chosen that $F(0) = 1$ and $(\partial_q)^i F(0)=0, i \ge 1$. Now, the integral of $\int_0^\infty f(q) dq$ is controlled by the parameter $\Lambda$, 
and goes back to the original form at the limit of $\Lambda \to \infty$.
Then
\begin{equation}
\begin{aligned} 
g^{(n)}(\rho, \theta; \rho, \theta)
& = \frac{1}{2\pi n}
\sum_{q=0} d_q  I_{q/n} (m\rho) K_{q/n}(m\rho) F(\frac{q}{n\Lambda}) \\
& =  \frac{1}{2 \pi n} \left[
2 \int_0^{\infty} I_{q/n} (m\rho) K_{q/n}(m\rho) F(\frac{q}{n\Lambda}) dq 
- 2 \frac{B_2}{2} \left. \frac{\partial}{\partial q} [I_{q/n} (m\rho) K_{q/n}(m\rho)] \right|_{q=0} \right] \\
& \qquad - \frac{1}{2\pi n} \sum_{k=2}^\infty \frac{B_{2k}}{(2k)!} \left. (\frac{\partial}{\partial q})^{2k-1} [I_{q/n} (m\rho) K_{q/n}(m\rho)] \right|_{q=0} \\
& = \frac{1}{2\pi n} \left[
2\int_0^{\infty} I_{q/n} (m\rho) K_{q/n}(m\rho) F(\frac{q}{n\Lambda}) dq 
+ \frac{1}{6n} [K_0(m\rho)]^2
\right] + \mathcal{O}(k=2) ,
\end{aligned} 
\end{equation}
where we have used $B_2 = \frac{1}{6}$, $\left. \frac{\partial I_\nu(z)}{\partial \nu} \right|_{\nu=0} = -K_0(z)$, $\left. \frac{\partial K_\nu(z)}{\partial \nu} \right|_{\nu=0} = 0$.
It should be noticed that although the higher-order derivatives of $[I_{q/n} (m\rho) K_{q/n}(m\rho)]$ at $q=0$ do not vanish, the their integral over $\rho$ all vanishes in the later trace over the plane, as
\begin{equation}
\begin{aligned}
\int_0^\infty \rho d\rho \frac{d^{k}}{dq^k} \left[ I_{q/n}(m\rho) K_{q/n}(m\rho) \right]
= \frac{d^k}{dq^k} \int_0^\infty \rho d\rho \left[ I_{q/n}(m\rho) K_{q/n}(m\rho) \right] 
= \frac{d^k}{dq^k} \left[ \frac{1}{2m^2} \frac{q}{n} \right] 
= 0 \quad {\rm for} \ k > 1.
\end{aligned}
\end{equation}
The above regularization of the replicated Green's function gives
\begin{equation}
g^{(n)}(\rho,\theta;\rho,\theta)
= g^{(1)}(\rho,\theta;\rho,\theta)
+ \frac{1-n^2}{12 \pi n^2} \left[ K_0(m\rho) \right]^2 .
\end{equation}
The first term is just the flat divergence of the Green's function at coincident points, and the second term is the contribution from the conical singularity. 
Moreover, in the calculation of the products of these Green's functions, one can first integrate out the integral measure on the vertex as taking the trace over whole plane, then the higher-order derivatives of $q$ all vanish as the same. 
This makes an approximation of $g^{(n)}$ for general two points in the $(\rho, \theta)$ plane
reasonable in the calculation of dimensional reduction method that is discussed in the main text.

\section{Calculation of constructing of $3$D replicated partition function from $2$D Green's function}\label{app:change_order}

In this appendix, we show that exchanging the order of summation over perturbation levels $l$ and the outside trace does not influence on the result of constructing higher-dimensional ($3$D) replicated partition function. 
Start from Eq. (23) in the main text, before integral out $k$, here we perform the summation over $l$. It gives 
\begin{equation}
\begin{aligned}
G^{(n)} - G^{(1)} 
& = \sum_{l=0}^\infty \left( -k_\perp^2 \right)^l 
\left[ P^{(n)}_l - P^{(1)}_{l} \right] 
= \frac{1-n^2}{12 \pi n^2} K_0(m\rho) 
\int \frac{d^2 \mathbf{k}_\parallel}{2\pi}
e^{i \mathbf{k}_\parallel \mathbf{r}_\parallel'}
\sum_{l=0}^\infty (l+1) \left( \frac{-k_\perp^2}{k_\parallel^2 + m^2} \right)^l \\
& = \frac{1-n^2}{12 \pi n^2} K_0(m\rho) 
\bigg[
K_0(\sqrt{m^2 + k_\perp^2} \rho') 
- \frac{\rho' k_\perp^2}{2 \sqrt{m^2 + k_\perp^2}} 
K_1(\sqrt{m^2 + k_\perp^2} \rho') 
\bigg] .
\end{aligned}
\end{equation}
The trace in $3$D replica spacetime is then given by 
\begin{equation}
\begin{aligned}
\Tr^{(n)} G^{(n)} - n \Tr^{(1)} G^{(1)} 
& = \Tr^{(n)} \left[ G^{(n)} - G^{(1)} \right] 
= \int dr_\perp 
\int \frac{dk_\perp}{2\pi} 
\int d^2 \mathbf{r}_\parallel 
\left[ G^{(n)} - G^{(1)} \right]  \\
& = \frac{1-n^2}{12 n} \int dr_\perp 
\int \frac{dk_\perp}{2\pi} \frac{1}{k_\perp^2 + m^2} .
\end{aligned}
\end{equation}
It is clear to see that the result of $\Tr G^{(n)}$
is identical to the calculation that is presented in the main text, 
and of course leads the same result of EE. 

\section{The solution of the replicated Green's function for $(1+1)$D massive free Dirac field}
\label{app:1p1_free_dirac}

\subsection{A direct derivation of the spinor Green's function on $2$D replica manifold}

In this section, we present a detailed derivation of the replicated Green's function for the $(1+1)$D free Dirac field. 
The Lagrangian density of Dirac field in $2$D (Euclidean) space is
\begin{equation}
\mathcal{L} = \overline{\Psi} (\gamma^\mu\partial_\mu+m) \Psi .
\end{equation}
We choose the representation of gamma matrices to be
\begin{equation}
\gamma^0 = \sigma_3 = \left(\begin{matrix} 1 & 0 \\ 0 & -1 \end{matrix}\right) ,
\gamma^1 = \sigma_1 = \left(\begin{matrix} 0 & 1 \\ 1 & 0 \end{matrix}\right) ,
\gamma^2 = \sigma_2 = \left(\begin{matrix} 0 & -i \\ i & 0 \end{matrix}\right) .
\end{equation}
By applying variation to the Lagrangian, we have the spinor Green's function satisfies
\begin{equation}
(\gamma^\mu\partial_\mu+m) G_D^{(n)}(r,r') = \delta^{[2]}(r-r') ,
\end{equation}
its explicit matrix form is
\begin{equation}
\left( \begin{matrix}
m & \partial_x - i\partial_y \\
\partial_x + i\partial_y & m
\end{matrix} \right)
\left( \begin{matrix}
G_{11} & G_{12} \\ G_{21} & G_{22}
\end{matrix} \right)
=  \left( \begin{matrix} \delta^{2} (r-r') & 0 \\ 0 & \delta^{2} (r-r') \end{matrix} \right)
\end{equation}

To calculate the Green's function we solve the eigenvalue problem
\begin{equation}
\gamma^\mu \partial_\mu \Psi = - \lambda \Psi .
\end{equation}
It is important to notice that the spinor differential operator 
$\gamma^\mu \partial_\nu$ is anti-hermitian, so that its eigenvalue 
is purely imaginary. For convenience, we rewrite the above equation to be 
\begin{equation}
\gamma^\mu \partial_\mu \Psi = - i \lambda \Psi ,
\end{equation}
with $\lambda$ real.

Write it explicitly in the matrix form we have
\begin{equation}
\left( \begin{matrix}
0 & \partial_x - i\partial_y \\
\partial_x + i\partial_y & 0
\end{matrix} \right)
\left( \begin{matrix}
\Psi_1 \\ \Psi_2
\end{matrix} \right)
= - i \lambda \left( \begin{matrix} \Psi_1 \\ \Psi_2 \end{matrix} \right) .
\end{equation}
In the polar coordinates, we have
\begin{equation}
\partial_x - i\partial_y = e^{-i\theta} \left[ \partial_\rho - \frac{i}{\rho} \partial_\theta \right]
\ , \qquad
\partial_x + i\partial_y = e^{i\theta} \left[ \partial_\rho + \frac{i}{\rho} \partial_\theta \right]
\ ,
\end{equation}
use this to translate the eigenvalue problem, it becomes
\begin{equation}
\begin{aligned}
e^{-i\theta} [\partial_\rho - \frac{i}{\rho} \partial_\theta] 
\Psi_2(\rho, \theta) = -i\lambda \Psi_1(\rho, \theta) , \\
e^{i\theta} [\partial_\rho + \frac{i}{\rho} \partial_\theta] 
\Psi_1(\rho, \theta) = -i\lambda \Psi_2(\rho, \theta) . \\
\end{aligned}
\end{equation}
Assuming the solution of $\Psi_1$ has the form
\begin{equation}
\Psi_1 (\rho, \theta) = A e^{i\nu\theta} R_1(\rho) ,
\end{equation}
where $\nu = q/n$ with integer $q$ (take both negative and non-negative values).
This form satisfies the periodic boundary condition in the angular direction $\Psi_1(\rho, \theta+2\pi n) = \Psi_1(\rho,\theta)$, and it gives
\begin{equation}
e^{i\theta} [\partial_\rho + \frac{i}{\rho} \partial_\theta] \Psi_1(\rho, \theta) 
= e^{i\theta} [\partial_\rho + \frac{i}{\rho} \partial_\theta]
\left[ A e^{i\nu\theta} R_1(\rho) \right]
= A e^{i(\nu+1)\theta} \left[ \partial_\rho R_1(\rho) - \frac{\nu}{\rho} R_1(\rho) \right] .
\end{equation}
According to this, we assume
\begin{equation}
\Psi_2 (\rho, \theta) = A e^{i(\nu+1)\theta} R_2(\rho) .
\end{equation}
We then have
\begin{equation}
-i\lambda R_1(\rho) = \left[ \frac{d}{d\rho} + \frac{\nu+1}{\rho} \right] R_2(\rho)
\ , \qquad \
-i\lambda R_2(\rho) = \left[ \frac{d}{d\rho} - \frac{\nu}{\rho} \right] R_1(\rho) .
\end{equation}
This gives
\begin{equation}\label{eq:differential_dirac}
\begin{aligned}
\rho^2 \frac{d^2 R_1(\rho)}{d\rho^2} + \rho \frac{d R_1(\rho)}{d\rho} + (\lambda^2 \rho^2 - \nu^2) R_1(\rho) = 0 ,
\\
\rho^2 \frac{d^2 R_2(\rho)}{d\rho^2} + \rho \frac{d R_2(\rho)}{d\rho} + (\lambda^2 \rho^2 - (\nu+1)^2) R_2(\rho) = 0 ,
\end{aligned}
\end{equation}
which is the $(\nu+1)$-th order Bessel equation, 
its solution is the Bessel function
\begin{equation}
R_1(\rho) = J_\nu(\lambda \rho) \ , \qquad R_2(\rho) = J_{\nu+1}(\lambda \rho) \ .
\end{equation}
Finally, we have the solution of the eigenvalue problem as
\begin{equation}\label{eq:dirac_wavefunc}
\Psi = A \left(
\begin{matrix}
e^{i\nu\theta} J_\nu(\lambda \rho) \\
e^{i(\nu+1)\theta} J_{\nu+1}(\lambda \rho) \\
\end{matrix}
\right) ,
\end{equation}
where $A$ is the normalization factor.
Note, there is still no constrain on the value that the eigenvalue $\lambda$ can take, and $\nu = q/n$ that $q$ runs over all integers (including negative and zero).

We then require the boundary condition that $R_1(\rho)$ vanishes at the boundary $\rho = L$, which means that the eigenvalues satisfy
\begin{equation}
\lambda_{\nu, i} = \frac{\alpha_{\nu, i}}{L} ,
\end{equation}
where $\alpha_{\nu, i}$ is the zeros of $\nu$-th order Bessel function of the first kind.
It is important to notice that the solution of eigenfunction has a ``particle-hole'' symmetry with respect to the sign of the eigenvalue $\lambda_{\nu, i}$ (it is also the sign of the ``angular momentum'' $\nu$). 
When we switch the sign of the eigenvalue $\lambda \to -\lambda$, the choice of $R_1(\rho) \to -R_1(\rho)$ makes the differential equation of Eq.~\eqref{eq:differential_dirac} unchanged.
Require the form of Eq.~\eqref{eq:dirac_wavefunc} is only valid for the positive eigenvalues, then for negative eigenvalues we have
\begin{equation}
\Psi^{-} = A \left(
\begin{matrix}
e^{i\nu\theta} J_\nu(\lambda \rho) \\
-e^{i(\nu+1)\theta} J_{\nu+1}(\lambda \rho) \\
\end{matrix}
\right) .
\end{equation}

The normalization factor $A_{\nu,i}$ can be calculated as
\begin{equation}
\begin{aligned}
1 & = \int_0^L \rho d\rho \int_0^{2\pi n} d\theta 
\Psi^\dagger(\rho,\theta) \Psi(\rho,\theta)
= |A_{\nu,i}|^2 \int_0^L \rho d\rho \int_0^{2\pi n} 
\left\{ \left[ J_{\nu}(\lambda_{\nu,i}\rho) \right]^2 
+ \left[ J_{\nu+1}(\lambda_{\nu,i}\rho) \right]^2 \right\} \\
& = |A_{\nu,i}|^2 2\pi n \left\{ 
\frac{L^2}{2} \left[ J_{\nu+1}(\lambda_{\nu,i}L) \right]^2
+ \frac{L^2}{2} \left[ J_{\nu+1}(\lambda_{\nu,i}L) \right]^2
\right\} \\
& \quad \quad \Longrightarrow \qquad \qquad \qquad \qquad
|A_{\nu,i}|^2 = \frac{1}{2\pi n L^2} \frac{1}{\left[ J_{\nu+1}(\lambda_{\nu,i}L) \right]^2}
\end{aligned}
\end{equation}
Note that the above equation has no typo, the two integrals of Bessel functions with different order indeed give the same result due to the nice properties that $\lambda_{\nu, i} L$ is the zero of $J_\nu$. This is ensured by the following fact.
First, recall that we have the recurrence relationship
\begin{equation}
\frac{d}{dx} \left[ \frac{J_\nu(x)}{x^\nu} \right] = - \frac{J_{\nu+1}(x)}{x^\nu} , \qquad
\frac{d}{dx} \left[ x^\nu J_\nu(x) \right] = x^\nu J_{\nu-1}(x) .
\end{equation}
If we take the zeros $x=\alpha_{\nu, i}$, it becomes
\begin{equation}
J'_\nu(\alpha_{\nu, i}) = -J_{\nu+1}(\alpha_{\nu, i}) , \qquad
J'_{\nu}(\alpha_{\nu, i}) = J_{\nu-1}(\alpha_{\nu, i}) .
\end{equation}
Now we see that the $(\nu+1)$-th and $(\nu-1)$-th Bessel functions are just different in a sign.
Second, we have the integral of the double product of Bessel functions as
\begin{equation}
\int t dt [J_\nu(at)]^2 = \frac{t^2}{2} \left\{
\left[ J_\nu(at) \right]^2 - J_{\nu+1}(at) J_{\nu-1}(at)
\right\} .
\end{equation}
By using
\begin{equation}
-J_{\nu+1}(\alpha_{\nu, i}) = J_{\nu-1}(\alpha_{\nu, i}) ,
\end{equation}
we have
\begin{equation}
\int_0^L t dt [J_\nu(\frac{\alpha_{\nu, i}}{L}t)]^2
= -\frac{L^2}{2} J_{\nu+1}(\alpha_{\nu, i}) J_{\nu-1}(\alpha_{\nu, i})
= \frac{L^2}{2} \left[ J_{\nu+1}(\alpha_{\nu, i}) \right]^2 .
\end{equation}
On the other hand
\begin{equation}
\int_0^L t dt [J_{\nu+1}(\frac{\alpha_{\nu, i}}{L}t)]^2
= \frac{L^2}{2} \left[ J_{\nu+1}(\alpha_{\nu, i}) \right]^2 .
\end{equation}

We move to the calculation of Green's function, let 
\begin{equation}
G_D^{(n)} (\rho, \theta; \rho', \theta')
= \sum_{\nu,i} C_{\nu, i, n} \psi_{\nu, i, n}(\rho, \theta) .
\end{equation}
Substitute it back to the Dirac equation, we have
\begin{equation}
\sum_{\nu,i} (i\lambda_{\nu,i} + m) C_{\nu,i,n} \psi_{\nu,i,n}
= \delta(\rho - \rho') \delta(\theta - \theta') ,
\end{equation}
which leads to
\begin{equation}
G_D^{(n)} (\rho, \theta; \rho', \theta')
= \sum_{\nu,i} \frac{1}{i\lambda_{\nu,i} + m} 
\psi_{\nu, i, n}(\rho, \theta)
\psi_{\nu, i, n}^{\dagger}(\rho', \theta') .
\end{equation}
Note that here the eigenvalues ${\lambda_{\nu, i}} = \alpha_{\nu, i}/L$ can take both negative and non-negative values.

Separation of the negative and non-negative will simplify the calculation of the Green's function, as
\begin{equation}
\begin{aligned}
& G_D^{(n)} (\rho, \theta; \rho', \theta') 
= \sum_{\nu,i} \frac{\psi_{\nu, i, n, +}(\rho, \theta) \psi_{\nu, i, n, +}^{\dagger}(\rho', \theta')}{i\lambda_{\nu,i} + m} 
+ \sum_{\nu,i} \frac{\psi_{\nu, i, n, -}(\rho, \theta) \psi_{\nu, i, n, -}^{\dagger}(\rho', \theta')}{-i\lambda_{\nu,i} + m} \\
= & \ \sum_{\nu,i} |A_{\nu, i}|^2 \frac{
	\left(
	\begin{matrix}
	e^{i\nu(\theta-\theta')} J_\nu(\lambda \rho) J_\nu(\lambda \rho') & \quad
	& e^{i\nu(\theta-\theta')} e^{-i\theta'} J_{\nu}(\lambda \rho) J_{\nu+1}(\lambda \rho') \\
	e^{i\nu(\theta-\theta')} e^{i\theta} J_{\nu+1}(\lambda \rho) J_{\nu}(\lambda \rho') &
	& e^{i(\nu+1)(\theta-\theta')} J_{\nu+1}(\lambda \rho) J_{\nu+1}(\lambda \rho') \\
	\end{matrix}
	\right)
}{i\lambda_{\nu,i} + m} \\
& + \sum_{\nu,i} |A_{\nu, i}|^2 \frac{
	\left(
	\begin{matrix}
	e^{i\nu(\theta-\theta')} J_\nu(\lambda \rho) J_\nu(\lambda \rho') & \quad
	& - e^{i\nu(\theta-\theta')} e^{-i\theta'} J_{\nu}(\lambda \rho) J_{\nu+1}(\lambda \rho') \\
	- e^{i\nu(\theta-\theta')} e^{i\theta} J_{\nu+1}(\lambda \rho) J_{\nu}(\lambda \rho') &
	& e^{i(\nu+1)(\theta-\theta')} J_{\nu+1}(\lambda \rho) J_{\nu+1}(\lambda \rho') \\
	\end{matrix}
	\right)
}{-i\lambda_{\nu,i} + m} \\
= & \ \sum_{\nu,i} \frac{|A_{\nu,i}|^2}{-\lambda_{\nu,i}^2 + m^2} 
\left(
\begin{matrix}
2m e^{i\nu(\theta-\theta')} J_\nu(\lambda \rho) J_\nu(\lambda \rho') & \quad
& 2i\lambda e^{i\nu(\theta-\theta')} e^{-i\theta'} J_{\nu}(\lambda \rho) J_{\nu+1}(\lambda \rho') \\
2i\lambda e^{i\nu(\theta-\theta')} e^{i\theta} J_{\nu+1}(\lambda \rho) J_{\nu}(\lambda \rho') &
& 2m e^{i(\nu+1)(\theta-\theta')} J_{\nu+1}(\lambda \rho) J_{\nu+1}(\lambda \rho') \\
\end{matrix}
\right)
\end{aligned}
\end{equation}
where we have set $\lambda_{\nu, i} \ge 0$. Note that here the index $\nu = q/n$, and $q$ takes both negative and non-negative integers.
Here we do not need to worry about the problem of double counting on the zero eigenvalue.
The double counted terms of $\lambda = 0$ is the zero of all non-zeroth order Bessel functions, 
so that all double counted terms vanish.

Now we extend the solution into thermodynamic limit $L \to \infty$. The normalization factor becomes 
\begin{equation}
\lim_{L\to\infty} |A_{\nu,i}|^2
= \lim_{L\to\infty} \frac{1}{2 \pi n L^2 \left[ J_{\nu+1} (\lambda L) \right]^2} 
= \frac{1}{2\pi n L^2 \frac{2}{\pi \lambda L}}
= \frac{\lambda}{4nL} ,
\end{equation}
and the value of $\lambda$ becomes continuous for each given order $\nu$.
This means that the summation of index $i$ changes into the integral over $\lambda$ via 
$\frac{1}{L} \sum_i \to \frac{1}{2\pi} \int_0^\infty d\lambda$,
then we have the spinor Green's function 
\begin{equation}
G_D^{(n)} (\rho, \theta; \rho', \theta') 
= \frac{1}{4\pi n} \sum_\nu e^{i\nu(\theta-\theta')}
\int_0^\infty d\lambda \frac{\lambda }{\lambda^2+m^2} 
\left(
\begin{matrix}
m J_\nu(\lambda \rho) J_\nu(\lambda \rho') & \quad
& i\lambda e^{-i\theta'} J_{\nu}(\lambda \rho) J_{\nu+1}(\lambda \rho') \\
i\lambda e^{i\theta} J_{\nu+1}(\lambda \rho) J_{\nu}(\lambda \rho') &
& m e^{i(\theta-\theta')} J_{\nu+1}(\lambda \rho) J_{\nu+1}(\lambda \rho') \\
\end{matrix}
\right) ,
\end{equation}
where $\nu=q/n$ with integer $q$. 

\subsection{Entanglement entropy of $(1+1)$D free Dirac field}

As long as the $2$D replicated Green's function is obtained, we can calculate the entanglement entropy of $(1+1)$D free Dirac field. The relation between the partition function and the Green's function is
\begin{equation}
- \frac{\partial \ln Z_D}{\partial m} = \Tr G_D .
\end{equation}
Here the trace contains the sum over diagonal components of the spinor Green's function, so that requires the explicit form of them. 
The integral of diagonal components has been already calculated in the free scalar field.
The sum over index $\nu$ has the following form for both $G_{11}$ and $G_{22}$
\begin{equation}
\begin{aligned}
& \frac{m}{4\pi n} \sum_{q=-\infty}^\infty e^{i\frac{q}{n}(\theta-\theta')} 
\int_0^\infty d\lambda \frac{\lambda}{\lambda^2+m^2} 
J_{q/n}(\lambda \rho) J_{q/n}(\lambda \rho') \\
= & \frac{m}{4\pi n} \sum_{q=-\infty}^\infty e^{i\frac{q}{n}(\theta-\theta')} 
\frac{\pi i}{2} \left[
\theta(\rho'-\rho) J_{q/n}(im \rho) H_{q/n}^{(1)}(im \rho') 
+ \theta(\rho-\rho') J_{q/n}(im \rho') H_{q/n}^{(1)}(im \rho) 
\right]
\\
= & \frac{m}{4\pi n} \sum_{q=-\infty}^\infty e^{i\frac{q}{n}(\theta-\theta')} 
\left[
\theta(\rho'-\rho) I_{q/n}(m\rho) K_{q/n}(m\rho')
+ \theta(\rho-\rho') I_{q/n}(m\rho') K_{q/n}(m\rho)
\right] .
\end{aligned}
\end{equation}

Then the trace of the Green's function becomes
\begin{equation}
\Tr G_D^{(n)}
= 2 \left(\frac{m}{4\pi n}\right) 
\int_0^{2\pi n} d\theta
\int_0^\infty \rho d\rho
\sum_{q=0}^\infty d_q 
I_{q/n}(m\rho) K_{q/n}(m\rho) 
= m 
\int_0^\infty \rho d\rho
\sum_{q=0}^\infty d_q 
I_{q/n}(m\rho) K_{q/n}(m\rho) ,
\end{equation}
The summation over $q$ is UV divergent, so that we introduce a renormalization function $F(\frac{q}{n\Lambda})$, which is chosen that $F(0)=1$ and $F^{(k)}(0)=0$.
Similar to the $2$D free scalar field, from the Euler-Maclaurin formula, we have
\begin{equation}
\sum_{q=0}^\infty d_q I_{q/n}(m\rho) K_{q/n}(m\rho) F(\frac{q}{n\Lambda}) 
= 2 \int_0^{\infty} I_{q/n}(m\rho) K_{q/n}(m\rho) F(\frac{q}{n\Lambda}) dq
+ \frac{1}{6n} [K_0(m\rho)]^2 .
\end{equation}
Then
\begin{equation}
\begin{aligned}
\Tr G_D^{(n)} & = m \int_0^\infty \rho d\rho
\left[
2 \int_0^{\infty} I_{q/n}(m\rho) K_{q/n}(m\rho) F(\frac{q}{n\Lambda}) dq
+ \frac{1}{6n} [K_0(m\rho)]^2
\right] \\
& = 2mn \int_0^\infty \rho d\rho \int_0^\infty 
I_{q}(m\rho) K_{q}(m\rho) F(\frac{q}{\Lambda}) dq
+ \frac{m}{6n} \int_0^\infty \rho d\rho [K_0(m\rho)]^2 \\
& = 2mn C(m) + \frac{m}{6n} \left[ \frac{1}{2m^2} \right]
= 2mn C(m) + \frac{1}{12mn}
\end{aligned}
\end{equation}
From this we have
\begin{equation}
-\frac{\partial}{\partial m} \ln \frac{Z_D^{(n)}}{\left[ Z_D^{(1)} \right]^n} 
= \Tr^{(n)} G_D^{(n)} - n \Tr^{(1)} G_D^{(1)} = \frac{1-n^2}{12mn} . 
\end{equation}
We intermediately notice that this will lead to the same partition function as the free scalar field
\begin{equation}
\ln \frac{Z_D^{(n)}}{\left[ Z_D^{(1)} \right]^n} = \frac{n^2-1}{12n} \ln m + C .
\end{equation}
By letting $C = \frac{n^2-1}{12n} \ln \epsilon$, it gives the entanglement entropy
\begin{equation}
S_D = - \left. \frac{\partial}{\partial n} 
\frac{Z_D^{(n)}}{\left[ Z_D^{(1)} \right]^n}  \right|_{n=1}
= - \frac{1}{6} \ln(m \epsilon) ,
\end{equation}
where $\epsilon$ plays the role of UV cutoff of the theory (lattice constant).

\subsection{Reduction to the usual flat spinor Green's function at the replica limit $n \to 1$}

In this section, we show that by taking the replica limit $n \to 1$, the above solved replicated Green's function is reduced to the usual flat one. 
The key is using the Addition Theorem of the Bessel function.

Starting from the diagonal components, they are 
\begin{equation}
G_{11}^{(1)} = G_{22}^{(1)} 
= \frac{m}{4\pi} \sum_q e^{iq(\theta-\theta')}
\int_0^\infty \frac{\lambda J_q(\lambda \rho) 
	J_q(\lambda \rho')}{\lambda^2+m^2} 
d\lambda 
\end{equation}
One can exchange the summation and the integral, it gives 
\begin{equation}
G_{11}^{(1)} = G_{22}^{(1)} 
=
\frac{m}{4\pi} \int_0^\infty d\lambda 
\frac{\lambda}{\lambda^2+m^2} 
\sum_q e^{iq(\theta-\theta')} 
J_q(\lambda \rho) 
J_q(\lambda \rho')
= \frac{m}{4\pi} \int_0^\infty \frac{\lambda J_0(\lambda R)}{\lambda^2 + m^2} d\lambda
= \frac{m}{4\pi} K_0(mR) ,
\end{equation}
where we have applied the (Neumann) Addition Theorem of the Bessel function
\begin{equation}
J_0(R) = \sum_0^\infty d_q \cos \left[ q (\theta - \theta') \right] J_q(\rho) J_q(\rho') ,
\end{equation} 
with $R = \sqrt{\rho^2 + \rho'^2 - 2\rho\rho'\cos(\theta-\theta')}$ is the distance between $(\rho,\theta)$ and $(\rho',\theta')$.

For the off-diagonal component, we have 
\begin{equation}
\begin{aligned}
G_{12}^{(1)} 
& = \frac{i}{4\pi} \sum_q e^{iq(\theta-\theta')} \int_0^\infty d\lambda 
\frac{\lambda^2}{\lambda^2 + m^2} 
e^{-i\theta'} J_q(\lambda \rho) J_{q+1}(\lambda \rho') \\
& = \frac{i}{4\pi} \int_0^\infty d\lambda \frac{\lambda^2}{\lambda^2 + m^2} 
e^{-i\theta'} \sum_q e^{iq(\theta-\theta')} 
J_q(\lambda \rho) J_{q+1}(\lambda \rho') 
=  \frac{i}{4\pi} \int_0^\infty d\lambda 
\frac{\lambda^2 J_1(\lambda R)}{\lambda^2 + m^2} 
e^{-i\theta'} \left[
\frac{\rho' - \rho e^{-i(\theta-\theta')}}
{\rho' - \rho e^{+i(\theta-\theta')}}
\right]^{\frac{1}{2}} \\
& = \frac{i}{4\pi} \int_0^\infty d\lambda 
\frac{\lambda^2 J_1(\lambda R)}{\lambda^2 + m^2} 
e^{-i\theta'} \frac{\rho' - \rho e^{-i(\theta-\theta')}}{R} 
= \frac{i}{4\pi} m K_1(m R) \frac{\rho' e^{-i\theta'} - \rho e^{-i\theta}}{R} .
\end{aligned}
\end{equation}
Similarly
\begin{equation}
\begin{aligned}
G_{21}^{(1)} 
& = \frac{i}{4\pi} \sum_q e^{iq(\theta-\theta')} \int_0^\infty d\lambda 
\frac{\lambda^2}{\lambda^2 + m^2} 
e^{i\theta} J_{q+1}(\lambda \rho) J_{q}(\lambda \rho') \\
& = \frac{i}{4\pi} \int_0^\infty d\lambda \frac{\lambda^2}{\lambda^2 + m^2} 
e^{i\theta'} \sum_q e^{iq(\theta-\theta')} 
J_{q}(\lambda \rho) J_{q-1}(\lambda \rho') 
= \frac{i}{4\pi} \int_0^\infty d\lambda 
\frac{\lambda^2 J_{-1}(\lambda R)}{\lambda^2 + m^2} 
e^{i\theta'} \left[
\frac{\rho' - \rho e^{-i(\theta-\theta')}}
{\rho' - \rho e^{+i(\theta-\theta')}}
\right]^{-\frac{1}{2}} \\ 
& = \frac{-i}{4\pi} \int_0^\infty d\lambda 
\frac{\lambda^2 J_1(\lambda R)}{\lambda^2 + m^2} 
e^{i\theta'} \frac{\rho' - \rho e^{+i(\theta-\theta')}}{R} 
= \frac{-i}{4\pi} m K_1(m R) \frac{\rho' e^{+i\theta'} - \rho e^{+i\theta}}{R} .
\end{aligned}
\end{equation}

This is identical to the solution of usual Green's function in flat Euclidean spacetime, which has the following form
\begin{equation}
G_{ab}(x,x') = \int \frac{d^2 k}{(2\pi)^2} e^{ik(x-x')} 
\frac{(i \gamma^i k_i + m)_{ab}}{k^2 + m^2} 
= \frac{m}{2\pi} \left(\begin{matrix}
K_0(mR) & e^{i \arctan \frac{R_1}{R_2}} K_1(m R) \\
e^{-i \arctan \frac{R_1}{R_2}} K_1(m R) & K_0(mR)
\end{matrix}\right) .
\end{equation}
The difference of a factor $\frac{1}{2}$ comes from the choice of normalization condition. In our approach, we choose to normalization the entire spinor. In the usual convention, the normalization is taken for each component of the spinor.

\section{$(2+1)$D Dirac fermions under a random magnetic field}\label{app:gauge_dirac}

\subsection{Preliminary results of the usual flat Green's function}\label{app:gauge_dirac_flat}

In this section, we present a detailed derivation of some known results of the Dirac field under a random static gauge field (magnetic field). The $2$D reduction of the Lagrangian in Minkowski spacetime is 
\begin{equation}
\mathcal{L}[\omega] = \overline{\Psi} [ i \gamma^\mu 
(\partial_\mu - i \sqrt{g_A} A_\mu) ] \Psi 
+ \overline{\Psi} (\omega \gamma^0) \Psi ,
\end{equation} 
where $\omega$ is the frequency (energy scale). Here we choose $g^{\mu}_{\nu} = {\text{diag}}(-1, +1, +1)$, $\gamma^5 = \gamma^0 = i\gamma^1 \gamma^2 = \sigma_3$ with $\gamma^1 = i\sigma_1$ and $\gamma^2 = i\sigma_2$.
For simplicity, the $2$D random static gauge field $A_\mu$ is chosen to be Gaussian-distributed 
\begin{equation}
\mathcal{P}(A_\mu) \propto 
e^{-\frac{1}{2} \int d^2\mathbf{r}_\parallel 
	A_\mu^2(\mathbf{r}_\parallel)} ,
\end{equation} 
with vanishing mean value and variance $g_A$. 

Let us start from the case of $\omega = 0$. 
First, we apply the Hodge decomposition for the $2$D static gauge field 
\begin{equation}
A_\mu = \epsilon \partial_\nu \Phi_1(\mathbf{r}_\parallel) + \partial_\mu \Phi_2(\mathbf{r}_\parallel),
\end{equation}
where $\epsilon_{\mu \nu}$ is the Levi-Cevita tensor, $\Phi_1$ and $\Phi_2$ are longitudinal and transverse components.
This gives 
\begin{equation}
\mathcal{L}[\omega = 0] = \overline{\Psi} \left\{
i\gamma^\mu \left[ \partial_\mu - i\sqrt{g_A}
(\epsilon_{\mu \nu} \partial_{\nu} \Phi_1 + \partial_\mu \Phi_2)
\right] \right\} \Psi 
\end{equation}
Second, we introduce the following axial gauge transformation 
\begin{equation}
\overline{\Psi}  = \overline{\Psi}' 
e^{\gamma^5 \sqrt{g_A} \Phi_1 - i \sqrt{g_A} \Phi_2} 
, \qquad
\Psi  = e^{\gamma^5 \sqrt{g_A} \Phi_1 + i \sqrt{g_A} \Phi_2} \Psi' .
\end{equation}
After some straightforward algebra, we have
\begin{equation}
\mathcal{L}_0 = \overline{\Psi}' (i\gamma^\mu \partial_\mu) \Psi' .
\end{equation}
To make it easy to be calculated, we further transform the theory into the chiral basis
\begin{equation}
\overline{\Psi} (i\gamma^\mu \partial_\mu) \Psi
= \Psi^\dagger \gamma^1 (i\gamma^\mu \partial_\mu) \Psi
= -i \Psi^\dagger (\partial_1 + \gamma^0 \partial_2) \Psi 
= -i \left( \begin{matrix}
\Psi^\dagger_+ & \Psi^\dagger_-
\end{matrix} \right)
\left( \begin{matrix}
2\partial_{\overline{z}} & 0 \\
0 & 2\partial_z
\end{matrix} \right) 
\left( \begin{matrix}
\Psi_+ \\ \Psi_-
\end{matrix} \right) ,
\end{equation}
where $z = x + i y$, $\overline{z} = x - i y$, and $\mathbf{r}_\parallel = (x,y)$.
In this representation, the filed operators can be written as
\begin{equation}
\Psi^\dagger_\pm = \Psi'^\dagger_\pm
e^{\mp\sqrt{g_A}\Phi_1-i\sqrt{g_A}\Phi_2} 
\ , \qquad
\Psi_\pm = e^{\pm\sqrt{g_A}\Phi_1+i\sqrt{g_A}\Phi_2} \Psi'_\pm \ ,
\end{equation}
and their two-point correlation function 
\begin{equation}
\begin{aligned}
& \left\langle \Psi_\pm(z,\overline{z}) 
\Psi^\dagger_\pm(w,\overline{w}) \right\rangle
= \left\langle 
e^{\pm \sqrt{g_A} \Phi_1(z,\overline{z}) + i \sqrt{g_A}\Phi_2(z,\overline{z})}
\Psi'_\pm(z,\overline{z})
\Psi'^\dagger_\pm(w,\overline{w}) 
e^{\mp \sqrt{g_A} \Phi_1(w,\overline{w}) - i \sqrt{g_A}\Phi_2(w,\overline{w})} 
\right\rangle \\
= & \ \left\langle \Psi'_\pm(z,\overline{z})
\Psi'^\dagger_\pm(w,\overline{w}) \right\rangle
\left\langle e^{\pm \sqrt{g_A} \Phi_1(z,\overline{z})}
e^{\mp \sqrt{g_A} \Phi_1(w,\overline{w})} \right\rangle
\left\langle e^{+ i \sqrt{g_A}\Phi_2(z,\overline{z})}
e^{- i \sqrt{g_A}\Phi_2(w,\overline{w})}
\right\rangle ,
\end{aligned}
\end{equation}
where the correlation function of the chiral Dirac field is
\begin{equation}
\left\langle \Psi'_+(z,\overline{z})
\Psi'^\dagger_+(w,\overline{w})
\right\rangle \sim
\frac{1}{2\pi} \frac{1}{z-w}
, \qquad 
\left\langle \Psi'_-(z,\overline{z})
\Psi'^\dagger_-(w,\overline{w})
\right\rangle \sim
\frac{1}{2\pi} \frac{1}{\overline{z}-\overline{w}} 
\end{equation}
and the correlation function of the axial field is simply the correlation function of the vertex operator of the free scalar field
\begin{equation}
\left\langle e^{+ i f\Phi(z,\overline{z})}
e^{- i f\Phi(w,\overline{w})}
\right\rangle \sim
|z-w|^{\frac{-2f^2}{4\pi}} .
\end{equation}
Interestingly, here we will see that, in $\left\langle \Psi_\pm(z,\overline{z}) 
\Psi^\dagger_\pm(w,\overline{w}) \right\rangle$ the contribution from the longitudinal field $\Phi_1$ cancels with the contribution from the transversal field $\Phi_2$, i.e. 
\begin{equation}
\left\langle \Psi_\pm(z,\overline{z}) 
\Psi^\dagger_\pm(w,\overline{w}) \right\rangle
= \left\langle \Psi'_\pm(z,\overline{z}) 
\Psi'^\dagger_\pm(w,\overline{w}) \right\rangle .
\end{equation}

We now turn to consider the case of $\omega \neq 0$, where the frequency term can be understood as an interaction 
\begin{equation}
\omega \overline{\Psi} \gamma^0 \Psi
= \omega \overline{\Psi}' e^{2\gamma^5\sqrt{g_A}\Phi_1} \gamma^0 \Psi'
\end{equation}
The $l$-th order tree level diagram is
\begin{equation}
\begin{aligned}
(\omega)^l P'_{l}(\mathbf{r}_{\parallel}, \mathbf{r}_{\parallel}') = &
(\omega)^l \int d^2\mathbf{r}_{\parallel,1} \cdots d^2\mathbf{r}_{\parallel,2}  
g'(\mathbf{r}_{\parallel}, \mathbf{r}_{\parallel,1}) 
e^{2\gamma^5 \sqrt{g_A} \Phi_1(\mathbf{r}_{\parallel, 1})} 
\left[ \gamma^0 g'(\mathbf{r}_{\parallel,1}, \mathbf{r}_{\parallel,2}) 
e^{2\gamma^5 \sqrt{g_A} \Phi_1(\mathbf{r}_{\parallel,2})} 
\right]   \cdots  \\ 
& \qquad  \quad  
\left[ \gamma^0 g'(\mathbf{r}_{\parallel,l-1}, \mathbf{r}_{\parallel,l}) 
e^{2\gamma^5 \sqrt{g_A} \Phi_1(\mathbf{r}_{\parallel,l})} \right] 
\left[ \gamma^0 g'(\mathbf{r}_{\parallel,l}, \mathbf{r}_{\parallel}') 
e^{2\gamma^5 \sqrt{g_A} \Phi_1(\mathbf{r}_{\parallel}')}  \right] .
\end{aligned}
\end{equation}
It is important to notice that the odd order perturbations vanish since the expectation value of the charged vertex operators vanishes. The even order perturbations contributes to the finial result as 
\begin{equation}
(\omega)^{2l} P'_{2l}(\mathbf{r}_{\parallel}, \mathbf{r}_{\parallel}') 
= (\omega)^{2l} \int \frac{d^2 \mathbf{k}}{(2\pi)^2} 
e^{i \mathbf{k} (\mathbf{r}_{\parallel} - \mathbf{r}_{\parallel}')}
\widetilde{g}(\mathbf{k}) 
\left[ \gamma^0 \widetilde{g}(\mathbf{k}) \right]^{2l} .
\end{equation}
Here we introduce the modified Green's function with gauge contribution as 
\begin{equation}
\widetilde{g}(\mathbf{r}_{\parallel}, \mathbf{r}_{\parallel}') 
= 
\left\langle e^{+\sqrt{g_A}\Phi_1(\mathbf{r}_{\parallel})} e^{-\sqrt{g_A}\Phi_1(\mathbf{r}_{\parallel}')} \right\rangle
g'(\mathbf{r}_{\parallel}, \mathbf{r}_{\parallel}') 
= 
r^{-1 + \frac{g_A}{2\pi}} 
\left(\begin{matrix}
0 & e^{i\theta} \\
e^{-i\theta}
\end{matrix}\right) , 
\end{equation}
where we have defined $r = |\mathbf{r}| = |\mathbf{r}_{\parallel} - \mathbf{r}_{\parallel}'|$ and $\theta = \arctan \frac{\mathbf{r}_y}{\mathbf{r}_x}$. 
Its Fourier transformation is
\begin{equation}
\begin{aligned}
\widetilde{g}(\mathbf k) 
& = \int_0^\infty rdr \int_0^{2\pi} d\theta 
e^{-ikr \sin(\theta + \arctan\frac{k_x}{k_y})} 
r^{-1 + \frac{g_A}{2\pi}} 
\left(\begin{matrix}
0 & e^{i\theta} \\
e^{-i\theta} & 0
\end{matrix}\right) \\
& = \int_0^\infty  r^{\frac{g}{2\pi}} J_1(kr)  dr 
\left(\begin{matrix}
0 & e^{-i\arctan\frac{k_x}{k_y}} \\
e^{+i\arctan\frac{k_x}{k_y}} & 0
\end{matrix}\right) .
\end{aligned}
\end{equation}
The integral is divergent when $\frac{g}{2\pi} \ge \frac{1}{2}$. 
For $0 \le \frac{g}{2\pi} < \frac{1}{2}$, we have
\begin{equation}
\widetilde{g}(\mathbf k) 
= 2^{\frac{g_A}{2\pi}}
\frac{\Gamma(1+\frac{g_A}{4\pi})}{\Gamma(1-\frac{g_A}{4\pi})}
k^{-(1+\frac{g_A}{2\pi})} 
\left(\begin{matrix}
0 & e^{-i\arctan\frac{k_x}{k_y}} \\
e^{+i\arctan\frac{k_x}{k_y}} & 0
\end{matrix}\right) 
= 2^{\frac{g_A}{2\pi}}
\frac{\Gamma(1+\frac{g_A}{4\pi})}{\Gamma(1-\frac{g_A}{4\pi})}
k^{-\frac{g_A}{2\pi}} g'(k) 
= C(g_A) k^{-\frac{g_A}{2\pi}} g'(k) .
\end{equation}
Then, for $0 < \frac{g_A}{2\pi} < \frac{1}{2}$, the summation over tree level diagrams then becomes
\begin{equation}\label{eq:3d_gauge_GreenFunction}
\begin{aligned}
G(\mathbf{r}_{\parallel}, \mathbf{r}_{\parallel}'; \omega)
& = \sum_{l=0}^\infty (\omega)^{2l} 
\int \frac{d^2\mathbf{k}}{(2\pi)^2}
e^{i \mathbf{k} (\mathbf{r}_{\parallel} - \mathbf{r}_{\parallel}')} 
\widetilde{g}(\mathbf{k})
\left[ \gamma^0 \widetilde{g}(\mathbf{k}) \right]^{2l}   \\ 
& = \sum_{l=0}^\infty (\omega)^{2l} 
\int \frac{d^2\mathbf{k}}{(2\pi)^2}
e^{i \mathbf{k} (\mathbf{r}_{\parallel} - \mathbf{r}_{\parallel}')} 
\left[ C(g_A) k^{-\frac{g_A}{2\pi}} \right]^{2l+1} 
g'(k) 
\left[ \gamma^0 g'(k) \right]^{2l}   \\
& = \int \frac{d^2\mathbf{k}}{(2\pi)^2}
e^{i \mathbf{k} (\mathbf{r}_{\parallel} - \mathbf{r}_{\parallel}')} 
\frac{C(g_A) k^{-\frac{g_A}{2\pi}} g'(k)}{1 - \left[ C(g_A) k^{-\frac{g_A}{2\pi}} 
	\gamma^0 g'(k) \omega \right]^2} \\
& = \int \frac{d^2\mathbf{k}}{(2\pi)^2}
e^{i \mathbf{k} (\mathbf{r}_{\parallel} - \mathbf{r}_{\parallel}')} 
\frac{- \gamma^i k_i \left[C(g_A)\right]^{-1} k^{\frac{g_A}{2\pi}}}
{\left[ C(g_A) \right]^{-2} k^{2+\frac{g_A}{\pi}} - \omega^2} . 
\end{aligned}
\end{equation}
The Green's function provides a lot of information of the theory. 
First, its poles are located at $\omega^2 = \left[ C(g_A) \right]^{-2} k^{2+\frac{g_A}{\pi}}$, which indicates the dispersion relation as
\begin{equation}\label{eq:dispersion}
E(k) \propto |k|^{1+\frac{g_A}{2\pi}} . 
\end{equation}
Meanwhile, as one of the most important physical quantity for single-electron physics, the density of states (DOS) can be calculated from the Green's function as 
\begin{equation}\label{eq:DOS}
\rho(\omega) 
= \frac{1}{2\pi i} 
\lim_{\mathbf{r}_{\parallel} \to \mathbf{r}_{\parallel}'} 
\left[
G(\mathbf{r}_{\parallel}, \mathbf{r}_{\parallel}'; \omega)_{\rm adv} 
- G(\mathbf{r}_{\parallel}, \mathbf{r}_{\parallel}'; \omega)_{\rm ret} 
\right] 
= \frac{1}{\pi i} \Im \left[ 
\lim_{\mathbf{r}_{\parallel} \to \mathbf{r}_{\parallel}'} 
G(\mathbf{r}_{\parallel}, \mathbf{r}_{\parallel}'; \omega)_{\rm adv} 
\right] 
\propto 
\omega^{\frac{1-{g_A}/{2\pi}}{1+{g_A}/{2\pi}}} . 
\end{equation}
This result is consistent with the previous calculations, e.g. \cite{Wen1996conformal,Ludwig1994_IQH_transition}(Note that here we have the difference on a factor of $2$ in the definition of $g_A$ with these literature.), which again validates the dimensional reduction approach.

\subsection{The replicated Green's function of Dirac field under a random static gauge field}\label{app:dirac_randomgauge_replica}

In the calculation of Green's function, it is convenient to impose a ``source'' term $M \overline{\Psi}' \Psi' $ into the Lagrangian
\begin{equation}\label{eq:Lagrangian_axial_mass}
\mathcal{L} = \overline{\Psi}' (\gamma^\mu \partial_\mu + M) \Psi' 
+ i\omega \overline{\Psi}' e^{2 \gamma^5 \sqrt{g_A} \Phi_1} \gamma^0 \Psi' .
\end{equation}
Here for Euclidean spacetime we choose $g^\mu_\nu = {\text{diag}}(+1,+1,+1)$, $\gamma^1 = \sigma_1$, $\gamma^2 = \sigma_2$, and $\gamma^0 = \gamma^5 = -i \gamma^1 \gamma^2 = \sigma_3$. 
For convenience, below we set a parameter $a = \frac{g_A}{2\pi}$. 

\subsubsection{The Fourier transformation of the corrected internal spinor propagator}

The presence of a random static gauge field leads to the corrected internal $2$D propagator of the Dirac field
\begin{equation}
\widetilde{g}_D^{(1)} 
(\mathbf{r}_{\parallel, 1}, \mathbf{r}_{\parallel, 2}) 
= {g}_D^{(1)} 
(\mathbf{r}_{\parallel, 1}, \mathbf{r}_{\parallel, 2}) 
r_{1,2}^{a} . 
\end{equation}
Evaluating its Fourier transformation is the key to access the higher-dimensional construction 
\begin{equation}
\widetilde{g}^{(1)}_D (k) 
= \int d^2 \mathbf{r}_{1,2} 
e^{i \mathbf{k} (\mathbf{r}_{\parallel,1} - \mathbf{r}_{\parallel,2})}
\widetilde{g}_D^{(1)} 
(\mathbf{r}_{\parallel, 1}, \mathbf{r}_{\parallel, 2}) .
\end{equation}
The diagonal components are
\begin{equation}
\begin{aligned}
\left[ \widetilde{g}^{(1)}_D (k) \right]_{11} 
= \left[ \widetilde{g}^{(1)}_D (k) \right]_{22} 
& = \int d^2 \mathbf{r} e^{-i \mathbf{k} \mathbf{r}} 
r^{a} \frac{M}{2\pi} K_0(M r) \\
& = \int_0^\infty rdr \int_0^{2\pi} d\theta e^{-ik r\cos\theta} 
r^a \frac{M}{2\pi} K_0(Mr) \\
& = M \int_0^\infty r^{1+a} J_0(kr) K_0(Mr) dr \\
& = M \Big( 2^a M^{-2-a} \Big)  \left[ \Gamma\left( 1 + \frac{a}{2} \right) \right]^2
{}_2F_1 \left( \frac{a}{2}+1, \frac{a}{2}+1; 1; -\frac{k^2}{M^2} \right) ,
\end{aligned}
\end{equation}
and the off-diagonal components are
\begin{equation}
\begin{aligned}
\left[ \widetilde{g}^{(1)}_D (k) \right]_{12} 
= - \left[ \widetilde{g}^{(1)}_D (k) \right]_{21}^* 
& = \int d^2 \mathbf{r} e^{-i \mathbf{k} \mathbf{r}} 
r^{a} (i) \frac{M}{2\pi} e^{i \arctan \frac{r_1}{r_2}} K_1(Mr) \\
& = \frac{iM}{2\pi} \int_0^\infty rdr \int_0^{2\pi} d\theta 
e^{-ikr\sin(\theta + \arctan \frac{k_2}{k_1})} 
e^{i\theta} K_1(Mr) r^a \\
& = \frac{iM}{2\pi} \int_0^\infty dr r^{1+a} K_1(Mr) 
\int_0^{2\pi} d\theta e^{i\theta} \sum_{n=-\infty}^{\infty} (i)^n J_n(-kr) 
e^{in(\theta - \frac{\pi}{2} + \arctan\frac{k_2}{k_1})} \\
& = \frac{iM}{2\pi} \int_0^\infty dr r^{1+a} K_1(Mr) 
(2\pi) (i)^{-1} J_{-1}(-kr) e^{-i(-\frac{\pi}{2} + \arctan \frac{k_2}{k_1})} \\
& = iM e^{-i\arctan\frac{k_2}{k_1}} 
\int_0^\infty r^{1+a} K_1(Mr) J_1(kr) dr \\
& = iM e^{-i\arctan\frac{k_2}{k_1}} 
2^a k M^{-3-a} 
\Gamma\left( 1 + \frac{a}{2} \right) 
\Gamma\left( 2 + \frac{a}{2} \right) 
{}_2F_1 \left( \frac{a}{2}+1, \frac{a}{2}+2; 2; -\frac{k^2}{M^2} \right) \\
& = (ik_1 + k_2) 2^a M^{-2-a} 
\Gamma\left( 1 + \frac{a}{2} \right) 
\Gamma\left( 2 + \frac{a}{2} \right) 
{}_2F_1 \left( \frac{a}{2}+1, \frac{a}{2}+2; 2; -\frac{k^2}{M^2} \right) .
\end{aligned}
\end{equation}

The above is the non-perturbative results of the internal propagator, 
we then expand them to the lowest order with respect to $a$. 
For that, on the one hand, the series of momentum-independent coefficients are
\begin{equation}
\begin{aligned}
2^a M^{-2-a} \left[ \Gamma\left( 1+ \frac{a}{2} \right) \right]^2 
= \frac{1}{M^2} + \frac{(-\gamma + \ln 2 - \ln M)a}{M^2} 
+ \mathcal{O}(a^2)
\end{aligned}
\end{equation}
and
\begin{equation}
2^a M^{-2-a} \Gamma\left( 1 + \frac{a}{2} \right) \Gamma\left( 2 + \frac{a}{2} \right)   
= \frac{1}{M^2} + \frac{(-\gamma + \ln 2 - \ln M)a}{M^2} 
+ \frac{a}{2M^2} + \mathcal{O}(a^2) ,
\end{equation}
where $\gamma \simeq 0.577216$ is the Euler's constant.
On the other hand, for expanding the hypergeometric function, we need to calculate its derivative with respect to the parameters~\cite{Ancarani_2010}
%
\begin{equation}
\begin{aligned}
& \quad
\left. \frac{\partial}{\partial \alpha} \ 
{}_2F_1 \left( \alpha, 1; 1; -\frac{k^2}{M^2} \right) 
\right|_{\alpha = 1} 
= \left. \frac{\partial}{\partial \beta} \ 
{}_2F_1 \left( 1, \beta; 1; -\frac{k^2}{M^2} \right) 
\right|_{\beta = 1} 
= \left. \frac{\partial}{\partial \alpha} \ 
{}_2F_1 \left( \alpha, 2; 2; -\frac{k^2}{M^2} \right) 
\right|_{\alpha = 1} 
\\
& = - \frac{k^2}{M^2} \sum_{n=0}^\infty \frac{(1)_n}{(2)_n} \frac{(1)_n (2)_n}{(2)_n} 
\frac{\left( - \frac{k^2}{M^2} \right)^n}{n!} \ 
{}_3F_2 \left( 1, n+2, n+2; n+2, n+2; -\frac{k^2}{M^2} \right) \\
& = - \frac{k^2}{M^2} \sum_{n=0}^\infty \frac{(1)_n (1)_n}{(2)_n} 
\frac{\left( - \frac{k^2}{M^2} \right)^n}{n!} \ 
\frac{1}{1+\frac{k^2}{M^2}} 
= - \frac{k^2}{k^2 + M^2} 
{}_2F_1 \left( 1, 1; 2; -\frac{k^2}{M^2} \right) \\
& = - \frac{k^2}{k^2 + M^2} 
\left( \frac{k^2}{M^2} \right)^{-1} 
\ln \left( 1 + \frac{k^2}{M^2} \right) 
= - \left( 1 + \frac{k^2}{M^2} \right)^{-1} 
\ln \left( 1 + \frac{k^2}{M^2} \right) ,
\end{aligned}
\end{equation}
and 
\begin{equation}
\begin{aligned}
& \quad
\left. \frac{\partial}{\partial \beta} \ 
{}_2F_1 \left( 1, \beta; 2; -\frac{k^2}{M^2} \right) 
\right|_{\beta = 2} \\
& = - \frac{k^2}{M^2} \frac{1}{2} 
\sum_{n=0}^\infty \frac{(1)_n}{(2)_n} \frac{(2)_n (2)_n}{(3)_n} 
\frac{\left( - \frac{k^2}{M^2} \right)^n}{n!} \ 
{}_3F_2 \left( 1, n+2, n+3; n+2, n+3; -\frac{k^2}{M^2} \right) \\
& = - \frac{1}{2} \frac{k^2}{k^2 + M^2} 
{}_2F_1 \left( 1, 2; 3; -\frac{k^2}{M^2} \right) 
= - \frac{1}{2} \frac{k^2}{k^2 + M^2} 
\frac{2\left[ k^2 M^2 - M^4 \ln\left( 1 + \frac{k^2}{M^2} \right) \right]}{k^4} \\
& = - \left( 1 + \frac{k^2}{M^2} \right)^{-1} 
+ \left( \frac{k^2}{M^2} \right)^{-1} 
\left( 1 + \frac{k^2}{M^2} \right)^{-1} 
\ln \left( 1 + \frac{k^2}{M^2} \right) ,
\end{aligned}
\end{equation}
where $(\lambda)_0 = 1$ and $(\lambda)_n = \Gamma(\lambda+n)/\Gamma(\lambda)$ for $n \ge 1$ is  the Pochhammer symbol. 

Meanwhile, we have the zero-th order contribution 
\begin{equation}
{}_2F_1 \left( 1, 1; 1; -\frac{k^2}{M^2} \right) 
= {}_2F_1 \left( 1, 2; 2; -\frac{k^2}{M^2} \right) 
= \left( 1 + \frac{k^2}{M^2} \right)^{-1} .
\end{equation}
These lead to the expansion of $\widetilde{g}^{(1)}_D (k)$ at the lowest-order with respect to $a$
\begin{equation}
\begin{aligned}
\left[ \widetilde{g}_D^{(1)} (k) \right]_{11} 
= \left[ \widetilde{g}_D^{(1)} (k) \right]_{22} 
& = M \left[ \frac{1}{M^2} + \frac{(-\gamma + \ln 2 - \ln M)a}{M^2} \right]
\left( 1 + \frac{k^2}{M^2} \right)^{-1} 
\left[ 1 - a \ln \left( 1 + \frac{k^2}{M^2} \right) \right] 
+ \mathcal{O}(a^2) \\ 
& = \frac{M}{k^2 + M^2} \bigg\{
1 + a \Big[
-\gamma + \ln 2 + \ln M - \ln \left( k^2 + M^2 \right)
\Big]
\bigg\} 
+ \mathcal{O}(a^2) \\ 
\end{aligned}
\end{equation}
and
\begin{equation}
\begin{aligned}
\left[ \widetilde{g}^{(1)}_D (k) \right]_{12} 
= - \left[ \widetilde{g}^{(1)}_D (k) \right]_{21}^* 
& =  \frac{ik_1 + k_2}{k^2 + M^2} 
\bigg\{
1 + a \Big[
-\gamma + \ln 2 - \frac{1}{2} \ln \left( k^2 + M^2 \right) 
+ \frac{M^2}{2k^2} \ln  \left( 1 + \frac{k^2}{M^2} \right) 
\Big] \bigg\} 
+ \mathcal{O}(a^2)
\end{aligned}
\end{equation}

Recall that the scalar nature of the double product of spinor propagators strongly simplifies the calculation of entanglement entropy for free Dirac field. 
Here, it is important to notice that the Dirac Green's function in presence of a random static gauge field has the same property, since the gauge correction is a scalar function. 
After some straightforward calculation, we have the double product of the modified Green's function as
\begin{equation}\label{eq:double_product_gauge}
\begin{aligned}
\gamma^0 \widetilde{g}_D^{(1)} (k) 
\gamma^0 \widetilde{g}_D^{(1)} (k) 
= \frac{1}{k^2 + M^2} 
\left\{
1 + 2a \left[
- \gamma + \ln 2 - \frac{1}{2} \ln \left( k^2 + M^2 \right)
\right] 
\right\} {\rm Id} 
+ \mathcal{O}(a^2) . 
\end{aligned}
\end{equation}

\subsubsection{The construction from the $2$D corrected propagator to $3$D theory}

The above calculation leads to the elements of the infinite series that are used to construct the higher-dimensional theory, namely the interacting two-point correlator on $2$D replica manifold
\begin{equation}\label{eq:dimensional_reduction_3d_gauge}
\begin{aligned}
& \quad 
\left( -\omega^2 \right)^l \left[
P^{(n)}_{D,2l,{\rm int}}(\mathbf{r}_\parallel, \mathbf{r}_\parallel') 
- P^{(1)}_{D,2l,{\rm int}}(\mathbf{r}_\parallel, \mathbf{r}_\parallel') 
\right]_{\text{diag}} \\ 
& = (l+1) (-\omega^2)^{l} 
\int d^2 \mathbf{r}_{\parallel,1} 
\cdots 
\int d^2 \mathbf{r}_{\parallel,2l} 
\left[
g^{(n)}_{D}(\mathbf{r}_\parallel, \mathbf{r}_{\parallel,1}) 
- g^{(1)}_{D}(\mathbf{r}_\parallel, \mathbf{r}_{\parallel,1}) 
\right]_{\text{diag}}
\gamma^0 
\widetilde{g}^{(1)}_{D}(\mathbf{r}_{\parallel,1}, \mathbf{r}_{\parallel,2}) 
\cdots
\gamma^0 
\widetilde{g}^{(1)}_{D}(\mathbf{r}_{\parallel,2l}, \mathbf{r}_{\parallel}') 
\\ 
& = (l+1) (-\omega^2)^l 
\frac{1-n^2}{6n^2} \frac{M}{2\pi} K_0(M\rho)
\int \frac{d^2 \mathbf{k}}{2\pi}
e^{i \mathbf{k} \mathbf{r}_\parallel'} 
\frac{\left[
	\gamma^0 \widetilde{g}_D^{(1)}(k) \gamma^0 
	\widetilde{g}_D^{(1)}(k) 
	\right]^l }
{ k^2 + M^2 } 
+ \mathcal{O}(a^2) \\
& = (l+1) (-\omega^2)^l 
\frac{1-n^2}{6n^2} \frac{M}{2\pi} K_0(M\rho)
\int \frac{d^2 \mathbf{k}}{2\pi}
e^{i \mathbf{k} \mathbf{r}_\parallel'} 
\left\{ 
\frac{1 + 2la \left[ - \gamma + \ln 2 - \frac{1}{2} \ln \left( k^2 + M^2 \right) \right]}
{\left( k^2 + M^2 \right)^{l+1}} 
\right\} 
+ \mathcal{O}(a^2) .
\end{aligned}
\end{equation}

\subsubsection{The entanglement entropy and partition function of constructed $3$D theory}\label{app:first_order_EE_gauge}

To calculate the partition function, we start from evaluating the trace of the corrected replicated Green's function $\left(-\omega^2\right)^l P^{(n)}_{D,2l,{\rm int}}$. 
In the previous section, we have presented the explicit evaluation of its lowest-order perturbation with respect to the randomness strength $a = \frac{g_A}{2\pi}$. 
We now separate it into three parts, the first one is 
\begin{equation}
\begin{aligned}
& \quad 
\left(-\omega^2\right)^l 
\Tr^{(n)}_{2D} 
\left[ P^{(n)}_{D,2l,{\rm int}} (\mathbf{r}_\parallel, \mathbf{r}_\parallel') 
- P^{(1)}_{D,2l,{\rm int}} (\mathbf{r}_\parallel, \mathbf{r}_\parallel') \right]_{\rm first} \\
& = \Tr^{(n)}_{2D} 
\left[ 
(l+1) (-\omega^2)^l \frac{1-n^2}{6n^2} \frac{M}{2\pi} K_0(M\rho) 
\int_0^\infty  \frac{k J_0(k\rho')}{(k^2 + M^2)^{l+1}} dk 
\right] \\ 
& = \Tr^{(n)}_{2D} \left[ 
(l+1) (-\omega^2)^l \frac{1-n^2}{6n^2} \frac{M}{2\pi} K_0(M\rho) 
\frac{\left( \rho' \right)^l M^{-l}}{2^l \Gamma(l+1)} K_l(M\rho')
\right] \\
& = 2 (l+1) (-\omega^2)^l \frac{1-n^2}{6n^2} 
\frac{M^{-l+1}}{2^l \Gamma(l+1)} 
\int_0^\infty \rho^{l+1} K_0(M\rho) K_l(M\rho) d\rho \\
& = 2 (l+1) (-\omega^2)^l \frac{1-n^2}{6n^2} 
\frac{M^{-l+1}}{2^l \Gamma(l+1)} 
\frac{\Gamma(l+1)}{l+1} 2^{l-1} M^{-l-2} 
= \frac{1-n^2}{6n} \frac{1}{M} 
\left( -\frac{\omega^2}{M^2} \right)^l .
\end{aligned}
\end{equation}
The summation over perturbation levels $l$ gives 
\begin{equation}
\Tr^{(n)}_{2D} 
\left[
G^{(n)}_{D,{\rm int}} - G^{(1)}_{D,{\rm int}}
\right]_{\rm first} 
= \sum_{l=0}^\infty \Tr^{(n)}_{2D} 
\left(-\omega^2\right)^l 
\Tr^{(n)}_{2D} 
\left[ P^{(n)}_{D,2l,{\rm int}} 
- P^{(1)}_{D,2l,{\rm int}} \right]_{\rm first}
= \frac{1-n^2}{6n} \frac{M}{\omega^2 + M^2} ,
\end{equation}
which leads to the free contribution of the EE 
\begin{equation}\label{eq:Sfirst}
S_{\rm first} = \frac{1}{6} \mathcal{A} (\epsilon^{-1} - M) .
\end{equation}

The second part is 
\begin{equation}
\begin{aligned}
& \quad
\left(-\omega^2\right)^l 
\Tr^{(n)}_{2D} 
\left[ P^{(n)}_{D,2l,{\rm int}} (\mathbf{r}_\parallel, \mathbf{r}_\parallel') 
- P^{(1)}_{D,2l,{\rm int}} (\mathbf{r}_\parallel, \mathbf{r}_\parallel') \right]_{\rm second} \\
& =  \Tr^{(n)}_{2D}
\left[ 
(l+1)
2l a \left( -\gamma + \ln 2 \right) 
(-\omega^2)^l \frac{1-n^2}{6n^2} \frac{M}{2\pi} K_0(M\rho) 
\int_0^\infty  \frac{k J_0(k\rho')}{(k^2 + M^2)^{l+1}} dk
\right] \\ 
& = 2a \left( -\gamma + \ln 2 \right) 
\frac{1-n^2}{6n} \frac{1}{M} 
\left[
l \left( - \frac{\omega^2}{M^2} \right)^l 
\right] .
\end{aligned}
\end{equation}
The summation over perturbation levels $l$ gives 
\begin{equation}
\begin{aligned}
\Tr^{(n)}_{2D} 
\left[
G^{(n)}_{D,{\rm int}} - G^{(1)}_{D,{\rm int}}
\right]_{\rm second} 
& = \sum_{l=0}^\infty 
\left(-\omega^2\right)^l 
\Tr^{(n)}_{2D} 
\left[ P^{(n)}_{D,2l,{\rm int}} 
- P^{(1)}_{D,2l,{\rm int}}  \right]_{\rm second} \\
& = \frac{1-n^2}{6n} 
2a \left( -\gamma + \ln 2 \right) 
\frac{-M \omega^2}{\left( \omega^2 + M^2 \right)^2} ,
\end{aligned}
\end{equation}
which leads to the correction to EE as
\begin{equation}\label{eq:Ssecond}
S_{\rm second} 
=  - a   
\left( -\gamma + \ln 2 \right) 
\frac{1}{6} \mathcal{A} (\epsilon^{-1} - M)  .
\end{equation}

The third part is much more complicated 
\begin{equation}
\begin{aligned}
& \quad
\left(-\omega^2\right)^l 
\Tr^{(n)}_{2D} 
\left[ P^{(n)}_{D,2l,{\rm int}} (\mathbf{r}_\parallel, \mathbf{r}_\parallel') 
- P^{(1)}_{D,2l,{\rm int}} (\mathbf{r}_\parallel, \mathbf{r}_\parallel') \right]_{\rm third} \\
& = - \frac{1}{2}  \Tr^{(n)}_{2D}
\left[ 
(l+1)
2l a  
(-\omega^2)^l \frac{1-n^2}{6n^2} \frac{M}{2\pi} K_0(M\rho) 
\int_0^\infty  \frac{k \ln \left( k^2 + M^2 \right) J_0(k\rho')}{(k^2 + M^2)^{l+1}} dk
\right] \\ 
& =  - 2a  
\frac{1-n^2}{6n^2} \frac{M}{2\pi} 
l(l+1) (-\omega^2)^l 
\int_0^\infty \rho d\rho \int_0^{2\pi n} d\theta
K_0(M\rho) 
\frac{1}{\Gamma(l+1)} 2^{-l} M^{-l} \rho^l \\
& \qquad 
\times \left\{
K_{-l}(M\rho) \left[ \ln 2 + \ln M - \ln \rho + \psi(l+1) \right] 
+ \left. 
\frac{\partial}{\partial \nu} K_\nu (M\rho) 
\right|_{\nu=-l}
\right\} \\ 
&=i)+ii)+iii). 
\end{aligned} 
\end{equation}
To evaluate this integral, we further separate it into three parts. 
The first part is
\begin{equation}
\begin{aligned}
i)=& \quad
\left(-\omega^2\right)^l 
\Tr^{(n)}_{2D} 
\left[ P^{(n)}_{D,2l,{\rm int}} (\mathbf{r}_\parallel, \mathbf{r}_\parallel') 
- P^{(1)}_{D,2l,{\rm int}} (\mathbf{r}_\parallel, \mathbf{r}_\parallel') \right]_{{\rm third},1} \\ 
& = - 2a \frac{1-n^2}{6n} 
\frac{l(l+1)}{\Gamma(l+1)} 
2^{-l} M^{-l+1} \left( - \omega^2 \right)^l 
\left[ \ln 2 + \ln M + \psi(l+1) \right]
\int_0^\infty \rho^{l+1} K_0(M\rho) K_l(M\rho) d\rho \\
& = - 2a \frac{1-n^2}{6n} 
\frac{l(l+1)}{\Gamma(l+1)} 
2^{-l} M^{-l+1} \left( - \omega^2 \right)^l 
\left[ \ln 2 + \ln M + \psi(l+1) \right] 
\frac{\Gamma(l+1)}{l+1} 2^{l-1} M^{-l-2} \\ 
& = - a \frac{1-n^2}{6n}  \frac{1}{M} 
l \left[ \ln 2 + \ln M + \psi(l+1) \right] 
\left( - \frac{\omega^2}{M^2} \right)^l  .
\end{aligned}
\end{equation}
Its summation over perturbation levels $l$ gives
\begin{equation}
\begin{aligned}
& \quad 
\Tr^{(n)}_{2D} 
\left[
G^{(n)}_{D,{\rm int}} - G^{(1)}_{D,{\rm int}}
\right]_{{\rm third},1} 
= \sum_{l=0}^\infty \left(-\omega^2\right)^l 
\Tr^{(n)}_{2D} 
\left[ P^{(n)}_{D,2l,{\rm int}} 
- P^{(1)}_{D,2l,{\rm int}} \right]_{{\rm third},1} \\ 
& = - a \frac{1-n^2}{6n}  \frac{1}{M} 
\sum_{l=0}^\infty l 
\left[ \ln 2 + \ln M + \psi(l+1) \right] 
\left( - \frac{\omega^2}{M^2} \right)^l \\
& = - a \frac{1-n^2}{6n}  \frac{1}{M} 
\left[ 
\left( \ln 2 + \ln M \right)
\sum_{l=0}^\infty l \left( - \frac{\omega^2}{M^2} \right)^l 
+ \sum_{l=0}^\infty l \psi(l+1) 
\left( - \frac{\omega^2}{M^2} \right)^l 
\right] \\ 
& = - a \frac{1-n^2}{6n}  \frac{1}{M} 
\left\{ 
\left( \ln 2 + \ln M \right) 
\frac{-M^2 \omega^2}{\left( M^2 + \omega^2 \right)^2} 
+ \frac{M^2 \omega^2 \left[ -1 + \gamma + 
	\ln \left( 1 + \frac{\omega^2}{M^2} \right) \right]}
{\left( M^2 + \omega^2 \right)^2}
\right\} 
\end{aligned}
\end{equation}
which leads to the correction to EE as
\begin{equation}\label{eq:Sthird1}
S_{{\rm third},1} 
= - \frac{a}{2} \left( 1 + \gamma + \ln 2 \right) 
\frac{1}{6} \mathcal{A} \left( \epsilon^{-1} - M \right) 
- \frac{a}{2} \frac{1}{6} \mathcal{A} 
\left( - \epsilon^{-1} \ln \epsilon^{-1} + M \ln M \right) . 
\end{equation}
The second part is
\begin{equation}
\begin{aligned}
ii)=& \quad
\left(-\omega^2\right)^l 
\Tr^{(n)}_{2D} 
\left[ P^{(n)}_{D,2l,{\rm int}} (\mathbf{r}_\parallel, \mathbf{r}_\parallel') 
- P^{(1)}_{D,2l,{\rm int}} (\mathbf{r}_\parallel, \mathbf{r}_\parallel') \right]_{{\rm third},2} \\ 
& = - 2a \frac{1-n^2}{6n} 
\frac{l(l+1)}{\Gamma(l+1)} 
2^{-l} M^{-l+1} \left( - \omega^2 \right)^l 
(-1)
\int_0^\infty \rho^{l+1} \ln \rho K_0(M\rho) K_l(M\rho) d\rho \\ 
& = 2a \frac{1-n^2}{6n} 
\frac{l(l+1)}{\Gamma(l+1)} 
2^{-l} M^{-l+1} \left( - \omega^2 \right)^l 
\left[
- \frac{\Gamma(l+1)}{(l+1)^2} 2^{l-1} M^{-l-2} 
+ \frac{\Gamma(l+1)}{l+1} 2^{l-1} M^{-l-2} 
\left( - \gamma + \ln 2 - \ln M \right)
\right] \\ 
& = a \frac{1-n^2}{6n} \frac{1}{M} 
\left[ - \frac{l}{l+1} \left( - \frac{\omega^2}{M^2} \right)^l 
+ \left( -\gamma + \ln 2 - \ln M \right) 
l \left( - \frac{\omega^2}{M^2} \right)^l 
\right] .
\end{aligned}
\end{equation}
Its summation over perturbation levels $l$ gives
\begin{equation}
\begin{aligned}
& \quad 
\left(-\omega^2\right)^l 
\Tr^{(n)}_{2D} 
\left[ P^{(n)}_{D,2l,{\rm int}} (\mathbf{r}_\parallel, \mathbf{r}_\parallel') 
- P^{(1)}_{D,2l,{\rm int}} (\mathbf{r}_\parallel, \mathbf{r}_\parallel') \right]_{{\rm third},2} \\ 
& = a \frac{1-n^2}{6n} \frac{1}{M} 
\left[
- \sum_{l=0}^\infty \frac{l}{l+1} \left( - \frac{\omega^2}{M^2} \right)^l 
+ \left( -\gamma + \ln 2 - \ln M \right) 
\sum_{l=0}^\infty l \left( - \frac{\omega^2}{M^2} \right)^l 
\right] \\ 
& = a \frac{1-n^2}{6n} \frac{1}{M} 
\left[
\frac{M^2\left[ -\omega^2 + (M^2 + \omega^2) 
	\ln \left( 1 + \frac{\omega^2}{M^2} \right) \right]}
{\omega^2 \left( M^2 + \omega^2 \right)} 
+ \left( -\gamma + \ln 2 - \ln M \right) 
\frac{- M^2 \omega^2}{\left( M^2 + \omega^2 \right)^2}
\right]
\end{aligned}
\end{equation}
which leads to the correction to EE as
\begin{equation}\label{eq:Sthird2}
S_{{\rm third},2} 
= \frac{a}{2} \left( 1 + \gamma - \ln 2 \right) 
\frac{1}{6} \mathcal{A} \left( \epsilon^{-1} - M \right) 
+ \frac{a}{2} \frac{1}{6} \mathcal{A} 
\left( \epsilon^{-1} \ln \epsilon^{-1} - M \ln M  \right)  .
\end{equation}
The third part is 
\begin{equation}
\begin{aligned}
iii)=& \quad
\left(-\omega^2\right)^l 
\Tr^{(n)}_{2D} 
\left[ P^{(n)}_{D,2l,{\rm int}} (\mathbf{r}_\parallel, \mathbf{r}_\parallel') 
- P^{(1)}_{D,2l,{\rm int}} (\mathbf{r}_\parallel, \mathbf{r}_\parallel') \right]_{{\rm third},3} \\ 
& = -2a 
\frac{1-n^2}{6n^2} \frac{M}{2\pi} 
l(l+1) (-\omega^2)^l 
\int_0^\infty \rho d\rho \int_0^{2\pi n} d\theta
K_0(M\rho) 
\frac{1}{\Gamma(l+1)} 2^{-l} M^{-l} \rho^l  
\left. 
\frac{\partial}{\partial \nu} K_\nu (M\rho) 
\right|_{\nu=-l}
\\ 
& = 2a \frac{1-n^2}{6n} \frac{M}{2} l (l+1) 
\left( - \frac{\omega^2}{M^2} \right)^l 
\int_0^\infty \rho d\rho 
\sum_{k=0}^{l-1} \frac{1}{k! (l-k)!}
2^{-k} M^k \rho^k K_k(M\rho) .
\end{aligned}
\end{equation}
By exchanging the summation of $k$ and the integral over $\rho$, we have
\begin{equation}
\begin{aligned}
& \quad
\left(-\omega^2\right)^l 
\Tr^{(n)}_{2D} 
\left[ P^{(n)}_{D,2l,{\rm int}} (\mathbf{r}_\parallel, \mathbf{r}_\parallel') 
- P^{(1)}_{D,2l,{\rm int}} (\mathbf{r}_\parallel, \mathbf{r}_\parallel') \right]_{{\rm third},3} \\ 
& = a \frac{1-n^2}{6n} M l (l+1) 
\left( - \frac{\omega^2}{M^2} \right)^l 
\sum_{k=0}^{l-1} \frac{1}{k! (l-k)!} 2^{-k} M^k  
\frac{\Gamma(k+1)}{k+1} 2^{k-1} M^{-k-2} \\ 
& = a \frac{1-n^2}{6n} \frac{1}{2M} 
l (l+1) 
\left( - \frac{\omega^2}{M^2} \right)^l 
\sum_{k=0}^{l-1} \frac{1}{(k+1) (l-k)!} .
\end{aligned}
\end{equation}
Its summation over perturbation levels $l$ gives
\begin{equation}
\begin{aligned}
& \quad
\Tr^{(n)}_{2D} 
\left[
G^{(n)}_{D,{\rm int}} - G^{(1)}_{D,{\rm int}}
\right]_{{\rm third},3} 
= \sum_{l=0}^\infty \left(-\omega^2\right)^l 
\Tr^{(n)}_{2D} 
\left[ P^{(n)}_{D,2l,{\rm int}} 
- P^{(1)}_{D,2l,{\rm int}} \right]_{{\rm third},3} \\ 
& = a \frac{1-n^2}{6n} \frac{1}{2M} 
\sum_{l=0}^\infty l(l+1) 
\left( - \frac{\omega^2}{M^2} \right)^l 
\sum_{k=0}^{l-1} \frac{1}{(k+1) (l-k)!} \\ 
& = a \frac{1-n^2}{6n} \frac{1}{2M} 
\left[
\sum_{l=0}^\infty \sum_{k=0}^l 
l (l+1) 
\frac{1}{(k+1)(l-k)!} 
\left( - \frac{\omega^2}{M^2} \right)^l 
- \sum_{l=0}^\infty 
l  \left( - \frac{\omega^2}{M^2} \right)^l 
\right] 
\end{aligned}
\end{equation}
Consider an auxiliary function 
\begin{equation}
f(x) = \sum_{l=0}^\infty \sum_{k=0}^l 
\frac{x^l}{(k+1)(l-k)!} 
= \left[ \sum_{l=0}^\infty \frac{1}{l+1} x^l \right] 
\left[ \sum_{k=0}^\infty \frac{1}{k!} x^k \right] 
= \frac{\ln (1-x)}{x} e^x . 
\end{equation}
Its derivatives are 
\begin{equation}
\begin{aligned}
f'(x)  
& = \sum_{l=0}^\infty \sum_{k=0}^l 
\frac{ l x^{l-1} }{(k+1)(l-k)!} 
= \frac{1}{x} \sum_{l=0}^\infty \sum_{k=0}^l 
\frac{l x^l}{(k+1)(l-k)!} , \\
%
%
f''(x) & = \sum_{l=0}^\infty \sum_{k=0}^l 
\frac{l(l-1) x^{l-2}}{(k+1)(l-k)!} 
= \frac{1}{x^2} \sum_{l=0}^\infty \sum_{k=0}^l 
\frac{ l(l-1) x^l }{(k+1)(l-k)!}  \\
\end{aligned}
\end{equation}
These gives 
\begin{equation}
\Tr^{(n)}_{2D} 
\left[
G^{(n)}_{D,{\rm int}} - G^{(1)}_{D,{\rm int}}
\right]_{{\rm third},3} 
= a \frac{1-n^2}{6n} \frac{1}{2M} 
\left\{
\left[ 
\left( - \frac{\omega^2}{M^2} \right)^2 f''\left( - \frac{\omega^2}{M^2} \right) 
+ 2 \left( - \frac{\omega^2}{M^2} \right) f'\left( - \frac{\omega^2}{M^2} \right) 
\right] 
- \frac{-M^2 w^2}{\left( M^2 + w^2 \right)^2}
\right\} ,
\end{equation}
which leads to the correction to EE as 
\begin{equation}\label{eq:Sthird3}
\begin{aligned}
S_{{\rm third},3} & = 
\frac{a}{4} 
\left[ \frac{\gamma}{\sqrt{\pi}} 
+ e {\rm Erfc}(1) 
- \sqrt{\pi} {\rm Erfi}(1) 
+ \frac{2}{\sqrt{\pi}} \ _2 F_2 \left( 1,1;\frac{3}{2},2;1 \right) 
+ \frac{2}{\sqrt{\pi}} \ln 2 
+ 1
\right]
\frac{1}{6} \mathcal{A} \left( \epsilon^{-1} + M \right) 
\\ & = \mu_{{\rm third},3} 
a \frac{1}{6} \mathcal{A} \left( \epsilon^{-1} + M \right) 
= \left( 0.310214 \dots \right)
a \frac{1}{6} \mathcal{A} \left( \epsilon^{-1} + M \right) 
,  
\end{aligned}
\end{equation}
where ${\rm Erfc}$ and ${\rm Erci}$ are the complementary and imaginary error functions. 

In summary, combining with Eq.~\eqref{eq:Sfirst},~\eqref{eq:Ssecond},~\eqref{eq:Sthird1},~\eqref{eq:Sthird2},~\eqref{eq:Sthird3}, we have
\begin{equation}\label{eq:massiveEE_3d_gauge}
\begin{aligned}
S & = S_{\rm first} + S_{\rm second} 
+ S_{{\rm third},1} + S_{{\rm third},2} + S_{{\rm third},3} \\
& = \frac{1}{6} \mathcal{A} \left( \epsilon^{-1} - M \right) 
\left[ 1 - a \left( - \gamma + 2 \ln 2 - \mu_{{\rm third},3} \right) \right] 
+ a \frac{1}{6} \mathcal{A} 
\left( \epsilon^{-1} \ln \epsilon^{-1} - M \ln M  \right) ,
\end{aligned} 
\end{equation}
where $a = \frac{g_A}{2\pi}$ represents the disorder strength.
The above expansion includes two parts of entanglement entropy: the leading UV terms and the subleading finite terms. 
By taking $M \to 0$ in Eq.~\eqref{eq:massiveEE_3d_gauge}, we have the leading terms as 
\begin{equation}\label{eq:massiveEE_3d_gauge_UV_1stOrder}
S_{\text{leading UV}} \approx  \frac{1}{6} {\mathcal{A}} \epsilon^{-1}
\left[	1 - \frac{g_A}{2 \pi}	\mu_{\rm gauge} 
+ \frac{g_A}{2 \pi} {\ln \epsilon^{-1}} +\mathcal{O}(g_A^2)	\right] ,
\end{equation}
where $\mu_{\rm gauge} = -{\gamma} + 2\ln 2 
- \mu_{{\rm third},3} \approx 0.5$ is a positive number, 
and $\epsilon$ is the UV cut-off. 
We notice that the area-law coefficient depends on the disorder strength $g_A$. 
This is the main result of the present study.

The subleading corrections to the entanglement entropy appears as letting $M$ to have a finite value. According to the above derivations, at the first order of $g_A$ they take the following form 
\begin{equation}\label{eq:massiveEE_3d_gauge_finite_1stOrder}
S_{\text{finite}} \approx  - \frac{1}{6} {\mathcal{A}} M
\left[	1 - \frac{g_A}{2 \pi}	\mu_{\rm gauge} 
+ \frac{g_A}{2 \pi} \ln M +\mathcal{O}(g_A^2)	\right] .
\end{equation}
One subtlety of this result is that the dimension of the inserted parameter $M$ is not identical to the inverse of the surface area $\mathcal{A}$, so that $\mathcal{A} M$ is not a dimensionless parameter that scales under RG transformation (Note that $\mathcal{A} \epsilon^{-1}$ is a dimensionless parameter, so the Eq.~\eqref{eq:massiveEE_3d_gauge_UV_1stOrder} is sufficient to identify the change on leading area-law term). 
In this sense, it is difficult to address the obtained finite terms of Eq.~\eqref{eq:massiveEE_3d_gauge_finite_1stOrder} with the universal subleading correction of EE. 
However, as we will show in the next section, a direct connection to the RG flows can be achieved by considering the full form of the subleading terms.

\subsection{The full form of the entanglement entropy of Dirac fermions exposed to a random static gauge field}\label{app:dirac_randomgauge_full_form}

In our dimensional reduction scheme, one necessary step for accessing the entanglement entropy of Dirac fermions, is to estimate the double product of lower-dimensional (flat) Green's function $\left[ \gamma^0 \widetilde{g}_D^{(1)}(k) \right]^2$. 
This double product contains the dominate correction of applying a random static gauge field, for which an estimation only requires the knowledge of performing ordinary perturbation theory for Green's function on flat spacetime.

However, in presence of disorders, an exact calculation of the double product is hard even for non-interacting models. 
In the previous section, we have presented a perturbative analysis of the double product, and use it to calculate the entanglement entropy at the lowest-order of randomness strength $g_A$. 
The calculation appears to be mathematically complicated, and a direct extension to higher orders meets the difficulty on summing/integrating certain special functions. 
Here we show that, instead of computing a perturbative series order by order, it would be more convenient to estimate the full form of the entanglement entropy.

Recall that we have derived explicitly the ordinary Green's function on flat spacetime in Appendix~\ref{app:gauge_dirac_flat}. 
While the non-perturbative results are obtained for the massless case, the influence of a inserted source term $M$ (used to calculate the partition function from the Green's function) can be naturally considered as shifting the momentum $k^2 \to k^2+M^2$. 
By applying this transformation to Eq.~\eqref{eq:3d_gauge_GreenFunction}, we can write down the non-perturbative double product as 
\begin{equation}
\begin{aligned}
\gamma^0 \widetilde{g}_D^{(1)}(k) 
\gamma^0 \widetilde{g}_D^{(1)}(k) 
= \frac{ \left[ C(g_A) \right]^{2} }
{\left( k^2 + M^2 \right)^{1+\frac{g_A}{2\pi}} }
{\rm Id} 
\qquad \text{with} \quad
C(g_A) = 2^{1+\frac{g_A}{2\pi}} 
\frac{\Gamma(1+\frac{g_A}{4\pi})}
{\Gamma(1-\frac{g_A}{4\pi})} , 
\end{aligned}
\end{equation}
It is easy to check that the first-order expansion of the above formula with respect to the disorder strength $g_A$ is identical to the result of Eq.~\eqref{eq:double_product_gauge} that derived in previous sections. 
With this non-perturbative result in hand, the interacting two-point correlator on $2$D replica manifold of Eq.~\eqref{eq:dimensional_reduction_3d_gauge} becomes 
\begin{equation}
\begin{aligned}
& \quad 
\left( -\omega^2 \right)^l \left[
P^{(n)}_{D,2l,{\rm int}}(\mathbf{r}_\parallel, \mathbf{r}_\parallel') 
- P^{(1)}_{D,2l,{\rm int}}(\mathbf{r}_\parallel, \mathbf{r}_\parallel') 
\right]_{\text{diag}} \\ 
& = (l+1) (-\omega^2)^l 
\frac{1-n^2}{6n^2} \frac{M}{2\pi} K_0(M\rho)
\int \frac{d^2 \mathbf{k}}{2\pi}
e^{i \mathbf{k} \mathbf{r}_\parallel'} 
\frac{ \left[ C(g_A) \right]^{2l} }
{\left( k^2 + M^2 \right)^{l(1+\frac{g_A}{2\pi})+1} }
.
\end{aligned}
\end{equation}
It can be calculated as follows
\begin{equation}
\begin{aligned}
& \quad 
(l+1) [-\omega^2 C^2(g_A)]^{l} 
\frac{1-n^2}{6n^2} \frac{M}{2\pi} K_0(M\rho) 
\int_0^\infty dk 
\frac{ k J_0(k\rho') }
{(k^2 + M^2)^{l(1+\frac{g_A}{2\pi})+1}} \\
& = (l+1) \left[ -\omega^2 C^2(g_A) \right]^{l} 
\frac{1-n^2}{6n^2} \frac{M}{2\pi} K_0(M\rho) 
\frac{1}
{
	\Gamma \left[ l(1+\frac{g_A}{2\pi}+1) \right]
}
\left( \frac{\rho'}{2M} \right)^{ l (1+\frac{g_A}{2\pi}) }
K_{l(1+\frac{g_A}{2\pi})} (M \rho') .
\end{aligned}
\end{equation}
Its trace over $2$D replica manifold gives 
\begin{equation}
\begin{aligned}
& \quad 
\Tr^{(n)}_{2D} \left( -\omega^2 \right)^l \left[
P^{(n)}_{D,2l,{\rm int}}(\mathbf{r}_\parallel, \mathbf{r}_\parallel') 
- P^{(1)}_{D,2l,{\rm int}}(\mathbf{r}_\parallel, \mathbf{r}_\parallel') 
\right] \\
& = 
2 \frac{ (l+1) [-\omega^2 C^2(g_A)]^{l} }
{
	\Gamma \left[ l(1+\frac{g_A}{2\pi}+1) \right]
}
\frac{1-n^2}{6n^2} \frac{M}{2\pi} \int_0^{2\pi n} d\theta 
\int_0^\infty \rho d\rho 
\left( \frac{\rho'}{2M} \right)^{ l (1+\frac{g_A}{2\pi}) }
K_0(M\rho) K_{l(1+\frac{g_A}{2\pi})} (M \rho') \\
& = \frac{1-n^2}{6n} 
\frac{l+1}{1+l(1+\frac{g_A}{2\pi})} 
[-\omega^2 C^2(g_A)]^{l} 
M^{-1-2l(1+\frac{g_A}{2\pi})} .
\end{aligned}
\end{equation}
Then a resummation of $l$ leads to the trace of corresponding $3$D Green's function 
\begin{equation}
\begin{aligned}
& \quad 
\Tr^{(n)}_{2D} 
\left[ G^{(n)}_{D,\text{int}} - G^{(1)}_{D,\text{int}} \right] 
= \sum_{l=0}^\infty \Tr^{(n)}_{2D} \left( -\omega^2 \right)^l 
\left[
P^{(n)}_{D,2l,{\rm int}}(\mathbf{r}_\parallel, \mathbf{r}_\parallel') 
- P^{(1)}_{D,2l,{\rm int}}(\mathbf{r}_\parallel, \mathbf{r}_\parallel') 
\right] \\
& = \frac{1-n^2}{6n} \sum_{l=0}^\infty 
\frac{l+1}{1 + l(1+\frac{g_A}{2\pi})} 
\left[ -\omega^2 C^2{g_A} \right]^l 
M^{-1-2l(1+\frac{g_A}{2\pi})} \\
& = \frac{1-n^2}{6n} \frac{1}{2M}
\Bigg\{ 
{}_2F_1 \left[ 1, \frac{1}{1+\frac{g_A}{2\pi}}; 
\frac{2+\frac{g_A}{2\pi}}{1+\frac{g_A}{2\pi}}; 
- \frac{C^2(g_A) \omega^2}{M^{2(1+\frac{g_A}{2\pi})}} \right] 
\\ & \qquad \qquad \qquad \quad 
+ 
\frac{1}{2+\frac{g_A}{2\pi}} 
\frac{C^2(g_A) \omega^2}{M^{2(1+\frac{g_A}{2\pi})}} 
{}_2F_1 \left[ 2, \frac{2+\frac{g_A}{2\pi}}{1+\frac{g_A}{2\pi}}; 
\frac{3+2\frac{g_A}{2\pi}}{1+\frac{g_A}{2\pi}}; 
- \frac{C^2(g_A) \omega^2}{M^{2(1+\frac{g_A}{2\pi})}} \right] 
\Bigg\} . 
\end{aligned}
\end{equation}
This results the full form of entanglement entropy 
\begin{equation}
\begin{aligned}
S & = \frac{1}{6} \mathcal{A} 
\left[ \left(1-\frac{g_A^2}{4\pi^2}\right) C(g_A) \right]^{-1} 
\left[ \epsilon^{-(1+\frac{g_A}{2\pi})} 
- M^{1+\frac{g_A}{2\pi}} \right] .
\\ & 
\end{aligned}
\end{equation}
Its lowest-order expansion in $g_A$ gives 
$S \approx \frac{1}{6} \mathcal{A} 
\left( 1 - \frac{g_A}{2\pi} \overline{\mu} \right) 
\left[
\epsilon^{-1} \left( 1 + \frac{g_A}{2\pi} \ln \epsilon^{-1} \right)
- M \left( 1 + \frac{g_A}{2\pi} \ln M \right)
\right] $, 
which is consistent with the obtained form in Eq.~\eqref{eq:massiveEE_3d_gauge}. 
The only difference is on the value of $\overline{\mu} = \ln2 - \gamma \approx 0.115932$ that could be caused by exchanging integral and summation in the computation represented in Appendix~\ref{app:first_order_EE_gauge}. 
However, this does not influence on the main result of present work. 
Moreover, for the finite term that associated with parameter $M$, 
here we have $\mathcal{A} M^{1+\frac{g_A}{2\pi}} \sim \mathcal{A}/\xi$ as a dimensionless parameter ($\xi$ is the correlation length), so that the finite term can be understood as the universal subleading correction of EE.

\section{Estimation of the entanglement entropy in the quasiparticle picture}\label{app:quasiparticle}

Here, we adopt a quasiparticle picture to describe the EE in scale-invariant fermionic systems~\cite{nahum2020_majorana_defect, tang2021nonunitary}, where the entanglement is considered to be produced by quasiparticle entangled-pairs in the system. 
The only control parameter in this picture is the distribution function of those pairs $P(r)$, which gives the EE 
\begin{equation}
S_A \sim \int_A dV_A \int_{\overline{A}} dV_{\overline{A}} P(r_{A, \overline{A}}) , 
\end{equation}
where $A$ and $\overline{A}$ are complementary to the total system, and $r_{A, \overline{A}}$ is the distance between the (lattice) points in the two subsystems $A$ and $\overline{A}$. 
Although the current case is a ground state that different from the dynamical steady state with excitations, it is still naturally to understand $P(r_{A, \overline{A}})$ as the squared two-point correlation function of the fermion operator, which gives a power-law decay of $P(r_{A, \overline{A}}) \propto r_{A, \overline{A}}^{-k}$ for scale-invariant systems. 

It is convenient to work in polar coordinate with disc geometry, then the above integral becomes
\begin{equation}
S_{A} \sim \int_0^{L_A} r' dr' \int_0^{2\pi} d\varphi' \int_{L_A+\epsilon}^{\infty} r dr \int_0^{2\pi} d\varphi \frac{1}{|\vec{r} - \vec{r}'|^k} . 
\end{equation}
There are two set of angular variable, and one of them can be always removed. For instance, we have
\begin{equation}
S_{A} \sim  2\pi  \int_0^{L_A} r' dr' 
\int_{L_A+\epsilon}^{\infty} r dr \int_0^{2\pi} d\varphi \frac{1}{|\vec{r} - \vec{r}'|^k} 
\sim 2\pi L_A^4 \int_{-1}^0 d u_x 
\int_{{\epsilon}/{L_A}}^{\infty} d v_x 
\int_{-\infty}^\infty dv_y 
\left\{ L_A^2
\left[ (u_x - v_x)^2 + v_y^2 \right]
\right\}^{-\frac{k}{2}} ,
\end{equation}
where we have used $\vec{r'} = L_A \vec{u}$ and $\vec{r} = L_A \vec{v}$. 
For $k > 3$, this integral gives a robust area-law EE 
\begin{equation}
S_A \sim 2\pi \left( L_A \right)^{4-k} 
\left( \frac{\epsilon}{L_A} \right)^{3-k} 
\propto {L_A} .
\end{equation}
For $k=3$, this integral gives a logarithmic violation of the area-law
\begin{equation}
S_A \sim - 4 \pi L_A \ln \frac{\epsilon}{L_A} \sim 4\pi L_A \ln L_A \propto L_A \ln L_A .
\end{equation}

\section{Entanglement entropy in the presence of interactions}


In the main text, an explicit calculation of entanglement entropy is performed for non-interacting theories. 
In this section, we discuss the calculation of the entanglement entropy for interacting theories. 
In general, in the calculation of the entanglement entropy, the starting point is two-point correlations. 
As shown in Appendix~\ref{app:gauge_dirac}, the main features of entanglement entropy are captured by the spatial scaling behavior of Green's function $g(\mathbf{r}_{\parallel}, \mathbf{r}_{\parallel}') \sim |\mathbf{r}_{\parallel} - \mathbf{r}_{\parallel}'|^{-1+\frac{g_A}{2\pi}}$. 
Here, by providing a renormalization group analysis, we would like to argue that such spatial scaling behavior is qualitatively unchanged in the presence of interactions. Therefore, the main results on entanglement entropy (such as the area-law scaling) are believed to be universal.

The stability/instability of the non-interacting disordered critical points was widely discussed in previous literature~\cite{LeeDH_1996_interactIQHE, Ye1999, Herbut_2001_Coulomb_zeq1, Wang_2002_interactIQHE, Stauber_2005_disorder_interaction, Herbut_2008_Coulomb_MinimalConduct, Vafek_2008_RG_Coulomb_RandGauge}. 
On one hand, short-range interactions or screened Coulomb interactions are found to be irrelevant to the non-interacting disordered critical points that induced by random static gauge fields~\cite{LeeDH_1996_interactIQHE, Wang_2002_interactIQHE},  
so that we expect the scaling behaviors of two-point correlations and consequently the entanglement entropy remain unchanged. 
On the other hand, the non-interacting disordered critical points are unstable under a true long-range Coulomb interaction and the system flows to a line of new fixed points~\cite{Ye1999, Herbut_2001_Coulomb_zeq1, Herbut_2008_Coulomb_MinimalConduct, Vafek_2008_RG_Coulomb_RandGauge}. 
Interestingly, the spatial scaling behavior of two-point correlators on this critical line is found to be similar with the non-interacting disordered critical points. (details will be present in a future coming work)
If one repeat the calculation of using the two-point Green's function at the interacting critical points, one should reach the same area-law scaling of the entanglement entropy.
Thus, we believe that the entanglement entropy should take the same form in disordered quantum critical points in the presence of (weak) interactions.

\end{widetext}

\bibliography{3dEE}

\begin{thebibliography}{139}%
\makeatletter
\providecommand \@ifxundefined [1]{%
 \@ifx{#1\undefined}
}%
\providecommand \@ifnum [1]{%
 \ifnum #1\expandafter \@firstoftwo
 \else \expandafter \@secondoftwo
 \fi
}%
\providecommand \@ifx [1]{%
 \ifx #1\expandafter \@firstoftwo
 \else \expandafter \@secondoftwo
 \fi
}%
\providecommand \natexlab [1]{#1}%
\providecommand \enquote  [1]{``#1''}%
\providecommand \bibnamefont  [1]{#1}%
\providecommand \bibfnamefont [1]{#1}%
\providecommand \citenamefont [1]{#1}%
\providecommand \href@noop [0]{\@secondoftwo}%
\providecommand \href [0]{\begingroup \@sanitize@url \@href}%
\providecommand \@href[1]{\@@startlink{#1}\@@href}%
\providecommand \@@href[1]{\endgroup#1\@@endlink}%
\providecommand \@sanitize@url [0]{\catcode `\\12\catcode `\$12\catcode
  `\&12\catcode `\#12\catcode `\^12\catcode `\_12\catcode `\%12\relax}%
\providecommand \@@startlink[1]{}%
\providecommand \@@endlink[0]{}%
\providecommand \url  [0]{\begingroup\@sanitize@url \@url }%
\providecommand \@url [1]{\endgroup\@href {#1}{\urlprefix }}%
\providecommand \urlprefix  [0]{URL }%
\providecommand \Eprint [0]{\href }%
\providecommand \doibase [0]{http://dx.doi.org/}%
\providecommand \selectlanguage [0]{\@gobble}%
\providecommand \bibinfo  [0]{\@secondoftwo}%
\providecommand \bibfield  [0]{\@secondoftwo}%
\providecommand \translation [1]{[#1]}%
\providecommand \BibitemOpen [0]{}%
\providecommand \bibitemStop [0]{}%
\providecommand \bibitemNoStop [0]{.\EOS\space}%
\providecommand \EOS [0]{\spacefactor3000\relax}%
\providecommand \BibitemShut  [1]{\csname bibitem#1\endcsname}%
\let\auto@bib@innerbib\@empty
\bibitem [{\citenamefont {Amico}\ \emph {et~al.}(2008)\citenamefont {Amico},
  \citenamefont {Fazio}, \citenamefont {Osterloh},\ and\ \citenamefont
  {Vedral}}]{Amico2008}%
  \BibitemOpen
  \bibfield  {author} {\bibinfo {author} {\bibfnamefont {Luigi}\ \bibnamefont
  {Amico}}, \bibinfo {author} {\bibfnamefont {Rosario}\ \bibnamefont {Fazio}},
  \bibinfo {author} {\bibfnamefont {Andreas}\ \bibnamefont {Osterloh}}, \ and\
  \bibinfo {author} {\bibfnamefont {Vlatko}\ \bibnamefont {Vedral}},\
  }\bibfield  {title} {\enquote {\bibinfo {title} {Entanglement in many-body
  systems},}\ }\href {\doibase 10.1103/RevModPhys.80.517} {\bibfield  {journal}
  {\bibinfo  {journal} {Rev. Mod. Phys.}\ }\textbf {\bibinfo {volume} {80}},\
  \bibinfo {pages} {517--576} (\bibinfo {year} {2008})}\BibitemShut {NoStop}%
\bibitem [{\citenamefont {Eisert}\ \emph {et~al.}(2010)\citenamefont {Eisert},
  \citenamefont {Cramer},\ and\ \citenamefont {Plenio}}]{Eisert2010}%
  \BibitemOpen
  \bibfield  {author} {\bibinfo {author} {\bibfnamefont {J.}~\bibnamefont
  {Eisert}}, \bibinfo {author} {\bibfnamefont {M.}~\bibnamefont {Cramer}}, \
  and\ \bibinfo {author} {\bibfnamefont {M.~B.}\ \bibnamefont {Plenio}},\
  }\bibfield  {title} {\enquote {\bibinfo {title} {Colloquium: Area laws for
  the entanglement entropy},}\ }\href {\doibase 10.1103/RevModPhys.82.277}
  {\bibfield  {journal} {\bibinfo  {journal} {Rev. Mod. Phys.}\ }\textbf
  {\bibinfo {volume} {82}},\ \bibinfo {pages} {277--306} (\bibinfo {year}
  {2010})}\BibitemShut {NoStop}%
\bibitem [{\citenamefont {Bombelli}\ \emph {et~al.}(1986)\citenamefont
  {Bombelli}, \citenamefont {Koul}, \citenamefont {Lee},\ and\ \citenamefont
  {Sorkin}}]{Bombelli1986_area}%
  \BibitemOpen
  \bibfield  {author} {\bibinfo {author} {\bibfnamefont {Luca}\ \bibnamefont
  {Bombelli}}, \bibinfo {author} {\bibfnamefont {Rabinder~K.}\ \bibnamefont
  {Koul}}, \bibinfo {author} {\bibfnamefont {Joohan}\ \bibnamefont {Lee}}, \
  and\ \bibinfo {author} {\bibfnamefont {Rafael~D.}\ \bibnamefont {Sorkin}},\
  }\bibfield  {title} {\enquote {\bibinfo {title} {Quantum source of entropy
  for black holes},}\ }\href {\doibase 10.1103/PhysRevD.34.373} {\bibfield
  {journal} {\bibinfo  {journal} {Phys. Rev. D}\ }\textbf {\bibinfo {volume}
  {34}},\ \bibinfo {pages} {373--383} (\bibinfo {year} {1986})}\BibitemShut
  {NoStop}%
\bibitem [{\citenamefont {Srednicki}(1993)}]{Srednicki1993_area}%
  \BibitemOpen
  \bibfield  {author} {\bibinfo {author} {\bibfnamefont {Mark}\ \bibnamefont
  {Srednicki}},\ }\bibfield  {title} {\enquote {\bibinfo {title} {Entropy and
  area},}\ }\href {\doibase 10.1103/PhysRevLett.71.666} {\bibfield  {journal}
  {\bibinfo  {journal} {Phys. Rev. Lett.}\ }\textbf {\bibinfo {volume} {71}},\
  \bibinfo {pages} {666--669} (\bibinfo {year} {1993})}\BibitemShut {NoStop}%
\bibitem [{\citenamefont {Callan}\ and\ \citenamefont
  {Wilczek}(1994)}]{Callan1994GeometricEntropy}%
  \BibitemOpen
  \bibfield  {author} {\bibinfo {author} {\bibfnamefont {Curtis}\ \bibnamefont
  {Callan}}\ and\ \bibinfo {author} {\bibfnamefont {Frank}\ \bibnamefont
  {Wilczek}},\ }\bibfield  {title} {\enquote {\bibinfo {title} {On geometric
  entropy},}\ }\href {\doibase https://doi.org/10.1016/0370-2693(94)91007-3}
  {\bibfield  {journal} {\bibinfo  {journal} {Physics Letters B}\ }\textbf
  {\bibinfo {volume} {333}},\ \bibinfo {pages} {55--61} (\bibinfo {year}
  {1994})}\BibitemShut {NoStop}%
\bibitem [{\citenamefont {Kabat}(1995)}]{KABAT1995}%
  \BibitemOpen
  \bibfield  {author} {\bibinfo {author} {\bibfnamefont {Daniel}\ \bibnamefont
  {Kabat}},\ }\bibfield  {title} {\enquote {\bibinfo {title} {Black hole
  entropy and entropy of entanglement},}\ }\href {\doibase
  https://doi.org/10.1016/0550-3213(95)00443-V} {\bibfield  {journal} {\bibinfo
   {journal} {Nuclear Physics B}\ }\textbf {\bibinfo {volume} {453}},\ \bibinfo
  {pages} {281--299} (\bibinfo {year} {1995})}\BibitemShut {NoStop}%
\bibitem [{\citenamefont {Ryu}\ and\ \citenamefont
  {Takayanagi}(2006{\natexlab{a}})}]{RTformula2006}%
  \BibitemOpen
  \bibfield  {author} {\bibinfo {author} {\bibfnamefont {Shinsei}\ \bibnamefont
  {Ryu}}\ and\ \bibinfo {author} {\bibfnamefont {Tadashi}\ \bibnamefont
  {Takayanagi}},\ }\bibfield  {title} {\enquote {\bibinfo {title} {Holographic
  derivation of entanglement entropy from the anti--de Sitter space/conformal
  field theory correspondence},}\ }\href {\doibase
  10.1103/PhysRevLett.96.181602} {\bibfield  {journal} {\bibinfo  {journal}
  {Phys. Rev. Lett.}\ }\textbf {\bibinfo {volume} {96}},\ \bibinfo {pages}
  {181602} (\bibinfo {year} {2006}{\natexlab{a}})}\BibitemShut {NoStop}%
\bibitem [{\citenamefont
  {Swingle}(2012{\natexlab{a}})}]{swingle2012holography}%
  \BibitemOpen
  \bibfield  {author} {\bibinfo {author} {\bibfnamefont {Brian}\ \bibnamefont
  {Swingle}},\ }\bibfield  {title} {\enquote {\bibinfo {title} {Entanglement
  renormalization and holography},}\ }\href {\doibase
  10.1103/PhysRevD.86.065007} {\bibfield  {journal} {\bibinfo  {journal} {Phys.
  Rev. D}\ }\textbf {\bibinfo {volume} {86}},\ \bibinfo {pages} {065007}
  (\bibinfo {year} {2012}{\natexlab{a}})}\BibitemShut {NoStop}%
\bibitem [{\citenamefont {Rangamani}\ and\ \citenamefont
  {Takayanagi}(2017)}]{takayanagi2017holographic}%
  \BibitemOpen
  \bibfield  {author} {\bibinfo {author} {\bibfnamefont {Mukund}\ \bibnamefont
  {Rangamani}}\ and\ \bibinfo {author} {\bibfnamefont {Tadashi}\ \bibnamefont
  {Takayanagi}},\ }\href@noop {} {\emph {\bibinfo {title} {Holographic
  Entanglement Entropy}}}\ (\bibinfo  {publisher} {Springer},\ \bibinfo {year}
  {2017})\BibitemShut {NoStop}%
\bibitem [{\citenamefont {Nishioka}(2018)}]{Tatsuma2018}%
  \BibitemOpen
  \bibfield  {author} {\bibinfo {author} {\bibfnamefont {Tatsuma}\ \bibnamefont
  {Nishioka}},\ }\bibfield  {title} {\enquote {\bibinfo {title} {Entanglement
  entropy: Holography and renormalization group},}\ }\href {\doibase
  10.1103/RevModPhys.90.035007} {\bibfield  {journal} {\bibinfo  {journal}
  {Rev. Mod. Phys.}\ }\textbf {\bibinfo {volume} {90}},\ \bibinfo {pages}
  {035007} (\bibinfo {year} {2018})}\BibitemShut {NoStop}%
\bibitem [{\citenamefont {Kitaev}\ and\ \citenamefont
  {Preskill}(2006)}]{kitaev2006TopoEE}%
  \BibitemOpen
  \bibfield  {author} {\bibinfo {author} {\bibfnamefont {Alexei}\ \bibnamefont
  {Kitaev}}\ and\ \bibinfo {author} {\bibfnamefont {John}\ \bibnamefont
  {Preskill}},\ }\bibfield  {title} {\enquote {\bibinfo {title} {Topological
  entanglement entropy},}\ }\href {\doibase 10.1103/PhysRevLett.96.110404}
  {\bibfield  {journal} {\bibinfo  {journal} {Phys. Rev. Lett.}\ }\textbf
  {\bibinfo {volume} {96}},\ \bibinfo {pages} {110404} (\bibinfo {year}
  {2006})}\BibitemShut {NoStop}%
\bibitem [{\citenamefont {Levin}\ and\ \citenamefont
  {Wen}(2006)}]{levin2006TopoEE}%
  \BibitemOpen
  \bibfield  {author} {\bibinfo {author} {\bibfnamefont {Michael}\ \bibnamefont
  {Levin}}\ and\ \bibinfo {author} {\bibfnamefont {Xiao-Gang}\ \bibnamefont
  {Wen}},\ }\bibfield  {title} {\enquote {\bibinfo {title} {Detecting
  topological order in a ground state wave function},}\ }\href {\doibase
  10.1103/PhysRevLett.96.110405} {\bibfield  {journal} {\bibinfo  {journal}
  {Phys. Rev. Lett.}\ }\textbf {\bibinfo {volume} {96}},\ \bibinfo {pages}
  {110405} (\bibinfo {year} {2006})}\BibitemShut {NoStop}%
\bibitem [{\citenamefont {Li}\ and\ \citenamefont
  {Haldane}(2008)}]{LiHaldane2008_ES}%
  \BibitemOpen
  \bibfield  {author} {\bibinfo {author} {\bibfnamefont {Hui}\ \bibnamefont
  {Li}}\ and\ \bibinfo {author} {\bibfnamefont {F.~D.~M.}\ \bibnamefont
  {Haldane}},\ }\bibfield  {title} {\enquote {\bibinfo {title} {Entanglement
  spectrum as a generalization of entanglement entropy: Identification of
  topological order in non-Abelian fractional quantum Hall effect states},}\
  }\href {\doibase 10.1103/PhysRevLett.101.010504} {\bibfield  {journal}
  {\bibinfo  {journal} {Phys. Rev. Lett.}\ }\textbf {\bibinfo {volume} {101}},\
  \bibinfo {pages} {010504} (\bibinfo {year} {2008})}\BibitemShut {NoStop}%
\bibitem [{\citenamefont {Qi}\ \emph {et~al.}(2012)\citenamefont {Qi},
  \citenamefont {Katsura},\ and\ \citenamefont {Ludwig}}]{Qi2012ES}%
  \BibitemOpen
  \bibfield  {author} {\bibinfo {author} {\bibfnamefont {Xiao-Liang}\
  \bibnamefont {Qi}}, \bibinfo {author} {\bibfnamefont {Hosho}\ \bibnamefont
  {Katsura}}, \ and\ \bibinfo {author} {\bibfnamefont {Andreas W.~W.}\
  \bibnamefont {Ludwig}},\ }\bibfield  {title} {\enquote {\bibinfo {title}
  {General relationship between the entanglement spectrum and the edge state
  spectrum of topological quantum states},}\ }\href {\doibase
  10.1103/PhysRevLett.108.196402} {\bibfield  {journal} {\bibinfo  {journal}
  {Phys. Rev. Lett.}\ }\textbf {\bibinfo {volume} {108}},\ \bibinfo {pages}
  {196402} (\bibinfo {year} {2012})}\BibitemShut {NoStop}%
\bibitem [{\citenamefont {Holzhey}\ \emph {et~al.}(1994)\citenamefont
  {Holzhey}, \citenamefont {Larsen},\ and\ \citenamefont
  {Wilczek}}]{HOLZHEY1994443}%
  \BibitemOpen
  \bibfield  {author} {\bibinfo {author} {\bibfnamefont {Christoph}\
  \bibnamefont {Holzhey}}, \bibinfo {author} {\bibfnamefont {Finn}\
  \bibnamefont {Larsen}}, \ and\ \bibinfo {author} {\bibfnamefont {Frank}\
  \bibnamefont {Wilczek}},\ }\bibfield  {title} {\enquote {\bibinfo {title}
  {Geometric and renormalized entropy in conformal field theory},}\ }\href
  {\doibase 10.1016/0550-3213(94)90402-2} {\bibfield  {journal} {\bibinfo
  {journal} {Nuclear Physics B}\ }\textbf {\bibinfo {volume} {424}},\ \bibinfo
  {pages} {443 -- 467} (\bibinfo {year} {1994})}\BibitemShut {NoStop}%
\bibitem [{\citenamefont {Vidal}\ \emph {et~al.}(2003)\citenamefont {Vidal},
  \citenamefont {Latorre}, \citenamefont {Rico},\ and\ \citenamefont
  {Kitaev}}]{vidal2003entanglement}%
  \BibitemOpen
  \bibfield  {author} {\bibinfo {author} {\bibfnamefont {G.}~\bibnamefont
  {Vidal}}, \bibinfo {author} {\bibfnamefont {J.~I.}\ \bibnamefont {Latorre}},
  \bibinfo {author} {\bibfnamefont {E.}~\bibnamefont {Rico}}, \ and\ \bibinfo
  {author} {\bibfnamefont {A.}~\bibnamefont {Kitaev}},\ }\bibfield  {title}
  {\enquote {\bibinfo {title} {Entanglement in quantum critical phenomena},}\
  }\href {\doibase 10.1103/PhysRevLett.90.227902} {\bibfield  {journal}
  {\bibinfo  {journal} {Phys. Rev. Lett.}\ }\textbf {\bibinfo {volume} {90}},\
  \bibinfo {pages} {227902} (\bibinfo {year} {2003})}\BibitemShut {NoStop}%
\bibitem [{\citenamefont {Calabrese}\ and\ \citenamefont
  {Cardy}(2004)}]{Calabrese_2004}%
  \BibitemOpen
  \bibfield  {author} {\bibinfo {author} {\bibfnamefont {Pasquale}\
  \bibnamefont {Calabrese}}\ and\ \bibinfo {author} {\bibfnamefont {John}\
  \bibnamefont {Cardy}},\ }\bibfield  {title} {\enquote {\bibinfo {title}
  {Entanglement entropy and quantum field theory},}\ }\href {\doibase
  10.1088/1742-5468/2004/06/p06002} {\bibfield  {journal} {\bibinfo  {journal}
  {J. Stat. Mech.}\ }\textbf {\bibinfo {volume} {2004}},\ \bibinfo {pages}
  {P06002} (\bibinfo {year} {2004})}\BibitemShut {NoStop}%
\bibitem [{\citenamefont {Fradkin}\ and\ \citenamefont
  {Moore}(2006)}]{fradkin2006EE2p1Critical}%
  \BibitemOpen
  \bibfield  {author} {\bibinfo {author} {\bibfnamefont {Eduardo}\ \bibnamefont
  {Fradkin}}\ and\ \bibinfo {author} {\bibfnamefont {Joel~E.}\ \bibnamefont
  {Moore}},\ }\bibfield  {title} {\enquote {\bibinfo {title} {Entanglement
  entropy of 2D conformal quantum critical points: Hearing the shape of a
  quantum drum},}\ }\href {\doibase 10.1103/PhysRevLett.97.050404} {\bibfield
  {journal} {\bibinfo  {journal} {Phys. Rev. Lett.}\ }\textbf {\bibinfo
  {volume} {97}},\ \bibinfo {pages} {050404} (\bibinfo {year}
  {2006})}\BibitemShut {NoStop}%
\bibitem [{\citenamefont {Bueno}\ \emph {et~al.}(2015)\citenamefont {Bueno},
  \citenamefont {Myers},\ and\ \citenamefont
  {Witczak-Krempa}}]{william2015cornerEE}%
  \BibitemOpen
  \bibfield  {author} {\bibinfo {author} {\bibfnamefont {Pablo}\ \bibnamefont
  {Bueno}}, \bibinfo {author} {\bibfnamefont {Robert~C.}\ \bibnamefont
  {Myers}}, \ and\ \bibinfo {author} {\bibfnamefont {William}\ \bibnamefont
  {Witczak-Krempa}},\ }\bibfield  {title} {\enquote {\bibinfo {title}
  {Universality of corner entanglement in conformal field theories},}\ }\href
  {\doibase 10.1103/PhysRevLett.115.021602} {\bibfield  {journal} {\bibinfo
  {journal} {Phys. Rev. Lett.}\ }\textbf {\bibinfo {volume} {115}},\ \bibinfo
  {pages} {021602} (\bibinfo {year} {2015})}\BibitemShut {NoStop}%
\bibitem [{\citenamefont {Calabrese}\ and\ \citenamefont
  {Cardy}(2005)}]{Calabrese_2005}%
  \BibitemOpen
  \bibfield  {author} {\bibinfo {author} {\bibfnamefont {Pasquale}\
  \bibnamefont {Calabrese}}\ and\ \bibinfo {author} {\bibfnamefont {John}\
  \bibnamefont {Cardy}},\ }\bibfield  {title} {\enquote {\bibinfo {title}
  {Evolution of entanglement entropy in one-dimensional systems},}\ }\href
  {\doibase 10.1088/1742-5468/2005/04/p04010} {\bibfield  {journal} {\bibinfo
  {journal} {J. Stat. Mech.}\ }\textbf {\bibinfo {volume} {2005}},\ \bibinfo
  {pages} {P04010} (\bibinfo {year} {2005})}\BibitemShut {NoStop}%
\bibitem [{\citenamefont {Calabrese}\ and\ \citenamefont
  {Cardy}(2009)}]{Calabrese_2009_review}%
  \BibitemOpen
  \bibfield  {author} {\bibinfo {author} {\bibfnamefont {Pasquale}\
  \bibnamefont {Calabrese}}\ and\ \bibinfo {author} {\bibfnamefont {John}\
  \bibnamefont {Cardy}},\ }\bibfield  {title} {\enquote {\bibinfo {title}
  {Entanglement entropy and conformal field theory},}\ }\href {\doibase
  10.1088/1751-8113/42/50/504005} {\bibfield  {journal} {\bibinfo  {journal}
  {J. Phys. A: Math. Theor.}\ }\textbf {\bibinfo {volume} {42}},\ \bibinfo
  {pages} {504005} (\bibinfo {year} {2009})}\BibitemShut {NoStop}%
\bibitem [{\citenamefont {Laflorencie}(2016)}]{laflorencie2016quantum}%
  \BibitemOpen
  \bibfield  {author} {\bibinfo {author} {\bibfnamefont {Nicolas}\ \bibnamefont
  {Laflorencie}},\ }\bibfield  {title} {\enquote {\bibinfo {title} {Quantum
  entanglement in condensed matter systems},}\ }\href {\doibase
  10.1016/j.physrep.2016.06.008} {\bibfield  {journal} {\bibinfo  {journal}
  {Phys. Rep.}\ }\textbf {\bibinfo {volume} {646}},\ \bibinfo {pages} {1--59}
  (\bibinfo {year} {2016})}\BibitemShut {NoStop}%
\bibitem [{\citenamefont {Wen}(2017)}]{XiaoGang2017RevModPhys}%
  \BibitemOpen
  \bibfield  {author} {\bibinfo {author} {\bibfnamefont {Xiao-Gang}\
  \bibnamefont {Wen}},\ }\bibfield  {title} {\enquote {\bibinfo {title}
  {Colloquium: Zoo of quantum-topological phases of matter},}\ }\href {\doibase
  10.1103/RevModPhys.89.041004} {\bibfield  {journal} {\bibinfo  {journal}
  {Rev. Mod. Phys.}\ }\textbf {\bibinfo {volume} {89}},\ \bibinfo {pages}
  {041004} (\bibinfo {year} {2017})}\BibitemShut {NoStop}%
\bibitem [{\citenamefont {von Keyserlingk}\ \emph {et~al.}(2018)\citenamefont
  {von Keyserlingk}, \citenamefont {Rakovszky}, \citenamefont {Pollmann},\ and\
  \citenamefont {Sondhi}}]{pollmann2018_OTOC_rand}%
  \BibitemOpen
  \bibfield  {author} {\bibinfo {author} {\bibfnamefont {C.~W.}\ \bibnamefont
  {von Keyserlingk}}, \bibinfo {author} {\bibfnamefont {Tibor}\ \bibnamefont
  {Rakovszky}}, \bibinfo {author} {\bibfnamefont {Frank}\ \bibnamefont
  {Pollmann}}, \ and\ \bibinfo {author} {\bibfnamefont {S.~L.}\ \bibnamefont
  {Sondhi}},\ }\bibfield  {title} {\enquote {\bibinfo {title} {Operator
  hydrodynamics, otocs, and entanglement growth in systems without conservation
  laws},}\ }\href {\doibase 10.1103/PhysRevX.8.021013} {\bibfield  {journal}
  {\bibinfo  {journal} {Phys. Rev. X}\ }\textbf {\bibinfo {volume} {8}},\
  \bibinfo {pages} {021013} (\bibinfo {year} {2018})}\BibitemShut {NoStop}%
\bibitem [{\citenamefont {Abanin}\ \emph {et~al.}(2019)\citenamefont {Abanin},
  \citenamefont {Altman}, \citenamefont {Bloch},\ and\ \citenamefont
  {Serbyn}}]{altman2019Review}%
  \BibitemOpen
  \bibfield  {author} {\bibinfo {author} {\bibfnamefont {Dmitry~A.}\
  \bibnamefont {Abanin}}, \bibinfo {author} {\bibfnamefont {Ehud}\ \bibnamefont
  {Altman}}, \bibinfo {author} {\bibfnamefont {Immanuel}\ \bibnamefont
  {Bloch}}, \ and\ \bibinfo {author} {\bibfnamefont {Maksym}\ \bibnamefont
  {Serbyn}},\ }\bibfield  {title} {\enquote {\bibinfo {title} {Colloquium:
  Many-body localization, thermalization, and entanglement},}\ }\href {\doibase
  10.1103/RevModPhys.91.021001} {\bibfield  {journal} {\bibinfo  {journal}
  {Rev. Mod. Phys.}\ }\textbf {\bibinfo {volume} {91}},\ \bibinfo {pages}
  {021001} (\bibinfo {year} {2019})}\BibitemShut {NoStop}%
\bibitem [{\citenamefont {Lewis-Swan}\ \emph {et~al.}(2019)\citenamefont
  {Lewis-Swan}, \citenamefont {Safavi-Naini}, \citenamefont {Kaufman},\ and\
  \citenamefont {Rey}}]{NatRevPhys2019dynamics}%
  \BibitemOpen
  \bibfield  {author} {\bibinfo {author} {\bibfnamefont {R.~J.}\ \bibnamefont
  {Lewis-Swan}}, \bibinfo {author} {\bibfnamefont {A.}~\bibnamefont
  {Safavi-Naini}}, \bibinfo {author} {\bibfnamefont {A.~M.}\ \bibnamefont
  {Kaufman}}, \ and\ \bibinfo {author} {\bibfnamefont {A.~M.}\ \bibnamefont
  {Rey}},\ }\bibfield  {title} {\enquote {\bibinfo {title} {Dynamics of quantum
  information},}\ }\href {\doibase 10.1038/s42254-019-0090-y} {\bibfield
  {journal} {\bibinfo  {journal} {Nat. Rev. Phys.}\ }\textbf {\bibinfo {volume}
  {1}},\ \bibinfo {pages} {627--634} (\bibinfo {year} {2019})}\BibitemShut
  {NoStop}%
\bibitem [{\citenamefont {Nahum}\ and\ \citenamefont
  {Skinner}(2020)}]{nahum2020_majorana_defect}%
  \BibitemOpen
  \bibfield  {author} {\bibinfo {author} {\bibfnamefont {Adam}\ \bibnamefont
  {Nahum}}\ and\ \bibinfo {author} {\bibfnamefont {Brian}\ \bibnamefont
  {Skinner}},\ }\bibfield  {title} {\enquote {\bibinfo {title} {Entanglement
  and dynamics of diffusion-annihilation processes with majorana defects},}\
  }\href {\doibase 10.1103/PhysRevResearch.2.023288} {\bibfield  {journal}
  {\bibinfo  {journal} {Phys. Rev. Research}\ }\textbf {\bibinfo {volume}
  {2}},\ \bibinfo {pages} {023288} (\bibinfo {year} {2020})}\BibitemShut
  {NoStop}%
\bibitem [{\citenamefont {Tang}\ \emph {et~al.}(2021)\citenamefont {Tang},
  \citenamefont {Chen},\ and\ \citenamefont {Zhu}}]{tang2021nonunitary}%
  \BibitemOpen
  \bibfield  {author} {\bibinfo {author} {\bibfnamefont {Qicheng}\ \bibnamefont
  {Tang}}, \bibinfo {author} {\bibfnamefont {Xiao}\ \bibnamefont {Chen}}, \
  and\ \bibinfo {author} {\bibfnamefont {W.}~\bibnamefont {Zhu}},\ }\bibfield
  {title} {\enquote {\bibinfo {title} {Quantum criticality in the nonunitary
  dynamics of $(2+1)$-dimensional free fermions},}\ }\href {\doibase
  10.1103/PhysRevB.103.174303} {\bibfield  {journal} {\bibinfo  {journal}
  {Phys. Rev. B}\ }\textbf {\bibinfo {volume} {103}},\ \bibinfo {pages}
  {174303} (\bibinfo {year} {2021})}\BibitemShut {NoStop}%
\bibitem [{\citenamefont {Hastings}(2007)}]{hastings2007EE}%
  \BibitemOpen
  \bibfield  {author} {\bibinfo {author} {\bibfnamefont {M.~B.}\ \bibnamefont
  {Hastings}},\ }\bibfield  {title} {\enquote {\bibinfo {title} {Entropy and
  entanglement in quantum ground states},}\ }\href {\doibase
  10.1103/PhysRevB.76.035114} {\bibfield  {journal} {\bibinfo  {journal} {Phys.
  Rev. B}\ }\textbf {\bibinfo {volume} {76}},\ \bibinfo {pages} {035114}
  (\bibinfo {year} {2007})}\BibitemShut {NoStop}%
\bibitem [{\citenamefont {Wolf}\ \emph {et~al.}(2008)\citenamefont {Wolf},
  \citenamefont {Verstraete}, \citenamefont {Hastings},\ and\ \citenamefont
  {Cirac}}]{cirac2008AreaLaw}%
  \BibitemOpen
  \bibfield  {author} {\bibinfo {author} {\bibfnamefont {Michael~M.}\
  \bibnamefont {Wolf}}, \bibinfo {author} {\bibfnamefont {Frank}\ \bibnamefont
  {Verstraete}}, \bibinfo {author} {\bibfnamefont {Matthew~B.}\ \bibnamefont
  {Hastings}}, \ and\ \bibinfo {author} {\bibfnamefont {J.~Ignacio}\
  \bibnamefont {Cirac}},\ }\bibfield  {title} {\enquote {\bibinfo {title} {Area
  laws in quantum systems: Mutual information and correlations},}\ }\href
  {\doibase 10.1103/PhysRevLett.100.070502} {\bibfield  {journal} {\bibinfo
  {journal} {Phys. Rev. Lett.}\ }\textbf {\bibinfo {volume} {100}},\ \bibinfo
  {pages} {070502} (\bibinfo {year} {2008})}\BibitemShut {NoStop}%
\bibitem [{\citenamefont {Brand{\~a}o}\ and\ \citenamefont
  {Horodecki}(2013)}]{brandao2013AreaLaw}%
  \BibitemOpen
  \bibfield  {author} {\bibinfo {author} {\bibfnamefont {Fernando G. S.~L.}\
  \bibnamefont {Brand{\~a}o}}\ and\ \bibinfo {author} {\bibfnamefont
  {Micha{\l}}\ \bibnamefont {Horodecki}},\ }\bibfield  {title} {\enquote
  {\bibinfo {title} {An area law for entanglement from exponential decay of
  correlations},}\ }\href {\doibase 10.1038/nphys2747} {\bibfield  {journal}
  {\bibinfo  {journal} {Nature Physics}\ }\textbf {\bibinfo {volume} {9}},\
  \bibinfo {pages} {721--726} (\bibinfo {year} {2013})}\BibitemShut {NoStop}%
\bibitem [{\citenamefont {Cho}(2018)}]{cho2018AreaLaw}%
  \BibitemOpen
  \bibfield  {author} {\bibinfo {author} {\bibfnamefont {Jaeyoon}\ \bibnamefont
  {Cho}},\ }\bibfield  {title} {\enquote {\bibinfo {title} {Realistic area-law
  bound on entanglement from exponentially decaying correlations},}\ }\href
  {\doibase 10.1103/PhysRevX.8.031009} {\bibfield  {journal} {\bibinfo
  {journal} {Phys. Rev. X}\ }\textbf {\bibinfo {volume} {8}},\ \bibinfo {pages}
  {031009} (\bibinfo {year} {2018})}\BibitemShut {NoStop}%
\bibitem [{\citenamefont {Casini}\ and\ \citenamefont
  {Huerta}(2009{\natexlab{a}})}]{CasiniHuerta_2009}%
  \BibitemOpen
  \bibfield  {author} {\bibinfo {author} {\bibfnamefont {H}~\bibnamefont
  {Casini}}\ and\ \bibinfo {author} {\bibfnamefont {M}~\bibnamefont {Huerta}},\
  }\bibfield  {title} {\enquote {\bibinfo {title} {Entanglement entropy in free
  quantum field theory},}\ }\href {\doibase 10.1088/1751-8113/42/50/504007}
  {\bibfield  {journal} {\bibinfo  {journal} {J. Phys. A: Math. Theor.}\
  }\textbf {\bibinfo {volume} {42}},\ \bibinfo {pages} {504007} (\bibinfo
  {year} {2009}{\natexlab{a}})}\BibitemShut {NoStop}%
\bibitem [{\citenamefont
  {Van~Raamsdonk}(2010)}]{raamsdonk2010_spacetime_entanglement}%
  \BibitemOpen
  \bibfield  {author} {\bibinfo {author} {\bibfnamefont {Mark}\ \bibnamefont
  {Van~Raamsdonk}},\ }\bibfield  {title} {\enquote {\bibinfo {title} {Building
  up spacetime with quantum entanglement},}\ }\href {\doibase
  10.1007/s10714-010-1034-0} {\bibfield  {journal} {\bibinfo  {journal} {Gen.
  Relativ. Gravit.}\ }\textbf {\bibinfo {volume} {42}},\ \bibinfo {pages}
  {2323--2329} (\bibinfo {year} {2010})}\BibitemShut {NoStop}%
\bibitem [{\citenamefont {Faulkner}\ \emph {et~al.}(2014)\citenamefont
  {Faulkner}, \citenamefont {Guica}, \citenamefont {Hartman}, \citenamefont
  {Myers},\ and\ \citenamefont {Van~Raamsdonk}}]{faulkner2014_holographic_CFT}%
  \BibitemOpen
  \bibfield  {author} {\bibinfo {author} {\bibfnamefont {Thomas}\ \bibnamefont
  {Faulkner}}, \bibinfo {author} {\bibfnamefont {Monica}\ \bibnamefont
  {Guica}}, \bibinfo {author} {\bibfnamefont {Thomas}\ \bibnamefont {Hartman}},
  \bibinfo {author} {\bibfnamefont {Robert~C.}\ \bibnamefont {Myers}}, \ and\
  \bibinfo {author} {\bibfnamefont {Mark}\ \bibnamefont {Van~Raamsdonk}},\
  }\bibfield  {title} {\enquote {\bibinfo {title} {Gravitation from
  entanglement in holographic CFTs},}\ }\href {\doibase
  10.1007/JHEP03(2014)051} {\bibfield  {journal} {\bibinfo  {journal} {J. High
  Energ. Phys.}\ }\textbf {\bibinfo {volume} {2014}},\ \bibinfo {pages} {51}
  (\bibinfo {year} {2014})}\BibitemShut {NoStop}%
\bibitem [{\citenamefont {Swingle}\ and\ \citenamefont
  {Raamsdonk}(2014)}]{swingle2014universality}%
  \BibitemOpen
  \bibfield  {author} {\bibinfo {author} {\bibfnamefont {Brian}\ \bibnamefont
  {Swingle}}\ and\ \bibinfo {author} {\bibfnamefont {Mark~Van}\ \bibnamefont
  {Raamsdonk}},\ }\href@noop {} {\enquote {\bibinfo {title} {Universality of
  gravity from entanglement},}\ } (\bibinfo {year} {2014}),\ \Eprint
  {http://arxiv.org/abs/1405.2933} {arXiv:1405.2933 [hep-th]} \BibitemShut
  {NoStop}%
\bibitem [{\citenamefont {White}(1992)}]{white1992dmrg}%
  \BibitemOpen
  \bibfield  {author} {\bibinfo {author} {\bibfnamefont {Steven~R.}\
  \bibnamefont {White}},\ }\bibfield  {title} {\enquote {\bibinfo {title}
  {Density matrix formulation for quantum renormalization groups},}\ }\href
  {\doibase 10.1103/PhysRevLett.69.2863} {\bibfield  {journal} {\bibinfo
  {journal} {Phys. Rev. Lett.}\ }\textbf {\bibinfo {volume} {69}},\ \bibinfo
  {pages} {2863--2866} (\bibinfo {year} {1992})}\BibitemShut {NoStop}%
\bibitem [{\citenamefont {Verstraete}\ and\ \citenamefont
  {Cirac}(2004)}]{verstraete2004PEPS}%
  \BibitemOpen
  \bibfield  {author} {\bibinfo {author} {\bibfnamefont {F.}~\bibnamefont
  {Verstraete}}\ and\ \bibinfo {author} {\bibfnamefont {J.~I.}\ \bibnamefont
  {Cirac}},\ }\href@noop {} {\enquote {\bibinfo {title} {Renormalization
  algorithms for quantum-many body systems in two and higher dimensions},}\ }
  (\bibinfo {year} {2004}),\ \Eprint {http://arxiv.org/abs/cond-mat/0407066}
  {arXiv:cond-mat/0407066 [cond-mat]} \BibitemShut {NoStop}%
\bibitem [{\citenamefont {Vidal}(2008)}]{vidal2008EfficientSimulate}%
  \BibitemOpen
  \bibfield  {author} {\bibinfo {author} {\bibfnamefont {G.}~\bibnamefont
  {Vidal}},\ }\bibfield  {title} {\enquote {\bibinfo {title} {Class of quantum
  many-body states that can be efficiently simulated},}\ }\href {\doibase
  10.1103/PhysRevLett.101.110501} {\bibfield  {journal} {\bibinfo  {journal}
  {Phys. Rev. Lett.}\ }\textbf {\bibinfo {volume} {101}},\ \bibinfo {pages}
  {110501} (\bibinfo {year} {2008})}\BibitemShut {NoStop}%
\bibitem [{\citenamefont {Schollwöck}(2011)}]{Schollwock2011_age_mps}%
  \BibitemOpen
  \bibfield  {author} {\bibinfo {author} {\bibfnamefont {Ulrich}\ \bibnamefont
  {Schollwöck}},\ }\bibfield  {title} {\enquote {\bibinfo {title} {The
  density-matrix renormalization group in the age of matrix product states},}\
  }\href {\doibase https://doi.org/10.1016/j.aop.2010.09.012} {\bibfield
  {journal} {\bibinfo  {journal} {Annals of Physics}\ }\textbf {\bibinfo
  {volume} {326}},\ \bibinfo {pages} {96--192} (\bibinfo {year} {2011})},\
  \bibinfo {note} {january 2011 Special Issue}\BibitemShut {NoStop}%
\bibitem [{\citenamefont {Chung}\ and\ \citenamefont
  {Peschel}(2001)}]{chung2001density}%
  \BibitemOpen
  \bibfield  {author} {\bibinfo {author} {\bibfnamefont {Ming-Chiang}\
  \bibnamefont {Chung}}\ and\ \bibinfo {author} {\bibfnamefont {Ingo}\
  \bibnamefont {Peschel}},\ }\bibfield  {title} {\enquote {\bibinfo {title}
  {Density-matrix spectra of solvable fermionic systems},}\ }\href {\doibase
  10.1103/PhysRevB.64.064412} {\bibfield  {journal} {\bibinfo  {journal} {Phys.
  Rev. B}\ }\textbf {\bibinfo {volume} {64}},\ \bibinfo {pages} {064412}
  (\bibinfo {year} {2001})}\BibitemShut {NoStop}%
\bibitem [{\citenamefont {Peschel}(2003)}]{Peschel_2003calculation}%
  \BibitemOpen
  \bibfield  {author} {\bibinfo {author} {\bibfnamefont {Ingo}\ \bibnamefont
  {Peschel}},\ }\bibfield  {title} {\enquote {\bibinfo {title} {Calculation of
  reduced density matrices from correlation functions},}\ }\href {\doibase
  10.1088/0305-4470/36/14/101} {\bibfield  {journal} {\bibinfo  {journal} {J.
  Phys. A: Math. Gen.}\ }\textbf {\bibinfo {volume} {36}},\ \bibinfo {pages}
  {L205--L208} (\bibinfo {year} {2003})}\BibitemShut {NoStop}%
\bibitem [{\citenamefont {Peschel}\ and\ \citenamefont
  {Eisler}(2009)}]{Peschel_2009reduced}%
  \BibitemOpen
  \bibfield  {author} {\bibinfo {author} {\bibfnamefont {Ingo}\ \bibnamefont
  {Peschel}}\ and\ \bibinfo {author} {\bibfnamefont {Viktor}\ \bibnamefont
  {Eisler}},\ }\bibfield  {title} {\enquote {\bibinfo {title} {Reduced density
  matrices and entanglement entropy in free lattice models},}\ }\href {\doibase
  10.1088/1751-8113/42/50/504003} {\bibfield  {journal} {\bibinfo  {journal}
  {J. Phys. A: Math. Theor.}\ }\textbf {\bibinfo {volume} {42}},\ \bibinfo
  {pages} {504003} (\bibinfo {year} {2009})}\BibitemShut {NoStop}%
\bibitem [{\citenamefont {Casini}\ and\ \citenamefont
  {Huerta}(2009{\natexlab{b}})}]{Casini_2009_resolvent_1p1dirac}%
  \BibitemOpen
  \bibfield  {author} {\bibinfo {author} {\bibfnamefont {H}~\bibnamefont
  {Casini}}\ and\ \bibinfo {author} {\bibfnamefont {M}~\bibnamefont {Huerta}},\
  }\bibfield  {title} {\enquote {\bibinfo {title} {Reduced density matrix and
  internal dynamics for multicomponent regions},}\ }\href {\doibase
  10.1088/0264-9381/26/18/185005} {\bibfield  {journal} {\bibinfo  {journal}
  {Class. Quantum Grav.}\ }\textbf {\bibinfo {volume} {26}},\ \bibinfo {pages}
  {185005} (\bibinfo {year} {2009}{\natexlab{b}})}\BibitemShut {NoStop}%
\bibitem [{\citenamefont {Casini}\ and\ \citenamefont
  {Huerta}(2010)}]{casini2010_EE_sphere}%
  \BibitemOpen
  \bibfield  {author} {\bibinfo {author} {\bibfnamefont {H.}~\bibnamefont
  {Casini}}\ and\ \bibinfo {author} {\bibfnamefont {M.}~\bibnamefont
  {Huerta}},\ }\bibfield  {title} {\enquote {\bibinfo {title} {Entanglement
  entropy for the n-sphere},}\ }\href {\doibase
  https://doi.org/10.1016/j.physletb.2010.09.054} {\bibfield  {journal}
  {\bibinfo  {journal} {Physics Letters B}\ }\textbf {\bibinfo {volume}
  {694}},\ \bibinfo {pages} {167--171} (\bibinfo {year} {2010})}\BibitemShut
  {NoStop}%
\bibitem [{\citenamefont {Solodukhin}(2011)}]{Solodukhin2011}%
  \BibitemOpen
  \bibfield  {author} {\bibinfo {author} {\bibfnamefont {Sergey~N.}\
  \bibnamefont {Solodukhin}},\ }\bibfield  {title} {\enquote {\bibinfo {title}
  {Entanglement entropy of black holes},}\ }\href {\doibase
  10.12942/lrr-2011-8} {\bibfield  {journal} {\bibinfo  {journal} {Living Rev.
  Relativ.}\ }\textbf {\bibinfo {volume} {14}},\ \bibinfo {pages} {8} (\bibinfo
  {year} {2011})}\BibitemShut {NoStop}%
\bibitem [{\citenamefont {Lewkowycz}\ \emph {et~al.}(2013)\citenamefont
  {Lewkowycz}, \citenamefont {Myers},\ and\ \citenamefont
  {Smolkin}}]{Myers2013_MassiveEE}%
  \BibitemOpen
  \bibfield  {author} {\bibinfo {author} {\bibfnamefont {Aitor}\ \bibnamefont
  {Lewkowycz}}, \bibinfo {author} {\bibfnamefont {Robert~C.}\ \bibnamefont
  {Myers}}, \ and\ \bibinfo {author} {\bibfnamefont {Michael}\ \bibnamefont
  {Smolkin}},\ }\bibfield  {title} {\enquote {\bibinfo {title} {Observations on
  entanglement entropy in massive QFT's},}\ }\href {\doibase
  10.1007/JHEP04(2013)017} {\bibfield  {journal} {\bibinfo  {journal} {J. High
  Energ. Phys.}\ }\textbf {\bibinfo {volume} {2013}},\ \bibinfo {pages} {17}
  (\bibinfo {year} {2013})}\BibitemShut {NoStop}%
\bibitem [{\citenamefont {Cardy}\ \emph {et~al.}(2008)\citenamefont {Cardy},
  \citenamefont {Castro-Alvaredo},\ and\ \citenamefont
  {Doyon}}]{cardy2008_form_factor}%
  \BibitemOpen
  \bibfield  {author} {\bibinfo {author} {\bibfnamefont {J.~L.}\ \bibnamefont
  {Cardy}}, \bibinfo {author} {\bibfnamefont {O.~A.}\ \bibnamefont
  {Castro-Alvaredo}}, \ and\ \bibinfo {author} {\bibfnamefont {B.}~\bibnamefont
  {Doyon}},\ }\bibfield  {title} {\enquote {\bibinfo {title} {Form factors of
  branch-point twist fields in quantum integrable models and entanglement
  entropy},}\ }\href {\doibase 10.1007/s10955-007-9422-x} {\bibfield  {journal}
  {\bibinfo  {journal} {J. Stat. Phys.}\ }\textbf {\bibinfo {volume} {130}},\
  \bibinfo {pages} {129--168} (\bibinfo {year} {2008})}\BibitemShut {NoStop}%
\bibitem [{\citenamefont {Casini}\ \emph {et~al.}(2011)\citenamefont {Casini},
  \citenamefont {Huerta},\ and\ \citenamefont {Myers}}]{casini2011towards}%
  \BibitemOpen
  \bibfield  {author} {\bibinfo {author} {\bibfnamefont {Horacio}\ \bibnamefont
  {Casini}}, \bibinfo {author} {\bibfnamefont {Marina}\ \bibnamefont {Huerta}},
  \ and\ \bibinfo {author} {\bibfnamefont {Robert~C.}\ \bibnamefont {Myers}},\
  }\bibfield  {title} {\enquote {\bibinfo {title} {Towards a derivation of
  holographic entanglement entropy},}\ }\href {\doibase
  10.1007/JHEP05(2011)036} {\bibfield  {journal} {\bibinfo  {journal} {J. High
  Energ. Phys.}\ }\textbf {\bibinfo {volume} {2011}},\ \bibinfo {pages} {36}
  (\bibinfo {year} {2011})}\BibitemShut {NoStop}%
\bibitem [{\citenamefont {Ryu}\ and\ \citenamefont
  {Takayanagi}(2006{\natexlab{b}})}]{RT2006Aspects}%
  \BibitemOpen
  \bibfield  {author} {\bibinfo {author} {\bibfnamefont {Shinsei}\ \bibnamefont
  {Ryu}}\ and\ \bibinfo {author} {\bibfnamefont {Tadashi}\ \bibnamefont
  {Takayanagi}},\ }\bibfield  {title} {\enquote {\bibinfo {title} {Aspects of
  holographic entanglement entropy},}\ }\href {\doibase
  10.1088/1126-6708/2006/08/045} {\bibfield  {journal} {\bibinfo  {journal} {J.
  High Energ. Phys.}\ }\textbf {\bibinfo {volume} {2006}},\ \bibinfo {pages}
  {045--045} (\bibinfo {year} {2006}{\natexlab{b}})}\BibitemShut {NoStop}%
\bibitem [{\citenamefont {Casini}\ and\ \citenamefont
  {Huerta}(2009{\natexlab{c}})}]{Casini_2009_EMI}%
  \BibitemOpen
  \bibfield  {author} {\bibinfo {author} {\bibfnamefont {H}~\bibnamefont
  {Casini}}\ and\ \bibinfo {author} {\bibfnamefont {M}~\bibnamefont {Huerta}},\
  }\bibfield  {title} {\enquote {\bibinfo {title} {Remarks on the entanglement
  entropy for disconnected regions},}\ }\href {\doibase
  10.1088/1126-6708/2009/03/048} {\bibfield  {journal} {\bibinfo  {journal} {J.
  High Energ. Phys.}\ }\textbf {\bibinfo {volume} {2009}},\ \bibinfo {pages}
  {048--048} (\bibinfo {year} {2009}{\natexlab{c}})}\BibitemShut {NoStop}%
\bibitem [{\citenamefont {Ag{\'o}n}\ \emph {et~al.}(2021)\citenamefont
  {Ag{\'o}n}, \citenamefont {Bueno},\ and\ \citenamefont
  {Casini}}]{pablo2021_EMI}%
  \BibitemOpen
  \bibfield  {author} {\bibinfo {author} {\bibfnamefont {C{\'e}sar~A.}\
  \bibnamefont {Ag{\'o}n}}, \bibinfo {author} {\bibfnamefont {Pablo}\
  \bibnamefont {Bueno}}, \ and\ \bibinfo {author} {\bibfnamefont {Horacio}\
  \bibnamefont {Casini}},\ }\bibfield  {title} {\enquote {\bibinfo {title} {Is
  the EMI model a QFT? An inquiry on the space of allowed entropy functions},}\
  }\href {\doibase 10.1007/JHEP08(2021)084} {\bibfield  {journal} {\bibinfo
  {journal} {J. High Energ. Phys.}\ }\textbf {\bibinfo {volume} {2021}},\
  \bibinfo {pages} {84} (\bibinfo {year} {2021})}\BibitemShut {NoStop}%
\bibitem [{\citenamefont {Ryu}\ and\ \citenamefont
  {Hatsugai}(2006)}]{Ryu2006EEandBerryPhase}%
  \BibitemOpen
  \bibfield  {author} {\bibinfo {author} {\bibfnamefont {S.}~\bibnamefont
  {Ryu}}\ and\ \bibinfo {author} {\bibfnamefont {Y.}~\bibnamefont {Hatsugai}},\
  }\bibfield  {title} {\enquote {\bibinfo {title} {Entanglement entropy and the
  Berry phase in the solid state},}\ }\href {\doibase
  10.1103/PhysRevB.73.245115} {\bibfield  {journal} {\bibinfo  {journal} {Phys.
  Rev. B}\ }\textbf {\bibinfo {volume} {73}},\ \bibinfo {pages} {245115}
  (\bibinfo {year} {2006})}\BibitemShut {NoStop}%
\bibitem [{\citenamefont {Casini}\ and\ \citenamefont
  {Huerta}(2005)}]{CasiniHuerta2005MassiveScalar}%
  \BibitemOpen
  \bibfield  {author} {\bibinfo {author} {\bibfnamefont {H}~\bibnamefont
  {Casini}}\ and\ \bibinfo {author} {\bibfnamefont {M}~\bibnamefont {Huerta}},\
  }\bibfield  {title} {\enquote {\bibinfo {title} {Entanglement and alpha
  entropies for a massive scalar field in two dimensions},}\ }\href {\doibase
  10.1088/1742-5468/2005/12/p12012} {\bibfield  {journal} {\bibinfo  {journal}
  {J. Stat. Mech.}\ }\textbf {\bibinfo {volume} {2005}},\ \bibinfo {pages}
  {P12012--P12012} (\bibinfo {year} {2005})}\BibitemShut {NoStop}%
\bibitem [{\citenamefont {Casini}\ and\ \citenamefont
  {Huerta}(2004)}]{CASINI2004_1p1_EE_RG}%
  \BibitemOpen
  \bibfield  {author} {\bibinfo {author} {\bibfnamefont {H.}~\bibnamefont
  {Casini}}\ and\ \bibinfo {author} {\bibfnamefont {M.}~\bibnamefont
  {Huerta}},\ }\bibfield  {title} {\enquote {\bibinfo {title} {A finite
  entanglement entropy and the c-theorem},}\ }\href {\doibase
  https://doi.org/10.1016/j.physletb.2004.08.072} {\bibfield  {journal}
  {\bibinfo  {journal} {Physics Letters B}\ }\textbf {\bibinfo {volume}
  {600}},\ \bibinfo {pages} {142--150} (\bibinfo {year} {2004})}\BibitemShut
  {NoStop}%
\bibitem [{\citenamefont {Myers}\ and\ \citenamefont
  {Sinha}(2010)}]{Myers2010_holographic_c_theorem}%
  \BibitemOpen
  \bibfield  {author} {\bibinfo {author} {\bibfnamefont {Robert~C.}\
  \bibnamefont {Myers}}\ and\ \bibinfo {author} {\bibfnamefont {Aninda}\
  \bibnamefont {Sinha}},\ }\bibfield  {title} {\enquote {\bibinfo {title}
  {Seeing a c-theorem with holography},}\ }\href {\doibase
  10.1103/PhysRevD.82.046006} {\bibfield  {journal} {\bibinfo  {journal} {Phys.
  Rev. D}\ }\textbf {\bibinfo {volume} {82}},\ \bibinfo {pages} {046006}
  (\bibinfo {year} {2010})}\BibitemShut {NoStop}%
\bibitem [{\citenamefont {Myers}\ and\ \citenamefont
  {Sinha}(2011)}]{Myers2011_holographic_c_theorem}%
  \BibitemOpen
  \bibfield  {author} {\bibinfo {author} {\bibfnamefont {Robert~C.}\
  \bibnamefont {Myers}}\ and\ \bibinfo {author} {\bibfnamefont {Aninda}\
  \bibnamefont {Sinha}},\ }\bibfield  {title} {\enquote {\bibinfo {title}
  {Holographic c-theorems in arbitrary dimensions},}\ }\href {\doibase
  10.1007/JHEP01(2011)125} {\bibfield  {journal} {\bibinfo  {journal} {J. High
  Energ. Phys.}\ }\textbf {\bibinfo {volume} {2011}},\ \bibinfo {pages} {125}
  (\bibinfo {year} {2011})}\BibitemShut {NoStop}%
\bibitem [{\citenamefont {Klebanov}\ \emph
  {et~al.}(2012{\natexlab{a}})\citenamefont {Klebanov}, \citenamefont {Pufu},
  \citenamefont {Sachdev},\ and\ \citenamefont {Safdi}}]{Klebanov2012_3dEE}%
  \BibitemOpen
  \bibfield  {author} {\bibinfo {author} {\bibfnamefont {Igor~R.}\ \bibnamefont
  {Klebanov}}, \bibinfo {author} {\bibfnamefont {Silviu~S.}\ \bibnamefont
  {Pufu}}, \bibinfo {author} {\bibfnamefont {Subir}\ \bibnamefont {Sachdev}}, \
  and\ \bibinfo {author} {\bibfnamefont {Benjamin~R.}\ \bibnamefont {Safdi}},\
  }\bibfield  {title} {\enquote {\bibinfo {title} {Entanglement entropy of 3-d
  conformal gauge theories with many flavors},}\ }\href {\doibase
  10.1007/JHEP05(2012)036} {\bibfield  {journal} {\bibinfo  {journal} {J. High
  Energ. Phys.}\ }\textbf {\bibinfo {volume} {2012}},\ \bibinfo {pages} {36}
  (\bibinfo {year} {2012}{\natexlab{a}})}\BibitemShut {NoStop}%
\bibitem [{\citenamefont {Casini}\ and\ \citenamefont
  {Huerta}(2012)}]{Casini2012_RG_EE_2p1}%
  \BibitemOpen
  \bibfield  {author} {\bibinfo {author} {\bibfnamefont {H.}~\bibnamefont
  {Casini}}\ and\ \bibinfo {author} {\bibfnamefont {M.}~\bibnamefont
  {Huerta}},\ }\bibfield  {title} {\enquote {\bibinfo {title} {Renormalization
  group running of the entanglement entropy of a circle},}\ }\href {\doibase
  10.1103/PhysRevD.85.125016} {\bibfield  {journal} {\bibinfo  {journal} {Phys.
  Rev. D}\ }\textbf {\bibinfo {volume} {85}},\ \bibinfo {pages} {125016}
  (\bibinfo {year} {2012})}\BibitemShut {NoStop}%
\bibitem [{\citenamefont {Liu}\ and\ \citenamefont
  {Mezei}(2013)}]{liu2013RefinementEE}%
  \BibitemOpen
  \bibfield  {author} {\bibinfo {author} {\bibfnamefont {Hong}\ \bibnamefont
  {Liu}}\ and\ \bibinfo {author} {\bibfnamefont {M{\'a}rk}\ \bibnamefont
  {Mezei}},\ }\bibfield  {title} {\enquote {\bibinfo {title} {A refinement of
  entanglement entropy and the number of degrees of freedom},}\ }\href
  {\doibase 10.1007/JHEP04(2013)162} {\bibfield  {journal} {\bibinfo  {journal}
  {J. High Energ. Phys.}\ }\textbf {\bibinfo {volume} {2013}},\ \bibinfo
  {pages} {162} (\bibinfo {year} {2013})}\BibitemShut {NoStop}%
\bibitem [{\citenamefont {Fei}\ \emph {et~al.}(2015)\citenamefont {Fei},
  \citenamefont {Giombi}, \citenamefont {Klebanov},\ and\ \citenamefont
  {Tarnopolsky}}]{Fei2015_generalized_F_theorem}%
  \BibitemOpen
  \bibfield  {author} {\bibinfo {author} {\bibfnamefont {Lin}\ \bibnamefont
  {Fei}}, \bibinfo {author} {\bibfnamefont {Simone}\ \bibnamefont {Giombi}},
  \bibinfo {author} {\bibfnamefont {Igor~R.}\ \bibnamefont {Klebanov}}, \ and\
  \bibinfo {author} {\bibfnamefont {Grigory}\ \bibnamefont {Tarnopolsky}},\
  }\bibfield  {title} {\enquote {\bibinfo {title} {Generalized F-theorem and
  the $\varepsilon$ expansion},}\ }\href {\doibase 10.1007/JHEP12(2015)155}
  {\bibfield  {journal} {\bibinfo  {journal} {J. High Energ. Phys.}\ }\textbf
  {\bibinfo {volume} {2015}},\ \bibinfo {pages} {1--37} (\bibinfo {year}
  {2015})}\BibitemShut {NoStop}%
\bibitem [{\citenamefont {Jafferis}\ \emph {et~al.}(2011)\citenamefont
  {Jafferis}, \citenamefont {Klebanov}, \citenamefont {Pufu},\ and\
  \citenamefont {Safdi}}]{Jafferis2011_F_theorem}%
  \BibitemOpen
  \bibfield  {author} {\bibinfo {author} {\bibfnamefont {Daniel~L.}\
  \bibnamefont {Jafferis}}, \bibinfo {author} {\bibfnamefont {Igor~R.}\
  \bibnamefont {Klebanov}}, \bibinfo {author} {\bibfnamefont {Silviu~S.}\
  \bibnamefont {Pufu}}, \ and\ \bibinfo {author} {\bibfnamefont {Benjamin~R.}\
  \bibnamefont {Safdi}},\ }\bibfield  {title} {\enquote {\bibinfo {title}
  {Towards the F-theorem: $\mathcal{N} = 2$ field theories on the
  three-sphere},}\ }\href {\doibase 10.1007/JHEP06(2011)102} {\bibfield
  {journal} {\bibinfo  {journal} {J. High Energ. Phys.}\ }\textbf {\bibinfo
  {volume} {2011}},\ \bibinfo {pages} {102} (\bibinfo {year}
  {2011})}\BibitemShut {NoStop}%
\bibitem [{\citenamefont {Metlitski}\ \emph {et~al.}(2009)\citenamefont
  {Metlitski}, \citenamefont {Fuertes},\ and\ \citenamefont
  {Sachdev}}]{Metlitski2009_ON}%
  \BibitemOpen
  \bibfield  {author} {\bibinfo {author} {\bibfnamefont {Max~A.}\ \bibnamefont
  {Metlitski}}, \bibinfo {author} {\bibfnamefont {Carlos~A.}\ \bibnamefont
  {Fuertes}}, \ and\ \bibinfo {author} {\bibfnamefont {Subir}\ \bibnamefont
  {Sachdev}},\ }\bibfield  {title} {\enquote {\bibinfo {title} {Entanglement
  entropy in the $O(N)$ model},}\ }\href {\doibase 10.1103/PhysRevB.80.115122}
  {\bibfield  {journal} {\bibinfo  {journal} {Phys. Rev. B}\ }\textbf {\bibinfo
  {volume} {80}},\ \bibinfo {pages} {115122} (\bibinfo {year}
  {2009})}\BibitemShut {NoStop}%
\bibitem [{\citenamefont {Hertzberg}\ and\ \citenamefont
  {Wilczek}(2011)}]{Hertzberg2011_FreeScalar}%
  \BibitemOpen
  \bibfield  {author} {\bibinfo {author} {\bibfnamefont {Mark~P.}\ \bibnamefont
  {Hertzberg}}\ and\ \bibinfo {author} {\bibfnamefont {Frank}\ \bibnamefont
  {Wilczek}},\ }\bibfield  {title} {\enquote {\bibinfo {title} {Some calculable
  contributions to entanglement entropy},}\ }\href {\doibase
  10.1103/PhysRevLett.106.050404} {\bibfield  {journal} {\bibinfo  {journal}
  {Phys. Rev. Lett.}\ }\textbf {\bibinfo {volume} {106}},\ \bibinfo {pages}
  {050404} (\bibinfo {year} {2011})}\BibitemShut {NoStop}%
\bibitem [{\citenamefont {Klebanov}\ \emph
  {et~al.}(2012{\natexlab{b}})\citenamefont {Klebanov}, \citenamefont {Pufu},
  \citenamefont {Sachdev},\ and\ \citenamefont {Safdi}}]{klebanov2012}%
  \BibitemOpen
  \bibfield  {author} {\bibinfo {author} {\bibfnamefont {Igor~R.}\ \bibnamefont
  {Klebanov}}, \bibinfo {author} {\bibfnamefont {Silviu~S.}\ \bibnamefont
  {Pufu}}, \bibinfo {author} {\bibfnamefont {Subir}\ \bibnamefont {Sachdev}}, \
  and\ \bibinfo {author} {\bibfnamefont {Benjamin~R.}\ \bibnamefont {Safdi}},\
  }\bibfield  {title} {\enquote {\bibinfo {title} {R{\'e}nyi entropies for free
  field theories},}\ }\href {\doibase 10.1007/JHEP04(2012)074} {\bibfield
  {journal} {\bibinfo  {journal} {J. High Energ. Phys.}\ }\textbf {\bibinfo
  {volume} {2012}},\ \bibinfo {pages} {74} (\bibinfo {year}
  {2012}{\natexlab{b}})}\BibitemShut {NoStop}%
\bibitem [{\citenamefont {Hertzberg}(2012)}]{Hertzberg2012_InteractingScalar}%
  \BibitemOpen
  \bibfield  {author} {\bibinfo {author} {\bibfnamefont {Mark~P}\ \bibnamefont
  {Hertzberg}},\ }\bibfield  {title} {\enquote {\bibinfo {title} {Entanglement
  entropy in scalar field theory},}\ }\href {\doibase
  10.1088/1751-8113/46/1/015402} {\bibfield  {journal} {\bibinfo  {journal} {J.
  Phys. A: Math. Theor.}\ }\textbf {\bibinfo {volume} {46}},\ \bibinfo {pages}
  {015402} (\bibinfo {year} {2012})}\BibitemShut {NoStop}%
\bibitem [{\citenamefont {Whitsitt}\ \emph {et~al.}(2017)\citenamefont
  {Whitsitt}, \citenamefont {Witczak-Krempa},\ and\ \citenamefont
  {Sachdev}}]{Whitsitt2017_LargeN_WF}%
  \BibitemOpen
  \bibfield  {author} {\bibinfo {author} {\bibfnamefont {Seth}\ \bibnamefont
  {Whitsitt}}, \bibinfo {author} {\bibfnamefont {William}\ \bibnamefont
  {Witczak-Krempa}}, \ and\ \bibinfo {author} {\bibfnamefont {Subir}\
  \bibnamefont {Sachdev}},\ }\bibfield  {title} {\enquote {\bibinfo {title}
  {Entanglement entropy of large-N Wilson-Fisher conformal field theory},}\
  }\href {\doibase 10.1103/PhysRevB.95.045148} {\bibfield  {journal} {\bibinfo
  {journal} {Phys. Rev. B}\ }\textbf {\bibinfo {volume} {95}},\ \bibinfo
  {pages} {045148} (\bibinfo {year} {2017})}\BibitemShut {NoStop}%
\bibitem [{\citenamefont {Hung}\ \emph {et~al.}(2017)\citenamefont {Hung},
  \citenamefont {Jiang},\ and\ \citenamefont {Wang}}]{Hung2017_ON}%
  \BibitemOpen
  \bibfield  {author} {\bibinfo {author} {\bibfnamefont {Ling-Yan}\
  \bibnamefont {Hung}}, \bibinfo {author} {\bibfnamefont {Yikun}\ \bibnamefont
  {Jiang}}, \ and\ \bibinfo {author} {\bibfnamefont {Yixu}\ \bibnamefont
  {Wang}},\ }\bibfield  {title} {\enquote {\bibinfo {title} {Area term of the
  entanglement entropy of a supersymmetric $O(N)$ vector model in three
  dimensions},}\ }\href {\doibase 10.1103/PhysRevD.95.085004} {\bibfield
  {journal} {\bibinfo  {journal} {Phys. Rev. D}\ }\textbf {\bibinfo {volume}
  {95}},\ \bibinfo {pages} {085004} (\bibinfo {year} {2017})}\BibitemShut
  {NoStop}%
\bibitem [{\citenamefont {Bhattacharyya}\ \emph {et~al.}(2017)\citenamefont
  {Bhattacharyya}, \citenamefont {Hung},\ and\ \citenamefont
  {Melby-Thompson}}]{Hung2017_Instanton}%
  \BibitemOpen
  \bibfield  {author} {\bibinfo {author} {\bibfnamefont {Arpan}\ \bibnamefont
  {Bhattacharyya}}, \bibinfo {author} {\bibfnamefont {Ling-Yan}\ \bibnamefont
  {Hung}}, \ and\ \bibinfo {author} {\bibfnamefont {Charles~M.}\ \bibnamefont
  {Melby-Thompson}},\ }\bibfield  {title} {\enquote {\bibinfo {title}
  {Instantons and entanglement entropy},}\ }\href {\doibase
  10.1007/JHEP10(2017)081} {\bibfield  {journal} {\bibinfo  {journal} {J. High
  Energ. Phys.}\ }\textbf {\bibinfo {volume} {2017}},\ \bibinfo {pages} {81}
  (\bibinfo {year} {2017})}\BibitemShut {NoStop}%
\bibitem [{\citenamefont {Chen}\ \emph {et~al.}(2020)\citenamefont {Chen},
  \citenamefont {Hackl}, \citenamefont {Kunjwal}, \citenamefont {Moradi},
  \citenamefont {Yazdi},\ and\ \citenamefont
  {Zilh{\~a}o}}]{chen2020EE_interacting_harmonic}%
  \BibitemOpen
  \bibfield  {author} {\bibinfo {author} {\bibfnamefont {Yangang}\ \bibnamefont
  {Chen}}, \bibinfo {author} {\bibfnamefont {Lucas}\ \bibnamefont {Hackl}},
  \bibinfo {author} {\bibfnamefont {Ravi}\ \bibnamefont {Kunjwal}}, \bibinfo
  {author} {\bibfnamefont {Heidar}\ \bibnamefont {Moradi}}, \bibinfo {author}
  {\bibfnamefont {Yasaman~K.}\ \bibnamefont {Yazdi}}, \ and\ \bibinfo {author}
  {\bibfnamefont {Miguel}\ \bibnamefont {Zilh{\~a}o}},\ }\bibfield  {title}
  {\enquote {\bibinfo {title} {Towards spacetime entanglement entropy for
  interacting theories},}\ }\href {\doibase 10.1007/JHEP11(2020)114} {\bibfield
   {journal} {\bibinfo  {journal} {J. High Energ. Phys.}\ }\textbf {\bibinfo
  {volume} {2020}},\ \bibinfo {pages} {114} (\bibinfo {year}
  {2020})}\BibitemShut {NoStop}%
\bibitem [{\citenamefont {Iso}\ \emph {et~al.}(2021{\natexlab{a}})\citenamefont
  {Iso}, \citenamefont {Mori},\ and\ \citenamefont {Sakai}}]{iso2021_EE_ZM}%
  \BibitemOpen
  \bibfield  {author} {\bibinfo {author} {\bibfnamefont {Satoshi}\ \bibnamefont
  {Iso}}, \bibinfo {author} {\bibfnamefont {Takato}\ \bibnamefont {Mori}}, \
  and\ \bibinfo {author} {\bibfnamefont {Katsuta}\ \bibnamefont {Sakai}},\
  }\bibfield  {title} {\enquote {\bibinfo {title} {Entanglement entropy in
  scalar field theory and ${\mathbb{Z}}_{M}$ gauge theory on Feynman
  diagrams},}\ }\href {\doibase 10.1103/PhysRevD.103.105010} {\bibfield
  {journal} {\bibinfo  {journal} {Phys. Rev. D}\ }\textbf {\bibinfo {volume}
  {103}},\ \bibinfo {pages} {105010} (\bibinfo {year}
  {2021}{\natexlab{a}})}\BibitemShut {NoStop}%
\bibitem [{\citenamefont {Iso}\ \emph {et~al.}(2021{\natexlab{b}})\citenamefont
  {Iso}, \citenamefont {Mori},\ and\ \citenamefont
  {Sakai}}]{iso2021_EE_composite}%
  \BibitemOpen
  \bibfield  {author} {\bibinfo {author} {\bibfnamefont {Satoshi}\ \bibnamefont
  {Iso}}, \bibinfo {author} {\bibfnamefont {Takato}\ \bibnamefont {Mori}}, \
  and\ \bibinfo {author} {\bibfnamefont {Katsuta}\ \bibnamefont {Sakai}},\
  }\bibfield  {title} {\enquote {\bibinfo {title} {Non-Gaussianity of
  entanglement entropy and correlations of composite operators},}\ }\href
  {\doibase 10.1103/PhysRevD.103.125019} {\bibfield  {journal} {\bibinfo
  {journal} {Phys. Rev. D}\ }\textbf {\bibinfo {volume} {103}},\ \bibinfo
  {pages} {125019} (\bibinfo {year} {2021}{\natexlab{b}})}\BibitemShut
  {NoStop}%
\bibitem [{\citenamefont {Refael}\ and\ \citenamefont
  {Moore}(2004)}]{Rafael2004}%
  \BibitemOpen
  \bibfield  {author} {\bibinfo {author} {\bibfnamefont {G.}~\bibnamefont
  {Refael}}\ and\ \bibinfo {author} {\bibfnamefont {J.~E.}\ \bibnamefont
  {Moore}},\ }\bibfield  {title} {\enquote {\bibinfo {title} {Entanglement
  entropy of random quantum critical points in one dimension},}\ }\href
  {\doibase 10.1103/PhysRevLett.93.260602} {\bibfield  {journal} {\bibinfo
  {journal} {Phys. Rev. Lett.}\ }\textbf {\bibinfo {volume} {93}},\ \bibinfo
  {pages} {260602} (\bibinfo {year} {2004})}\BibitemShut {NoStop}%
\bibitem [{\citenamefont {Lin}\ \emph {et~al.}(2007)\citenamefont {Lin},
  \citenamefont {Igl\'oi},\ and\ \citenamefont {Rieger}}]{lin2007EE_InfRand}%
  \BibitemOpen
  \bibfield  {author} {\bibinfo {author} {\bibfnamefont {Yu-Cheng}\
  \bibnamefont {Lin}}, \bibinfo {author} {\bibfnamefont {Ferenc}\ \bibnamefont
  {Igl\'oi}}, \ and\ \bibinfo {author} {\bibfnamefont {Heiko}\ \bibnamefont
  {Rieger}},\ }\bibfield  {title} {\enquote {\bibinfo {title} {Entanglement
  entropy at infinite-randomness fixed points in higher dimensions},}\ }\href
  {\doibase 10.1103/PhysRevLett.99.147202} {\bibfield  {journal} {\bibinfo
  {journal} {Phys. Rev. Lett.}\ }\textbf {\bibinfo {volume} {99}},\ \bibinfo
  {pages} {147202} (\bibinfo {year} {2007})}\BibitemShut {NoStop}%
\bibitem [{\citenamefont {Yu}\ \emph {et~al.}(2008)\citenamefont {Yu},
  \citenamefont {Saleur},\ and\ \citenamefont {Haas}}]{rong2008_rand_ising}%
  \BibitemOpen
  \bibfield  {author} {\bibinfo {author} {\bibfnamefont {Rong}\ \bibnamefont
  {Yu}}, \bibinfo {author} {\bibfnamefont {Hubert}\ \bibnamefont {Saleur}}, \
  and\ \bibinfo {author} {\bibfnamefont {Stephan}\ \bibnamefont {Haas}},\
  }\bibfield  {title} {\enquote {\bibinfo {title} {Entanglement entropy in the
  two-dimensional random transverse field Ising model},}\ }\href {\doibase
  10.1103/PhysRevB.77.140402} {\bibfield  {journal} {\bibinfo  {journal} {Phys.
  Rev. B}\ }\textbf {\bibinfo {volume} {77}},\ \bibinfo {pages} {140402}
  (\bibinfo {year} {2008})}\BibitemShut {NoStop}%
\bibitem [{\citenamefont {Harris}(1974)}]{Harris_1974}%
  \BibitemOpen
  \bibfield  {author} {\bibinfo {author} {\bibfnamefont {A~B}\ \bibnamefont
  {Harris}},\ }\bibfield  {title} {\enquote {\bibinfo {title} {Effect of random
  defects on the critical behaviour of Ising models},}\ }\href {\doibase
  10.1088/0022-3719/7/9/009} {\bibfield  {journal} {\bibinfo  {journal} {J.
  Phys. C: Solid State Phys.}\ }\textbf {\bibinfo {volume} {7}},\ \bibinfo
  {pages} {1671--1692} (\bibinfo {year} {1974})}\BibitemShut {NoStop}%
\bibitem [{\citenamefont {Levine}\ \emph {et~al.}(1983)\citenamefont {Levine},
  \citenamefont {Libby},\ and\ \citenamefont
  {Pruisken}}]{Pruisken1983Delocalization}%
  \BibitemOpen
  \bibfield  {author} {\bibinfo {author} {\bibfnamefont {Herbert}\ \bibnamefont
  {Levine}}, \bibinfo {author} {\bibfnamefont {Stephen~B.}\ \bibnamefont
  {Libby}}, \ and\ \bibinfo {author} {\bibfnamefont {Adrianus M.~M.}\
  \bibnamefont {Pruisken}},\ }\bibfield  {title} {\enquote {\bibinfo {title}
  {Electron delocalization by a magnetic field in two dimensions},}\ }\href
  {\doibase 10.1103/PhysRevLett.51.1915} {\bibfield  {journal} {\bibinfo
  {journal} {Phys. Rev. Lett.}\ }\textbf {\bibinfo {volume} {51}},\ \bibinfo
  {pages} {1915--1918} (\bibinfo {year} {1983})}\BibitemShut {NoStop}%
\bibitem [{\citenamefont {Fisher}(1992)}]{Daniel1992randomTFI}%
  \BibitemOpen
  \bibfield  {author} {\bibinfo {author} {\bibfnamefont {Daniel~S.}\
  \bibnamefont {Fisher}},\ }\bibfield  {title} {\enquote {\bibinfo {title}
  {Random transverse field Ising spin chains},}\ }\href {\doibase
  10.1103/PhysRevLett.69.534} {\bibfield  {journal} {\bibinfo  {journal} {Phys.
  Rev. Lett.}\ }\textbf {\bibinfo {volume} {69}},\ \bibinfo {pages} {534--537}
  (\bibinfo {year} {1992})}\BibitemShut {NoStop}%
\bibitem [{\citenamefont {Nersesyan}\ \emph {et~al.}(1994)\citenamefont
  {Nersesyan}, \citenamefont {Tsvelik},\ and\ \citenamefont
  {Wenger}}]{wenger1994_disorder_Dwave}%
  \BibitemOpen
  \bibfield  {author} {\bibinfo {author} {\bibfnamefont {A.~A.}\ \bibnamefont
  {Nersesyan}}, \bibinfo {author} {\bibfnamefont {A.~M.}\ \bibnamefont
  {Tsvelik}}, \ and\ \bibinfo {author} {\bibfnamefont {F.}~\bibnamefont
  {Wenger}},\ }\bibfield  {title} {\enquote {\bibinfo {title} {Disorder effects
  in two-dimensional d-wave superconductors},}\ }\href {\doibase
  10.1103/PhysRevLett.72.2628} {\bibfield  {journal} {\bibinfo  {journal}
  {Phys. Rev. Lett.}\ }\textbf {\bibinfo {volume} {72}},\ \bibinfo {pages}
  {2628--2631} (\bibinfo {year} {1994})}\BibitemShut {NoStop}%
\bibitem [{\citenamefont {Ludwig}\ \emph {et~al.}(1994)\citenamefont {Ludwig},
  \citenamefont {Fisher}, \citenamefont {Shankar},\ and\ \citenamefont
  {Grinstein}}]{Ludwig1994_IQH_transition}%
  \BibitemOpen
  \bibfield  {author} {\bibinfo {author} {\bibfnamefont {Andreas W.~W.}\
  \bibnamefont {Ludwig}}, \bibinfo {author} {\bibfnamefont {Matthew P.~A.}\
  \bibnamefont {Fisher}}, \bibinfo {author} {\bibfnamefont {R.}~\bibnamefont
  {Shankar}}, \ and\ \bibinfo {author} {\bibfnamefont {G.}~\bibnamefont
  {Grinstein}},\ }\bibfield  {title} {\enquote {\bibinfo {title} {Integer
  quantum Hall transition: An alternative approach and exact results},}\ }\href
  {\doibase 10.1103/PhysRevB.50.7526} {\bibfield  {journal} {\bibinfo
  {journal} {Phys. Rev. B}\ }\textbf {\bibinfo {volume} {50}},\ \bibinfo
  {pages} {7526--7552} (\bibinfo {year} {1994})}\BibitemShut {NoStop}%
\bibitem [{\citenamefont {Goswami}\ \emph {et~al.}(2017)\citenamefont
  {Goswami}, \citenamefont {Goldman},\ and\ \citenamefont
  {Raghu}}]{goldman2017QED3}%
  \BibitemOpen
  \bibfield  {author} {\bibinfo {author} {\bibfnamefont {Pallab}\ \bibnamefont
  {Goswami}}, \bibinfo {author} {\bibfnamefont {Hart}\ \bibnamefont {Goldman}},
  \ and\ \bibinfo {author} {\bibfnamefont {S.}~\bibnamefont {Raghu}},\
  }\bibfield  {title} {\enquote {\bibinfo {title} {Metallic phases from
  disordered (2+1)-dimensional quantum electrodynamics},}\ }\href {\doibase
  10.1103/PhysRevB.95.235145} {\bibfield  {journal} {\bibinfo  {journal} {Phys.
  Rev. B}\ }\textbf {\bibinfo {volume} {95}},\ \bibinfo {pages} {235145}
  (\bibinfo {year} {2017})}\BibitemShut {NoStop}%
\bibitem [{\citenamefont {Thomson}\ and\ \citenamefont
  {Sachdev}(2017)}]{sachdev2017QED3}%
  \BibitemOpen
  \bibfield  {author} {\bibinfo {author} {\bibfnamefont {Alex}\ \bibnamefont
  {Thomson}}\ and\ \bibinfo {author} {\bibfnamefont {Subir}\ \bibnamefont
  {Sachdev}},\ }\bibfield  {title} {\enquote {\bibinfo {title} {Quantum
  electrodynamics in 2+1 dimensions with quenched disorder: Quantum critical
  states with interactions and disorder},}\ }\href {\doibase
  10.1103/PhysRevB.95.235146} {\bibfield  {journal} {\bibinfo  {journal} {Phys.
  Rev. B}\ }\textbf {\bibinfo {volume} {95}},\ \bibinfo {pages} {235146}
  (\bibinfo {year} {2017})}\BibitemShut {NoStop}%
\bibitem [{\citenamefont {Yerzhakov}\ and\ \citenamefont
  {Maciejko}(2018)}]{yerzhakov2018disordered}%
  \BibitemOpen
  \bibfield  {author} {\bibinfo {author} {\bibfnamefont {Hennadii}\
  \bibnamefont {Yerzhakov}}\ and\ \bibinfo {author} {\bibfnamefont {Joseph}\
  \bibnamefont {Maciejko}},\ }\bibfield  {title} {\enquote {\bibinfo {title}
  {Disordered fermionic quantum critical points},}\ }\href {\doibase
  10.1103/PhysRevB.98.195142} {\bibfield  {journal} {\bibinfo  {journal} {Phys.
  Rev. B}\ }\textbf {\bibinfo {volume} {98}},\ \bibinfo {pages} {195142}
  (\bibinfo {year} {2018})}\BibitemShut {NoStop}%
\bibitem [{\citenamefont {Goldman}\ \emph {et~al.}(2020)\citenamefont
  {Goldman}, \citenamefont {Thomson}, \citenamefont {Nie},\ and\ \citenamefont
  {Bi}}]{goldman2020interplay}%
  \BibitemOpen
  \bibfield  {author} {\bibinfo {author} {\bibfnamefont {Hart}\ \bibnamefont
  {Goldman}}, \bibinfo {author} {\bibfnamefont {Alex}\ \bibnamefont {Thomson}},
  \bibinfo {author} {\bibfnamefont {Laimei}\ \bibnamefont {Nie}}, \ and\
  \bibinfo {author} {\bibfnamefont {Zhen}\ \bibnamefont {Bi}},\ }\bibfield
  {title} {\enquote {\bibinfo {title} {Interplay of interactions and disorder
  at the superfluid-insulator transition: A dirty two-dimensional quantum
  critical point},}\ }\href {\doibase 10.1103/PhysRevB.101.144506} {\bibfield
  {journal} {\bibinfo  {journal} {Phys. Rev. B}\ }\textbf {\bibinfo {volume}
  {101}},\ \bibinfo {pages} {144506} (\bibinfo {year} {2020})}\BibitemShut
  {NoStop}%
\bibitem [{\citenamefont {Narovlansky}\ and\ \citenamefont
  {Aharony}(2018)}]{narovlansky2018_disorder_RG}%
  \BibitemOpen
  \bibfield  {author} {\bibinfo {author} {\bibfnamefont {Vladimir}\
  \bibnamefont {Narovlansky}}\ and\ \bibinfo {author} {\bibfnamefont {Ofer}\
  \bibnamefont {Aharony}},\ }\bibfield  {title} {\enquote {\bibinfo {title}
  {Renormalization group in field theories with quantum quenched disorder},}\
  }\href {\doibase 10.1103/PhysRevLett.121.071601} {\bibfield  {journal}
  {\bibinfo  {journal} {Phys. Rev. Lett.}\ }\textbf {\bibinfo {volume} {121}},\
  \bibinfo {pages} {071601} (\bibinfo {year} {2018})}\BibitemShut {NoStop}%
\bibitem [{\citenamefont {Mudry}\ \emph {et~al.}(1996)\citenamefont {Mudry},
  \citenamefont {Chamon},\ and\ \citenamefont {Wen}}]{Wen1996conformal}%
  \BibitemOpen
  \bibfield  {author} {\bibinfo {author} {\bibfnamefont {Christopher}\
  \bibnamefont {Mudry}}, \bibinfo {author} {\bibfnamefont {Claudio}\
  \bibnamefont {Chamon}}, \ and\ \bibinfo {author} {\bibfnamefont {Xiao-Gang}\
  \bibnamefont {Wen}},\ }\bibfield  {title} {\enquote {\bibinfo {title}
  {Two-dimensional conformal field theory for disordered systems at
  criticality},}\ }\href {\doibase
  https://doi.org/10.1016/0550-3213(96)00128-9} {\bibfield  {journal} {\bibinfo
   {journal} {Nuclear Physics B}\ }\textbf {\bibinfo {volume} {466}},\ \bibinfo
  {pages} {383--443} (\bibinfo {year} {1996})}\BibitemShut {NoStop}%
\bibitem [{\citenamefont {Chamon}\ \emph {et~al.}(1996)\citenamefont {Chamon},
  \citenamefont {Mudry},\ and\ \citenamefont {Wen}}]{Wen1996Multifractality}%
  \BibitemOpen
  \bibfield  {author} {\bibinfo {author} {\bibfnamefont {Claudio de~C.}\
  \bibnamefont {Chamon}}, \bibinfo {author} {\bibfnamefont {Christopher}\
  \bibnamefont {Mudry}}, \ and\ \bibinfo {author} {\bibfnamefont {Xiao-Gang}\
  \bibnamefont {Wen}},\ }\bibfield  {title} {\enquote {\bibinfo {title}
  {Localization in two dimensions, Gaussian field theories, and
  multifractality},}\ }\href {\doibase 10.1103/PhysRevLett.77.4194} {\bibfield
  {journal} {\bibinfo  {journal} {Phys. Rev. Lett.}\ }\textbf {\bibinfo
  {volume} {77}},\ \bibinfo {pages} {4194--4197} (\bibinfo {year}
  {1996})}\BibitemShut {NoStop}%
\bibitem [{\citenamefont {Evers}\ \emph {et~al.}(2001)\citenamefont {Evers},
  \citenamefont {Mildenberger},\ and\ \citenamefont
  {Mirlin}}]{Mirlin2001Multifractality}%
  \BibitemOpen
  \bibfield  {author} {\bibinfo {author} {\bibfnamefont {F.}~\bibnamefont
  {Evers}}, \bibinfo {author} {\bibfnamefont {A.}~\bibnamefont {Mildenberger}},
  \ and\ \bibinfo {author} {\bibfnamefont {A.~D.}\ \bibnamefont {Mirlin}},\
  }\bibfield  {title} {\enquote {\bibinfo {title} {Multifractality of wave
  functions at the quantum Hall transition revisited},}\ }\href {\doibase
  10.1103/PhysRevB.64.241303} {\bibfield  {journal} {\bibinfo  {journal} {Phys.
  Rev. B}\ }\textbf {\bibinfo {volume} {64}},\ \bibinfo {pages} {241303}
  (\bibinfo {year} {2001})}\BibitemShut {NoStop}%
\bibitem [{\citenamefont {Foster}\ \emph {et~al.}(2009)\citenamefont {Foster},
  \citenamefont {Ryu},\ and\ \citenamefont {Ludwig}}]{Ryu2009Multifractal}%
  \BibitemOpen
  \bibfield  {author} {\bibinfo {author} {\bibfnamefont {Matthew~S.}\
  \bibnamefont {Foster}}, \bibinfo {author} {\bibfnamefont {Shinsei}\
  \bibnamefont {Ryu}}, \ and\ \bibinfo {author} {\bibfnamefont {Andreas W.~W.}\
  \bibnamefont {Ludwig}},\ }\bibfield  {title} {\enquote {\bibinfo {title}
  {Termination of typical wave-function multifractal spectra at the Anderson
  metal-insulator transition: Field theory description using the functional
  renormalization group},}\ }\href {\doibase 10.1103/PhysRevB.80.075101}
  {\bibfield  {journal} {\bibinfo  {journal} {Phys. Rev. B}\ }\textbf {\bibinfo
  {volume} {80}},\ \bibinfo {pages} {075101} (\bibinfo {year}
  {2009})}\BibitemShut {NoStop}%
\bibitem [{\citenamefont {Vassilevich}(2003)}]{VASSILEVICH2003HeatKernel}%
  \BibitemOpen
  \bibfield  {author} {\bibinfo {author} {\bibfnamefont {D.V.}\ \bibnamefont
  {Vassilevich}},\ }\bibfield  {title} {\enquote {\bibinfo {title} {Heat kernel
  expansion: user's manual},}\ }\href {\doibase
  https://doi.org/10.1016/j.physrep.2003.09.002} {\bibfield  {journal}
  {\bibinfo  {journal} {Physics Reports}\ }\textbf {\bibinfo {volume} {388}},\
  \bibinfo {pages} {279--360} (\bibinfo {year} {2003})}\BibitemShut {NoStop}%
\bibitem [{\citenamefont {Nesterov}\ and\ \citenamefont
  {Solodukhin}(2010)}]{Nesterov2010_modified_HeatKernel}%
  \BibitemOpen
  \bibfield  {author} {\bibinfo {author} {\bibfnamefont {Dmitry}\ \bibnamefont
  {Nesterov}}\ and\ \bibinfo {author} {\bibfnamefont {Sergey~N.}\ \bibnamefont
  {Solodukhin}},\ }\bibfield  {title} {\enquote {\bibinfo {title}
  {Short-distance regularity of Green's function and UV divergences in
  entanglement entropy},}\ }\href {\doibase 10.1007/JHEP09(2010)041} {\bibfield
   {journal} {\bibinfo  {journal} {J. High Energ. Phys.}\ }\textbf {\bibinfo
  {volume} {2010}},\ \bibinfo {pages} {41} (\bibinfo {year}
  {2010})}\BibitemShut {NoStop}%
\bibitem [{Note1()}]{Note1}%
  \BibitemOpen
  \bibinfo {note} {This is only for free scalar field, and for free Dirac field
  this relation becomes $\partial _\mu \protect \qopname \relax o{ln}Z^{(n)} =
  - \protect \Tr G^{(n)}$.}\BibitemShut {Stop}%
\bibitem [{\citenamefont {Dowker}(1977)}]{Dowker_1977}%
  \BibitemOpen
  \bibfield  {author} {\bibinfo {author} {\bibfnamefont {J~S}\ \bibnamefont
  {Dowker}},\ }\bibfield  {title} {\enquote {\bibinfo {title} {Quantum field
  theory on a cone},}\ }\href {\doibase 10.1088/0305-4470/10/1/023} {\bibfield
  {journal} {\bibinfo  {journal} {J. Phys. A: Math. Gen.}\ }\textbf {\bibinfo
  {volume} {10}},\ \bibinfo {pages} {115--124} (\bibinfo {year}
  {1977})}\BibitemShut {NoStop}%
\bibitem [{\citenamefont {Dowker}(1978)}]{dowker1978}%
  \BibitemOpen
  \bibfield  {author} {\bibinfo {author} {\bibfnamefont {J.~S.}\ \bibnamefont
  {Dowker}},\ }\bibfield  {title} {\enquote {\bibinfo {title} {Thermal
  properties of Green's functions in Rindler, de Sitter, and Schwarzschild
  spaces},}\ }\href {\doibase 10.1103/PhysRevD.18.1856} {\bibfield  {journal}
  {\bibinfo  {journal} {Phys. Rev. D}\ }\textbf {\bibinfo {volume} {18}},\
  \bibinfo {pages} {1856--1860} (\bibinfo {year} {1978})}\BibitemShut {NoStop}%
\bibitem [{\citenamefont {Guimar{\~a}es}\ and\ \citenamefont
  {Linet}(1994)}]{Linet1994_scalar}%
  \BibitemOpen
  \bibfield  {author} {\bibinfo {author} {\bibfnamefont {M.~E.~X.}\
  \bibnamefont {Guimar{\~a}es}}\ and\ \bibinfo {author} {\bibfnamefont
  {B.}~\bibnamefont {Linet}},\ }\bibfield  {title} {\enquote {\bibinfo {title}
  {Scalar Green's functions in an Euclidean space with a conical-type line
  singularity},}\ }\href {\doibase 10.1007/BF02099773} {\bibfield  {journal}
  {\bibinfo  {journal} {Commun.Math. Phys.}\ }\textbf {\bibinfo {volume}
  {165}},\ \bibinfo {pages} {297--310} (\bibinfo {year} {1994})}\BibitemShut
  {NoStop}%
\bibitem [{\citenamefont {Linet}(1995)}]{Linet1995_spinor}%
  \BibitemOpen
  \bibfield  {author} {\bibinfo {author} {\bibfnamefont {B.}~\bibnamefont
  {Linet}},\ }\bibfield  {title} {\enquote {\bibinfo {title} {Euclidean spinor
  Green’s functions in the space–time of a straight cosmic string},}\
  }\href {\doibase 10.1063/1.530991} {\bibfield  {journal} {\bibinfo  {journal}
  {J. Math. Phys.}\ }\textbf {\bibinfo {volume} {36}},\ \bibinfo {pages}
  {3694--3703} (\bibinfo {year} {1995})}\BibitemShut {NoStop}%
\bibitem [{\citenamefont {Dowker}\ and\ \citenamefont
  {Chang}(1992)}]{dowker1992}%
  \BibitemOpen
  \bibfield  {author} {\bibinfo {author} {\bibfnamefont {J.~S.}\ \bibnamefont
  {Dowker}}\ and\ \bibinfo {author} {\bibfnamefont {Peter}\ \bibnamefont
  {Chang}},\ }\bibfield  {title} {\enquote {\bibinfo {title} {Polyhedral cosmic
  strings},}\ }\href {\doibase 10.1103/PhysRevD.46.3458} {\bibfield  {journal}
  {\bibinfo  {journal} {Phys. Rev. D}\ }\textbf {\bibinfo {volume} {46}},\
  \bibinfo {pages} {3458--3464} (\bibinfo {year} {1992})}\BibitemShut {NoStop}%
\bibitem [{\citenamefont {Chang}\ and\ \citenamefont
  {Dowker}(1993)}]{CHANG1993}%
  \BibitemOpen
  \bibfield  {author} {\bibinfo {author} {\bibfnamefont {Peter}\ \bibnamefont
  {Chang}}\ and\ \bibinfo {author} {\bibfnamefont {J.S.}\ \bibnamefont
  {Dowker}},\ }\bibfield  {title} {\enquote {\bibinfo {title} {Vacuum energy on
  orbifold factors of spheres},}\ }\href {\doibase
  https://doi.org/10.1016/0550-3213(93)90223-C} {\bibfield  {journal} {\bibinfo
   {journal} {Nuclear Physics B}\ }\textbf {\bibinfo {volume} {395}},\ \bibinfo
  {pages} {407--432} (\bibinfo {year} {1993})}\BibitemShut {NoStop}%
\bibitem [{\citenamefont {Fursaev}\ and\ \citenamefont
  {Miele}(1994)}]{fursaev1994}%
  \BibitemOpen
  \bibfield  {author} {\bibinfo {author} {\bibfnamefont {D.~V.}\ \bibnamefont
  {Fursaev}}\ and\ \bibinfo {author} {\bibfnamefont {G.}~\bibnamefont
  {Miele}},\ }\bibfield  {title} {\enquote {\bibinfo {title}
  {Finite-temperature scalar field theory in static de Sitter space},}\ }\href
  {\doibase 10.1103/PhysRevD.49.987} {\bibfield  {journal} {\bibinfo  {journal}
  {Phys. Rev. D}\ }\textbf {\bibinfo {volume} {49}},\ \bibinfo {pages}
  {987--998} (\bibinfo {year} {1994})}\BibitemShut {NoStop}%
\bibitem [{\citenamefont {Cognola}\ \emph {et~al.}(1994)\citenamefont
  {Cognola}, \citenamefont {Kirsten},\ and\ \citenamefont
  {Vanzo}}]{Cognola1994}%
  \BibitemOpen
  \bibfield  {author} {\bibinfo {author} {\bibfnamefont {Guido}\ \bibnamefont
  {Cognola}}, \bibinfo {author} {\bibfnamefont {Klaus}\ \bibnamefont
  {Kirsten}}, \ and\ \bibinfo {author} {\bibfnamefont {Luciano}\ \bibnamefont
  {Vanzo}},\ }\bibfield  {title} {\enquote {\bibinfo {title} {Free and
  self-interacting scalar fields in the presence of conical singularities},}\
  }\href {\doibase 10.1103/PhysRevD.49.1029} {\bibfield  {journal} {\bibinfo
  {journal} {Phys. Rev. D}\ }\textbf {\bibinfo {volume} {49}},\ \bibinfo
  {pages} {1029--1038} (\bibinfo {year} {1994})}\BibitemShut {NoStop}%
\bibitem [{\citenamefont {Aurell}\ and\ \citenamefont
  {Salomonson}(1994)}]{functional_determinant}%
  \BibitemOpen
  \bibfield  {author} {\bibinfo {author} {\bibfnamefont {Erik}\ \bibnamefont
  {Aurell}}\ and\ \bibinfo {author} {\bibfnamefont {Per}\ \bibnamefont
  {Salomonson}},\ }\bibfield  {title} {\enquote {\bibinfo {title} {On
  functional determinants of Laplacians in polygons and simplicial
  complexes},}\ }\href {\doibase 10.1007/BF02099770} {\bibfield  {journal}
  {\bibinfo  {journal} {Commun.Math. Phys.}\ }\textbf {\bibinfo {volume}
  {165}},\ \bibinfo {pages} {233--259} (\bibinfo {year} {1994})}\BibitemShut
  {NoStop}%
\bibitem [{\citenamefont {Bisognano}\ and\ \citenamefont
  {Wichmann}(1975)}]{bisognano1975}%
  \BibitemOpen
  \bibfield  {author} {\bibinfo {author} {\bibfnamefont {Joseph~J.}\
  \bibnamefont {Bisognano}}\ and\ \bibinfo {author} {\bibfnamefont {Eyvind~H.}\
  \bibnamefont {Wichmann}},\ }\bibfield  {title} {\enquote {\bibinfo {title}
  {On the duality condition for a Hermitian scalar field},}\ }\href {\doibase
  10.1063/1.522605} {\bibfield  {journal} {\bibinfo  {journal} {J. Math.
  Phys.}\ }\textbf {\bibinfo {volume} {16}},\ \bibinfo {pages} {985--1007}
  (\bibinfo {year} {1975})}\BibitemShut {NoStop}%
\bibitem [{\citenamefont {Bisognano}\ and\ \citenamefont
  {Wichmann}(1976)}]{bisognano1976}%
  \BibitemOpen
  \bibfield  {author} {\bibinfo {author} {\bibfnamefont {Joseph~J.}\
  \bibnamefont {Bisognano}}\ and\ \bibinfo {author} {\bibfnamefont {Eyvind~H.}\
  \bibnamefont {Wichmann}},\ }\bibfield  {title} {\enquote {\bibinfo {title}
  {On the duality condition for quantum fields},}\ }\href {\doibase
  10.1063/1.522898} {\bibfield  {journal} {\bibinfo  {journal} {J. Math.
  Phys.}\ }\textbf {\bibinfo {volume} {17}},\ \bibinfo {pages} {303--321}
  (\bibinfo {year} {1976})}\BibitemShut {NoStop}%
\bibitem [{\citenamefont {Hislop}\ and\ \citenamefont
  {Longo}(1982)}]{hislop1982_modular_scalar_CFT}%
  \BibitemOpen
  \bibfield  {author} {\bibinfo {author} {\bibfnamefont {Peter~D.}\
  \bibnamefont {Hislop}}\ and\ \bibinfo {author} {\bibfnamefont {Roberto}\
  \bibnamefont {Longo}},\ }\bibfield  {title} {\enquote {\bibinfo {title}
  {Modular structure of the local algebras associated with the free massless
  scalar field theory},}\ }\href {\doibase 10.1007/BF01208372} {\bibfield
  {journal} {\bibinfo  {journal} {Commun.Math. Phys.}\ }\textbf {\bibinfo
  {volume} {84}},\ \bibinfo {pages} {71--85} (\bibinfo {year}
  {1982})}\BibitemShut {NoStop}%
\bibitem [{\citenamefont {Maldacena}(1999)}]{Maldacena1999AdSCFT}%
  \BibitemOpen
  \bibfield  {author} {\bibinfo {author} {\bibfnamefont {Juan}\ \bibnamefont
  {Maldacena}},\ }\bibfield  {title} {\enquote {\bibinfo {title} {The large-N
  limit of superconformal field theories and supergravity},}\ }\href {\doibase
  10.1023/A:1026654312961} {\bibfield  {journal} {\bibinfo  {journal}
  {International Journal of Theoretical Physics}\ }\textbf {\bibinfo {volume}
  {38}},\ \bibinfo {pages} {1113--1133} (\bibinfo {year} {1999})}\BibitemShut
  {NoStop}%
\bibitem [{\citenamefont {Casini}\ \emph {et~al.}(2015)\citenamefont {Casini},
  \citenamefont {Huerta}, \citenamefont {Myers},\ and\ \citenamefont
  {Yale}}]{casini2015_EMI}%
  \BibitemOpen
  \bibfield  {author} {\bibinfo {author} {\bibfnamefont {Horacio}\ \bibnamefont
  {Casini}}, \bibinfo {author} {\bibfnamefont {Marina}\ \bibnamefont {Huerta}},
  \bibinfo {author} {\bibfnamefont {Robert~C.}\ \bibnamefont {Myers}}, \ and\
  \bibinfo {author} {\bibfnamefont {Alexandre}\ \bibnamefont {Yale}},\
  }\bibfield  {title} {\enquote {\bibinfo {title} {Mutual information and the
  F-theorem},}\ }\href {\doibase 10.1007/JHEP10(2015)003} {\bibfield  {journal}
  {\bibinfo  {journal} {J. High Energ. Phys.}\ }\textbf {\bibinfo {volume}
  {2015}},\ \bibinfo {pages} {3} (\bibinfo {year} {2015})}\BibitemShut
  {NoStop}%
\bibitem [{\citenamefont {Bueno}\ \emph {et~al.}(2019)\citenamefont {Bueno},
  \citenamefont {Casini},\ and\ \citenamefont
  {Witczak-Krempa}}]{bueno2019_EMI}%
  \BibitemOpen
  \bibfield  {author} {\bibinfo {author} {\bibfnamefont {Pablo}\ \bibnamefont
  {Bueno}}, \bibinfo {author} {\bibfnamefont {Horacio}\ \bibnamefont {Casini}},
  \ and\ \bibinfo {author} {\bibfnamefont {William}\ \bibnamefont
  {Witczak-Krempa}},\ }\bibfield  {title} {\enquote {\bibinfo {title}
  {Generalizing the entanglement entropy of singular regions in conformal field
  theories},}\ }\href {\doibase 10.1007/JHEP08(2019)069} {\bibfield  {journal}
  {\bibinfo  {journal} {J. High Energ. Phys.}\ }\textbf {\bibinfo {volume}
  {2019}},\ \bibinfo {pages} {69} (\bibinfo {year} {2019})}\BibitemShut
  {NoStop}%
\bibitem [{\citenamefont {Bueno}\ \emph {et~al.}(2021)\citenamefont {Bueno},
  \citenamefont {Casini}, \citenamefont {Andino},\ and\ \citenamefont
  {Moreno}}]{bueno2021_EMI}%
  \BibitemOpen
  \bibfield  {author} {\bibinfo {author} {\bibfnamefont {Pablo}\ \bibnamefont
  {Bueno}}, \bibinfo {author} {\bibfnamefont {Horacio}\ \bibnamefont {Casini}},
  \bibinfo {author} {\bibfnamefont {Oscar~Lasso}\ \bibnamefont {Andino}}, \
  and\ \bibinfo {author} {\bibfnamefont {Javier}\ \bibnamefont {Moreno}},\
  }\bibfield  {title} {\enquote {\bibinfo {title} {Disks globally maximize the
  entanglement entropy in 2 + 1 dimensions},}\ }\href {\doibase
  10.1007/JHEP10(2021)179} {\bibfield  {journal} {\bibinfo  {journal} {J. High
  Energ. Phys.}\ }\textbf {\bibinfo {volume} {2021}},\ \bibinfo {pages} {179}
  (\bibinfo {year} {2021})}\BibitemShut {NoStop}%
\bibitem [{\citenamefont
  {Swingle}(2012{\natexlab{b}})}]{swingle2012_fermi_liquid_EE}%
  \BibitemOpen
  \bibfield  {author} {\bibinfo {author} {\bibfnamefont {Brian}\ \bibnamefont
  {Swingle}},\ }\bibfield  {title} {\enquote {\bibinfo {title} {Conformal field
  theory approach to Fermi liquids and other highly entangled states},}\ }\href
  {\doibase 10.1103/PhysRevB.86.035116} {\bibfield  {journal} {\bibinfo
  {journal} {Phys. Rev. B}\ }\textbf {\bibinfo {volume} {86}},\ \bibinfo
  {pages} {035116} (\bibinfo {year} {2012}{\natexlab{b}})}\BibitemShut
  {NoStop}%
\bibitem [{\citenamefont {Murciano}\ \emph {et~al.}(2020)\citenamefont
  {Murciano}, \citenamefont {Ruggiero},\ and\ \citenamefont
  {Calabrese}}]{Murciano_2020}%
  \BibitemOpen
  \bibfield  {author} {\bibinfo {author} {\bibfnamefont {Sara}\ \bibnamefont
  {Murciano}}, \bibinfo {author} {\bibfnamefont {Paola}\ \bibnamefont
  {Ruggiero}}, \ and\ \bibinfo {author} {\bibfnamefont {Pasquale}\ \bibnamefont
  {Calabrese}},\ }\bibfield  {title} {\enquote {\bibinfo {title} {Symmetry
  resolved entanglement in two-dimensional systems via dimensional
  reduction},}\ }\href {\doibase 10.1088/1742-5468/aba1e5} {\bibfield
  {journal} {\bibinfo  {journal} {J. Stat. Mech.}\ }\textbf {\bibinfo {volume}
  {2020}},\ \bibinfo {pages} {083102} (\bibinfo {year} {2020})}\BibitemShut
  {NoStop}%
\bibitem [{\citenamefont {Shankar}(1994)}]{shankar1994RG}%
  \BibitemOpen
  \bibfield  {author} {\bibinfo {author} {\bibfnamefont {R.}~\bibnamefont
  {Shankar}},\ }\bibfield  {title} {\enquote {\bibinfo {title}
  {Renormalization-group approach to interacting fermions},}\ }\href {\doibase
  10.1103/RevModPhys.66.129} {\bibfield  {journal} {\bibinfo  {journal} {Rev.
  Mod. Phys.}\ }\textbf {\bibinfo {volume} {66}},\ \bibinfo {pages} {129--192}
  (\bibinfo {year} {1994})}\BibitemShut {NoStop}%
\bibitem [{\citenamefont {Casini}\ and\ \citenamefont
  {Huerta}(2007)}]{CASINI2007_2p1scalar}%
  \BibitemOpen
  \bibfield  {author} {\bibinfo {author} {\bibfnamefont {H.}~\bibnamefont
  {Casini}}\ and\ \bibinfo {author} {\bibfnamefont {M.}~\bibnamefont
  {Huerta}},\ }\bibfield  {title} {\enquote {\bibinfo {title} {Universal terms
  for the entanglement entropy in 2+1 dimensions},}\ }\href {\doibase
  https://doi.org/10.1016/j.nuclphysb.2006.12.012} {\bibfield  {journal}
  {\bibinfo  {journal} {Nuclear Physics B}\ }\textbf {\bibinfo {volume}
  {764}},\ \bibinfo {pages} {183--201} (\bibinfo {year} {2007})}\BibitemShut
  {NoStop}%
\bibitem [{\citenamefont {Casini}\ \emph {et~al.}(2009)\citenamefont {Casini},
  \citenamefont {Huerta},\ and\ \citenamefont {Leitao}}]{CASINI2009_2p1dirac}%
  \BibitemOpen
  \bibfield  {author} {\bibinfo {author} {\bibfnamefont {H.}~\bibnamefont
  {Casini}}, \bibinfo {author} {\bibfnamefont {M.}~\bibnamefont {Huerta}}, \
  and\ \bibinfo {author} {\bibfnamefont {L.}~\bibnamefont {Leitao}},\
  }\bibfield  {title} {\enquote {\bibinfo {title} {Entanglement entropy for a
  Dirac fermion in three dimensions: Vertex contribution},}\ }\href {\doibase
  https://doi.org/10.1016/j.nuclphysb.2009.02.003} {\bibfield  {journal}
  {\bibinfo  {journal} {Nuclear Physics B}\ }\textbf {\bibinfo {volume}
  {814}},\ \bibinfo {pages} {594--609} (\bibinfo {year} {2009})}\BibitemShut
  {NoStop}%
\bibitem [{\citenamefont {Ye}(1999)}]{Ye1999}%
  \BibitemOpen
  \bibfield  {author} {\bibinfo {author} {\bibfnamefont {Jinwu}\ \bibnamefont
  {Ye}},\ }\bibfield  {title} {\enquote {\bibinfo {title} {Effects of weak
  disorders on quantum Hall critical points},}\ }\href {\doibase
  10.1103/PhysRevB.60.8290} {\bibfield  {journal} {\bibinfo  {journal} {Phys.
  Rev. B}\ }\textbf {\bibinfo {volume} {60}},\ \bibinfo {pages} {8290--8303}
  (\bibinfo {year} {1999})}\BibitemShut {NoStop}%
\bibitem [{\citenamefont {Fradkin}(1986)}]{fradkin1986disorder}%
  \BibitemOpen
  \bibfield  {author} {\bibinfo {author} {\bibfnamefont {Eduardo}\ \bibnamefont
  {Fradkin}},\ }\bibfield  {title} {\enquote {\bibinfo {title} {Critical
  behavior of disordered degenerate semiconductors. i. models, symmetries, and
  formalism},}\ }\href {\doibase 10.1103/PhysRevB.33.3257} {\bibfield
  {journal} {\bibinfo  {journal} {Phys. Rev. B}\ }\textbf {\bibinfo {volume}
  {33}},\ \bibinfo {pages} {3257--3262} (\bibinfo {year} {1986})}\BibitemShut
  {NoStop}%
\bibitem [{\citenamefont {Castillo}\ \emph {et~al.}(1997)\citenamefont
  {Castillo}, \citenamefont {de~C.~Chamon}, \citenamefont {Fradkin},
  \citenamefont {Goldbart},\ and\ \citenamefont
  {Mudry}}]{Fradkin1997multifractal}%
  \BibitemOpen
  \bibfield  {author} {\bibinfo {author} {\bibfnamefont {Horacio~E.}\
  \bibnamefont {Castillo}}, \bibinfo {author} {\bibfnamefont {Claudio}\
  \bibnamefont {de~C.~Chamon}}, \bibinfo {author} {\bibfnamefont {Eduardo}\
  \bibnamefont {Fradkin}}, \bibinfo {author} {\bibfnamefont {Paul~M.}\
  \bibnamefont {Goldbart}}, \ and\ \bibinfo {author} {\bibfnamefont
  {Christopher}\ \bibnamefont {Mudry}},\ }\bibfield  {title} {\enquote
  {\bibinfo {title} {Exact calculation of multifractal exponents of the
  critical wave function of Dirac fermions in a random magnetic field},}\
  }\href {\doibase 10.1103/PhysRevB.56.10668} {\bibfield  {journal} {\bibinfo
  {journal} {Phys. Rev. B}\ }\textbf {\bibinfo {volume} {56}},\ \bibinfo
  {pages} {10668--10677} (\bibinfo {year} {1997})}\BibitemShut {NoStop}%
\bibitem [{\citenamefont {Anderson}(1958)}]{anderson1958local}%
  \BibitemOpen
  \bibfield  {author} {\bibinfo {author} {\bibfnamefont {P.~W.}\ \bibnamefont
  {Anderson}},\ }\bibfield  {title} {\enquote {\bibinfo {title} {Absence of
  diffusion in certain random lattices},}\ }\href {\doibase
  10.1103/PhysRev.109.1492} {\bibfield  {journal} {\bibinfo  {journal} {Phys.
  Rev.}\ }\textbf {\bibinfo {volume} {109}},\ \bibinfo {pages} {1492--1505}
  (\bibinfo {year} {1958})}\BibitemShut {NoStop}%
\bibitem [{\citenamefont {Wegner}(1979)}]{wegner1979disorder}%
  \BibitemOpen
  \bibfield  {author} {\bibinfo {author} {\bibfnamefont {Franz~J.}\
  \bibnamefont {Wegner}},\ }\bibfield  {title} {\enquote {\bibinfo {title}
  {Disordered system with $n$ orbitals per site: $n=\ensuremath{\infty}$
  limit},}\ }\href {\doibase 10.1103/PhysRevB.19.783} {\bibfield  {journal}
  {\bibinfo  {journal} {Phys. Rev. B}\ }\textbf {\bibinfo {volume} {19}},\
  \bibinfo {pages} {783--792} (\bibinfo {year} {1979})}\BibitemShut {NoStop}%
\bibitem [{\citenamefont {Abrahams}\ \emph {et~al.}(1979)\citenamefont
  {Abrahams}, \citenamefont {Anderson}, \citenamefont {Licciardello},\ and\
  \citenamefont {Ramakrishnan}}]{abrahams1979localization}%
  \BibitemOpen
  \bibfield  {author} {\bibinfo {author} {\bibfnamefont {E.}~\bibnamefont
  {Abrahams}}, \bibinfo {author} {\bibfnamefont {P.~W.}\ \bibnamefont
  {Anderson}}, \bibinfo {author} {\bibfnamefont {D.~C.}\ \bibnamefont
  {Licciardello}}, \ and\ \bibinfo {author} {\bibfnamefont {T.~V.}\
  \bibnamefont {Ramakrishnan}},\ }\bibfield  {title} {\enquote {\bibinfo
  {title} {Scaling theory of localization: Absence of quantum diffusion in two
  dimensions},}\ }\href {\doibase 10.1103/PhysRevLett.42.673} {\bibfield
  {journal} {\bibinfo  {journal} {Phys. Rev. Lett.}\ }\textbf {\bibinfo
  {volume} {42}},\ \bibinfo {pages} {673--676} (\bibinfo {year}
  {1979})}\BibitemShut {NoStop}%
\bibitem [{\citenamefont {Kramer}\ and\ \citenamefont
  {MacKinnon}(1993)}]{Kramer1993Localization}%
  \BibitemOpen
  \bibfield  {author} {\bibinfo {author} {\bibfnamefont {B}~\bibnamefont
  {Kramer}}\ and\ \bibinfo {author} {\bibfnamefont {A}~\bibnamefont
  {MacKinnon}},\ }\bibfield  {title} {\enquote {\bibinfo {title} {Localization:
  theory and experiment},}\ }\href {\doibase 10.1088/0034-4885/56/12/001}
  {\bibfield  {journal} {\bibinfo  {journal} {Rep. Prog. Phys.}\ }\textbf
  {\bibinfo {volume} {56}},\ \bibinfo {pages} {1469--1564} (\bibinfo {year}
  {1993})}\BibitemShut {NoStop}%
\bibitem [{\citenamefont {Evers}\ and\ \citenamefont
  {Mirlin}(2008)}]{mirlin2008review_AndersonTransition}%
  \BibitemOpen
  \bibfield  {author} {\bibinfo {author} {\bibfnamefont {Ferdinand}\
  \bibnamefont {Evers}}\ and\ \bibinfo {author} {\bibfnamefont {Alexander~D.}\
  \bibnamefont {Mirlin}},\ }\bibfield  {title} {\enquote {\bibinfo {title}
  {Anderson transitions},}\ }\href {\doibase 10.1103/RevModPhys.80.1355}
  {\bibfield  {journal} {\bibinfo  {journal} {Rev. Mod. Phys.}\ }\textbf
  {\bibinfo {volume} {80}},\ \bibinfo {pages} {1355--1417} (\bibinfo {year}
  {2008})}\BibitemShut {NoStop}%
\bibitem [{\citenamefont {MacKinnon}\ and\ \citenamefont
  {Kramer}(1981)}]{kramer1981localization}%
  \BibitemOpen
  \bibfield  {author} {\bibinfo {author} {\bibfnamefont {A.}~\bibnamefont
  {MacKinnon}}\ and\ \bibinfo {author} {\bibfnamefont {B.}~\bibnamefont
  {Kramer}},\ }\bibfield  {title} {\enquote {\bibinfo {title} {One-parameter
  scaling of localization length and conductance in disordered systems},}\
  }\href {\doibase 10.1103/PhysRevLett.47.1546} {\bibfield  {journal} {\bibinfo
   {journal} {Phys. Rev. Lett.}\ }\textbf {\bibinfo {volume} {47}},\ \bibinfo
  {pages} {1546--1549} (\bibinfo {year} {1981})}\BibitemShut {NoStop}%
\bibitem [{\citenamefont {Pruisken}\ and\ \citenamefont
  {Sch\"afer}(1981)}]{Pruisken1981Anderson}%
  \BibitemOpen
  \bibfield  {author} {\bibinfo {author} {\bibfnamefont {Adrianus M.~M.}\
  \bibnamefont {Pruisken}}\ and\ \bibinfo {author} {\bibfnamefont {Lothar}\
  \bibnamefont {Sch\"afer}},\ }\bibfield  {title} {\enquote {\bibinfo {title}
  {Field theory and the Anderson model for disordered electronic systems},}\
  }\href {\doibase 10.1103/PhysRevLett.46.490} {\bibfield  {journal} {\bibinfo
  {journal} {Phys. Rev. Lett.}\ }\textbf {\bibinfo {volume} {46}},\ \bibinfo
  {pages} {490--492} (\bibinfo {year} {1981})}\BibitemShut {NoStop}%
\bibitem [{\citenamefont {Savary}\ and\ \citenamefont
  {Balents}(2016)}]{savary2016_spin_liquid}%
  \BibitemOpen
  \bibfield  {author} {\bibinfo {author} {\bibfnamefont {Lucile}\ \bibnamefont
  {Savary}}\ and\ \bibinfo {author} {\bibfnamefont {Leon}\ \bibnamefont
  {Balents}},\ }\bibfield  {title} {\enquote {\bibinfo {title} {Quantum spin
  liquids: A review},}\ }\href {\doibase 10.1088/0034-4885/80/1/016502}
  {\bibfield  {journal} {\bibinfo  {journal} {Rep. Prog. Phys.}\ }\textbf
  {\bibinfo {volume} {80}},\ \bibinfo {pages} {016502} (\bibinfo {year}
  {2016})}\BibitemShut {NoStop}%
\bibitem [{\citenamefont {Vozmediano}\ \emph {et~al.}(2010)\citenamefont
  {Vozmediano}, \citenamefont {Katsnelson},\ and\ \citenamefont
  {Guinea}}]{guinea2010graphene}%
  \BibitemOpen
  \bibfield  {author} {\bibinfo {author} {\bibfnamefont {M.A.H.}\ \bibnamefont
  {Vozmediano}}, \bibinfo {author} {\bibfnamefont {M.I.}\ \bibnamefont
  {Katsnelson}}, \ and\ \bibinfo {author} {\bibfnamefont {F.}~\bibnamefont
  {Guinea}},\ }\bibfield  {title} {\enquote {\bibinfo {title} {Gauge fields in
  graphene},}\ }\href {\doibase https://doi.org/10.1016/j.physrep.2010.07.003}
  {\bibfield  {journal} {\bibinfo  {journal} {Physics Reports}\ }\textbf
  {\bibinfo {volume} {496}},\ \bibinfo {pages} {109--148} (\bibinfo {year}
  {2010})}\BibitemShut {NoStop}%
\bibitem [{\citenamefont {Mudry}\ \emph {et~al.}(2003)\citenamefont {Mudry},
  \citenamefont {Ryu},\ and\ \citenamefont
  {Furusaki}}]{mudry2003_random_hopping}%
  \BibitemOpen
  \bibfield  {author} {\bibinfo {author} {\bibfnamefont {C.}~\bibnamefont
  {Mudry}}, \bibinfo {author} {\bibfnamefont {S.}~\bibnamefont {Ryu}}, \ and\
  \bibinfo {author} {\bibfnamefont {A.}~\bibnamefont {Furusaki}},\ }\bibfield
  {title} {\enquote {\bibinfo {title} {Density of states for the
  $\ensuremath{\pi}$-flux state with bipartite real random hopping only: A weak
  disorder approach},}\ }\href {\doibase 10.1103/PhysRevB.67.064202} {\bibfield
   {journal} {\bibinfo  {journal} {Phys. Rev. B}\ }\textbf {\bibinfo {volume}
  {67}},\ \bibinfo {pages} {064202} (\bibinfo {year} {2003})}\BibitemShut
  {NoStop}%
\bibitem [{\citenamefont {Ryu}\ and\ \citenamefont
  {Hatsugai}(2001)}]{ryu2001_random_flux}%
  \BibitemOpen
  \bibfield  {author} {\bibinfo {author} {\bibfnamefont {Shinsei}\ \bibnamefont
  {Ryu}}\ and\ \bibinfo {author} {\bibfnamefont {Yasuhiro}\ \bibnamefont
  {Hatsugai}},\ }\bibfield  {title} {\enquote {\bibinfo {title} {Singular
  density of states of disordered Dirac fermions in chiral models},}\ }\href
  {\doibase 10.1103/PhysRevB.65.033301} {\bibfield  {journal} {\bibinfo
  {journal} {Phys. Rev. B}\ }\textbf {\bibinfo {volume} {65}},\ \bibinfo
  {pages} {033301} (\bibinfo {year} {2001})}\BibitemShut {NoStop}%
\bibitem [{\citenamefont
  {Zamolodchikov}(1986)}]{Zamolodchikov1986Irreversibility}%
  \BibitemOpen
  \bibfield  {author} {\bibinfo {author} {\bibfnamefont {Alexander~B.}\
  \bibnamefont {Zamolodchikov}},\ }\bibfield  {title} {\enquote {\bibinfo
  {title} {“Irreversibility” of the flux of the renormalization group in a
  2d field theory},}\ }\href
  {http://www.jetpletters.ru/ps/1413/article_21504.shtml} {\bibfield  {journal}
  {\bibinfo  {journal} {JETP Letters}\ }\textbf {\bibinfo {volume} {43}},\
  \bibinfo {pages} {730--732} (\bibinfo {year} {1986})}\BibitemShut {NoStop}%
\bibitem [{\citenamefont {Cardy}(1988)}]{CARDY1988_4d_c_theorem}%
  \BibitemOpen
  \bibfield  {author} {\bibinfo {author} {\bibfnamefont {John~L.}\ \bibnamefont
  {Cardy}},\ }\bibfield  {title} {\enquote {\bibinfo {title} {Is there a
  c-theorem in four dimensions?}}\ }\href {\doibase
  https://doi.org/10.1016/0370-2693(88)90054-8} {\bibfield  {journal} {\bibinfo
   {journal} {Physics Letters B}\ }\textbf {\bibinfo {volume} {215}},\ \bibinfo
  {pages} {749--752} (\bibinfo {year} {1988})}\BibitemShut {NoStop}%
\bibitem [{\citenamefont {Giombi}\ \emph {et~al.}(2016)\citenamefont {Giombi},
  \citenamefont {Klebanov},\ and\ \citenamefont {Tarnopolsky}}]{Giombi_2016}%
  \BibitemOpen
  \bibfield  {author} {\bibinfo {author} {\bibfnamefont {Simone}\ \bibnamefont
  {Giombi}}, \bibinfo {author} {\bibfnamefont {Igor~R}\ \bibnamefont
  {Klebanov}}, \ and\ \bibinfo {author} {\bibfnamefont {Grigory}\ \bibnamefont
  {Tarnopolsky}},\ }\bibfield  {title} {\enquote {\bibinfo {title} {Conformal
  {QED${}_d$}, F-theorem and the $\varepsilon$ expansion},}\ }\href {\doibase
  10.1088/1751-8113/49/13/135403} {\bibfield  {journal} {\bibinfo  {journal}
  {J. Phys. A: Math. Theor.}\ }\textbf {\bibinfo {volume} {49}},\ \bibinfo
  {pages} {135403} (\bibinfo {year} {2016})}\BibitemShut {NoStop}%
\bibitem [{\citenamefont {Grover}\ \emph {et~al.}(2011)\citenamefont {Grover},
  \citenamefont {Turner},\ and\ \citenamefont
  {Vishwanath}}]{Grover2011_3dEE_topo}%
  \BibitemOpen
  \bibfield  {author} {\bibinfo {author} {\bibfnamefont {Tarun}\ \bibnamefont
  {Grover}}, \bibinfo {author} {\bibfnamefont {Ari~M.}\ \bibnamefont {Turner}},
  \ and\ \bibinfo {author} {\bibfnamefont {Ashvin}\ \bibnamefont
  {Vishwanath}},\ }\bibfield  {title} {\enquote {\bibinfo {title} {Entanglement
  entropy of gapped phases and topological order in three dimensions},}\ }\href
  {\doibase 10.1103/PhysRevB.84.195120} {\bibfield  {journal} {\bibinfo
  {journal} {Phys. Rev. B}\ }\textbf {\bibinfo {volume} {84}},\ \bibinfo
  {pages} {195120} (\bibinfo {year} {2011})}\BibitemShut {NoStop}%
\bibitem [{\citenamefont {Grover}(2014)}]{Grover2014_3dEE}%
  \BibitemOpen
  \bibfield  {author} {\bibinfo {author} {\bibfnamefont {Tarun}\ \bibnamefont
  {Grover}},\ }\bibfield  {title} {\enquote {\bibinfo {title} {Entanglement
  monotonicity and the stability of gauge theories in three spacetime
  dimensions},}\ }\href {\doibase 10.1103/PhysRevLett.112.151601} {\bibfield
  {journal} {\bibinfo  {journal} {Phys. Rev. Lett.}\ }\textbf {\bibinfo
  {volume} {112}},\ \bibinfo {pages} {151601} (\bibinfo {year}
  {2014})}\BibitemShut {NoStop}%
\bibitem [{\citenamefont {Ancarani}\ and\ \citenamefont
  {Gasaneo}(2010)}]{Ancarani_2010}%
  \BibitemOpen
  \bibfield  {author} {\bibinfo {author} {\bibfnamefont {L~U}\ \bibnamefont
  {Ancarani}}\ and\ \bibinfo {author} {\bibfnamefont {G}~\bibnamefont
  {Gasaneo}},\ }\bibfield  {title} {\enquote {\bibinfo {title} {Derivatives of
  any order of the hypergeometric function ${}_pF_q(a_1, \dots,
  a_p;b_1,\dots,b_q;z)$ with respect to the parameters $a_i$ and $b_i$},}\
  }\href {\doibase 10.1088/1751-8113/43/8/085210} {\bibfield  {journal}
  {\bibinfo  {journal} {J. Phys. A: Math. Theor.}\ }\textbf {\bibinfo {volume}
  {43}},\ \bibinfo {pages} {085210} (\bibinfo {year} {2010})}\BibitemShut
  {NoStop}%
\bibitem [{\citenamefont {Lee}\ and\ \citenamefont
  {Wang}(1996)}]{LeeDH_1996_interactIQHE}%
  \BibitemOpen
  \bibfield  {author} {\bibinfo {author} {\bibfnamefont {Dung-Hai}\
  \bibnamefont {Lee}}\ and\ \bibinfo {author} {\bibfnamefont {Ziqiang}\
  \bibnamefont {Wang}},\ }\bibfield  {title} {\enquote {\bibinfo {title}
  {Effects of electron-electron interactions on the integer quantum Hall
  transitions},}\ }\href {\doibase 10.1103/PhysRevLett.76.4014} {\bibfield
  {journal} {\bibinfo  {journal} {Phys. Rev. Lett.}\ }\textbf {\bibinfo
  {volume} {76}},\ \bibinfo {pages} {4014--4017} (\bibinfo {year}
  {1996})}\BibitemShut {NoStop}%
\bibitem [{\citenamefont {Herbut}(2001)}]{Herbut_2001_Coulomb_zeq1}%
  \BibitemOpen
  \bibfield  {author} {\bibinfo {author} {\bibfnamefont {Igor~F.}\ \bibnamefont
  {Herbut}},\ }\bibfield  {title} {\enquote {\bibinfo {title} {Quantum critical
  points with the Coulomb interaction and the dynamical exponent: When and why
  $\mathit{z} = 1$},}\ }\href {\doibase 10.1103/PhysRevLett.87.137004}
  {\bibfield  {journal} {\bibinfo  {journal} {Phys. Rev. Lett.}\ }\textbf
  {\bibinfo {volume} {87}},\ \bibinfo {pages} {137004} (\bibinfo {year}
  {2001})}\BibitemShut {NoStop}%
\bibitem [{\citenamefont {Wang}\ and\ \citenamefont
  {Xiong}(2002)}]{Wang_2002_interactIQHE}%
  \BibitemOpen
  \bibfield  {author} {\bibinfo {author} {\bibfnamefont {Ziqiang}\ \bibnamefont
  {Wang}}\ and\ \bibinfo {author} {\bibfnamefont {Shanhui}\ \bibnamefont
  {Xiong}},\ }\bibfield  {title} {\enquote {\bibinfo {title} {Electron-electron
  interactions, quantum Coulomb gap, and dynamical scaling near integer quantum
  Hall transitions},}\ }\href {\doibase 10.1103/PhysRevB.65.195316} {\bibfield
  {journal} {\bibinfo  {journal} {Phys. Rev. B}\ }\textbf {\bibinfo {volume}
  {65}},\ \bibinfo {pages} {195316} (\bibinfo {year} {2002})}\BibitemShut
  {NoStop}%
\bibitem [{\citenamefont {Stauber}\ \emph {et~al.}(2005)\citenamefont
  {Stauber}, \citenamefont {Guinea},\ and\ \citenamefont
  {Vozmediano}}]{Stauber_2005_disorder_interaction}%
  \BibitemOpen
  \bibfield  {author} {\bibinfo {author} {\bibfnamefont {T.}~\bibnamefont
  {Stauber}}, \bibinfo {author} {\bibfnamefont {F.}~\bibnamefont {Guinea}}, \
  and\ \bibinfo {author} {\bibfnamefont {M.~A.~H.}\ \bibnamefont
  {Vozmediano}},\ }\bibfield  {title} {\enquote {\bibinfo {title} {Disorder and
  interaction effects in two-dimensional graphene sheets},}\ }\href {\doibase
  10.1103/PhysRevB.71.041406} {\bibfield  {journal} {\bibinfo  {journal} {Phys.
  Rev. B}\ }\textbf {\bibinfo {volume} {71}},\ \bibinfo {pages} {041406}
  (\bibinfo {year} {2005})}\BibitemShut {NoStop}%
\bibitem [{\citenamefont {Herbut}\ \emph {et~al.}(2008)\citenamefont {Herbut},
  \citenamefont {Juri\ifmmode \check{c}\else \v{c}\fi{}i\ifmmode~\acute{c}\else
  \'{c}\fi{}},\ and\ \citenamefont
  {Vafek}}]{Herbut_2008_Coulomb_MinimalConduct}%
  \BibitemOpen
  \bibfield  {author} {\bibinfo {author} {\bibfnamefont {Igor~F.}\ \bibnamefont
  {Herbut}}, \bibinfo {author} {\bibfnamefont {Vladimir}\ \bibnamefont
  {Juri\ifmmode \check{c}\else \v{c}\fi{}i\ifmmode~\acute{c}\else \'{c}\fi{}}},
  \ and\ \bibinfo {author} {\bibfnamefont {Oskar}\ \bibnamefont {Vafek}},\
  }\bibfield  {title} {\enquote {\bibinfo {title} {Coulomb interaction,
  ripples, and the minimal conductivity of graphene},}\ }\href {\doibase
  10.1103/PhysRevLett.100.046403} {\bibfield  {journal} {\bibinfo  {journal}
  {Phys. Rev. Lett.}\ }\textbf {\bibinfo {volume} {100}},\ \bibinfo {pages}
  {046403} (\bibinfo {year} {2008})}\BibitemShut {NoStop}%
\bibitem [{\citenamefont {Vafek}\ and\ \citenamefont
  {Case}(2008)}]{Vafek_2008_RG_Coulomb_RandGauge}%
  \BibitemOpen
  \bibfield  {author} {\bibinfo {author} {\bibfnamefont {Oskar}\ \bibnamefont
  {Vafek}}\ and\ \bibinfo {author} {\bibfnamefont {Matthew~J.}\ \bibnamefont
  {Case}},\ }\bibfield  {title} {\enquote {\bibinfo {title} {Renormalization
  group approach to two-dimensional Coulomb interacting Dirac fermions with
  random gauge potential},}\ }\href {\doibase 10.1103/PhysRevB.77.033410}
  {\bibfield  {journal} {\bibinfo  {journal} {Phys. Rev. B}\ }\textbf {\bibinfo
  {volume} {77}},\ \bibinfo {pages} {033410} (\bibinfo {year}
  {2008})}\BibitemShut {NoStop}%
\end{thebibliography}%

\end{document}